\documentclass[11pt]{article}
\pdfoutput=1 

\usepackage{jcappub} 

\usepackage[T1]{fontenc} 
\usepackage{appendix}

\usepackage{pgfplots}
\usepackage{physics}
\usepackage{amsmath}
\usepackage{bbm}
\usepackage{xcolor}
\usepackage[normalem]{ulem}

\renewcommand{\k}{\mathbf{k}}

\allowdisplaybreaks[4]

\title{\boldmath One-point matter PDFs beyond TopHat filters}

\author[a]{Anton Chudaykin,}
\author[b]{Alexander M. Kayssi,}
\author[b,c]{Sergey Sibiryakov}

\affiliation[a]{D\'epartement de Physique Th\'eorique and Center for Astroparticle Physics,\\
Universit\'e de Gen\`eve, 24 quai Ernest  Ansermet, 1211 Gen\`eve 4, Switzerland}
\affiliation[b]{Department of Physics \& Astronomy, McMaster University, \\ 1280 Main Street West, Hamilton, ON L8S 4M1, Canada}
\affiliation[c]{Perimeter Institute for Theoretical Physics, \\
31 Caroline Street North, Waterloo, ON N2L 2Y5, Canada}

\emailAdd{anton.chudaykin@unige.ch}
\emailAdd{kayssia@mcmaster.ca}
\emailAdd{ssibiryakov@perimeterinstitute.ca}

\abstract{We study the one-point probability distribution function (PDF) for matter densities averaged with an arbitrary spherically symmetric window function. The PDF is analytically modeled within the path integral framework, enabling a non-perturbative description of large-scale structure. It contains a leading order contribution controlled by the spherically symmetric gravitational collapse dynamics, as well as an order-one factor arising from aspherical fluctuations. 
We develop a numerical pipeline to compute the leading spherical-collapse part of the PDF and apply it to a family of window functions interpolating between the TopHat and Gaussian filters in coordinate space, as well as to a window function with non-monotonic radial dependence.
We find that the PDF weakly depends on the choice of the filter, provided the width of the filter is normalized to yield a fixed linear averaged density variance. For each filter and each value of the averaged density, our pipeline gives the most probable density profile. We find that these profiles vastly differ for different filters in the case of overdensities, but closely follow a universal curve at underdensities. We obtain a perturbative expression for the aspherical part of the PDF valid at small density contrasts. We find from it that all PDFs are equally sensitive to the effective field theory (EFT) corrections accounting for short-scale clustering, regardless of how smooth the filter's boundary is. 
We test our PDF model against the results of high-resolution N-body simulations. The agreement is excellent for filters with widths $\gtrsim 10\,{\rm Mpc}/h$. Small discrepancies at a few percent level arise for narrower filters and are interpreted as higher-order perturbative corrections.
}

\begin{document}
\maketitle
\flushbottom
\section{Introduction}
\label{sec:intro}
Recently, there has been an abundance of observational data coming from the growing number of cosmological surveys mapping the Large-Scale-Structure (LSS) of the Universe~\cite{DES:2021wwk,Dalal:2023olq,Wright:2025xka,DESI:2024hhd,DESI:2025zgx}. This data carry information on cosmological parameters, dark matter properties, and fundamental physics occurring in the early universe. The importance of this data becomes even more apparent as we are entering the era of high-precision cosmology where uncertainties in measurements are at the sub-percent level \cite{turner}. In order to extract useful information from the data, one needs an accurate understanding of the non-linear collapse dynamics of matter within the specified cosmology. 

While numerical simulations have commonly been used to address this question, progress into analytic methods has been developing at a quick rate in an attempt to reduce computation strain and provide a fundamental understanding of the physics at play. 
Effective field theory of large scale structure (EFTofLSS) has been a highly successful approach capable of doing this \cite{baumann,Carrasco:2012cv} (see \cite{EFT_ivanov} for a review). It extends the standard cosmological perturbation theory \cite{bernardeauSPT} (SPT) by including the effects of highly non-linear dynamics at short scales through a systematic derivative expansion \cite{snowmass}.  
This allows one to get reliable analytical predictions at mildly non-linear scales corresponding to comoving wavenumbers $k_{\text{NL}}\lesssim 0.3\,h\text{Mpc}^{-1}$. Currently, EFTofLSS has become a well-developed framework with established renormalization procedure \cite{renorm1, renorm2,twoloop}. It has been applied to the full-shape analysis of the power spectrum data~\cite{Carrasco:2012cv,Carrasco:2013mua,Baldauf:2015aha}, bispectrum~\cite{Angulo:2014tfa,Baldauf:2014qfa,Steele:2020tak}, and higher-order correlators~\cite{Bertolini:2016bmt,Steele:2021lnz}.~\footnote{For a recent application of the EFTofLSS approach to the DESI full-shape data including galaxy bispectrum see Ref.~\cite{Chudaykin:2025aux,Chudaykin:2025lww,Chudaykin:2025vdh,Ivanov:2026dvl,Chudaykin:2026nls,NovellMasot:2025fju}.} 
However, the complexity of EFTofLSS for higher-order correlation functions grows rapidly. 
This underscores the importance for new methods to assess the non-Gaussian information encoded in the non-linear clustering regime probed by current and future surveys. 

In this paper, we examine a non-perturbative approach to compute one-point probability distribution function (PDF) of the matter density field. This observable is closely related to counts-in-cells (CiC) statistics \cite{peebles}. In this formalism, one divides up the density field into spherical cells of radius $r_*$ then computes the averages of the field in each cell. Given a collection of cell averaged densities, a PDF for measuring a given density can be constructed. Notably, the deviation of the averaged density from the mean density of the universe does not need to be small, hence the PDF carries information about non-perturbative dynamics, and encodes in a cumulative way information about all higher-order correlation functions \cite{ivanov,anton}.

Up to now, most studies have focused on the PDF for the total mass inside the cell, which corresponds to averaging the density with a TopHat filter. A path integral approach to the 1-point PDF was initiated in Refs.~\cite{Valageas:2001zr,Valageas:2001td} and developed in
\cite{ivanov,anton}. In this approach, the PDF is described as a path integral over all possible density configurations, which is evaluated in the saddle-point approximation. When working order-by-order, it was found that the variance of the density field within a cell plays the role of the expansion parameter, while the saddle-point solution dictates the exponential behavior of the PDF. Next-to-leading order corrections contribute to the prefactor of the PDF. In the language of perturbation theory, they corresponding to the one-loop corrections in the background of the saddle-point configuration. A similar but methodologically distinct approach~\cite{Uhlemann:2015npz} is based on applying a logarithmic transformation to the density field in order to improve the accuracy of the method. Although the validity of this transformation is not strictly defined, it has been shown to successfully reproduce the results of simulations~\cite{Uhlemann:2019gni,Boyle:2020bqn,Friedrich:2021xff}.

To access a more detailed information about the matter distribution across the cell, one can generalize the PDF statistics by considering a weighted average of the density field in the cell \cite{generalW}.
For example, applying a Gaussian weight would prioritize the matter distribution closer to the center of the cell, and less near its boundary. 
On the other hand, a filter peaking at the boundary, will increase the sensitivity to wall-like structures surrounding the cell.
Besides, for a general filter, the PDF will not only depend on the matter content within the cell, but also on the density in its neighborhood, as seen in the Gaussian case, thereby characterizing the structure in connection to its environment. 

There are several other motivations to go beyond the TopHat filter. 
The latter introduces a sharp boundary of the cell in position space and thus contains a broad spectrum of momenta when expanded in Fourier modes. This complicates connection to 
other well-known approaches to non-linear structure formation, such as excursion set theory \cite{bond,zentner} which relies on filters with sharp cutoff in the momentum space, and halo clustering models \cite{halo_gauss} which prefer Gaussians. 
Recent works studying the effects of different window functions on the mass function of primordial black holes \cite{BH_mass} and galaxy parameters \cite{galaxy_window} also suggest an interest in using filters other than the TopHat. 

Finally, as discussed in \cite{ivanov,anton}, the TopHat PDF is significantly affected\footnote{Up to $\sim 30\%$ corrections at the tails of the distribution.} 
by the short-distance physics. In the spirit of EFTofLSS, this is taken into account by introducing free {\it counterterm} parameters that must be fitted from the data. It is worth studying how different choices of the filter affect the EFT corrections and whether some choices can reduce their magnitude.

The purpose of this paper is to generalize the path integral approach of \cite{ivanov,anton} to the case of an arbitrary filter. While the TopHat case
greatly simplifies the calculations, resulting in an analytic expression for the saddle-point configuration, this is no longer true for general filters. We develop a numerical pipeline to find the saddle-point solution given an arbitrary filter and determine the leading exponential behavior of the PDF. We also numerically evaluate the dominant contribution into the prefactor coming from spherically symmetric fluctuations (we refer to this contribution as `monopole prefactor' in what follows). We validate our code by comparing it to the analytical results in the TopHat case, and apply it to a
variety of other filters. These include the Gaussian filter, a family of filters with smeared cell boundary, and a filter peaked at the boundary of the cell.

We do not attempt to calculate exactly the part of the prefactor 
due to fluctuations with higher multipoles 
(called `aspherical prefactor' in \cite{ivanov,anton}). Instead, we estimate it using two approaches.
One is the analytic perturbative method valid at small to moderate density contrasts. Another is comparison of the non-perturbative `spherical PDF' obtained using our numerical pipeline with the PDF extracted from 
N-body simulations. We find that the aspherical prefactor for various filters closely resembles the aspherical prefactor in the TopHat case. However, there are also some notable differences which suggest sizable 2-loop corrections.
Estimating the sensitivity of the PDF to the EFT corrections we find it to be largely independent of the shape of the filter, as long as the density variance in the cell is kept fixed.

The paper is organized as follows. In Sec.~\ref{sec:review} we review the path integral approach to the 1-point PDF. In Sec.~\ref{sec:numeric} we introduce the numerical scheme to find the PDF for arbitrary filters.
In Sec.~\ref{sec:results} we validate our pipeline and construct PDFs (modulo the aspherical prefactor) for sample non-TopHat filters. In Sec.~\ref{sec:pert}, we discuss an analytic, perturbative model for the aspherical prefactor, as well as the filter sensitivity to short-scale physics coming from EFT corrections.
 In Sec.~\ref{sec:asp} we compare our results with N-body simulations to estimate the aspherical prefactor and compare with our analytic results from the previous section. 
Sec.~\ref{sec:discuss} is dedicated to the discussion of our results and future directions. 

Several appendices contain supplementary material. Appendix~\ref{app:convention} lists our conventions. Appendix~\ref{app:kernal} contains derivation of the results used in Secs.~\ref{sec:review} and \ref{sec:numeric}. 
Appendix~\ref{app:C} discusses the analytic expansion of the aspherical prefactor at small density contrasts. Appendix~\ref{app:figs} provides additional details on the validation of the numerical code. Appendix~\ref{app:DWell} contains analytic derivation of the PDF for a non-monotonic filter peaked at the cell boundary.

\section{Path integral for the PDF}
\label{sec:review}

In this section we review the path integral approach to the 1-point PDF, following Refs.~\cite{ivanov,anton}.
We use the notations and conventions summarized in Appendix \ref{app:convention}.

Consider a matter density contrast averaged over a sphere of radius $r_*$, 
\begin{equation}\label{dw}
    \bar{\delta}_W=\int \frac{d^3x}{r_*^3} \Tilde{W}\left( r/r_*\right) \delta(\textbf{x})=\int_{\textbf{k}}W(kr_*)\,\delta(\textbf{k}), 
\end{equation}
where $\delta(\textbf{x})=\delta \rho_{\rm mat}(\textbf{x})/\langle\rho_{\text{mat}}\rangle$ with $\langle\rho_{\text{mat}}\rangle$ being the average matter density of the universe, $\Tilde{{W}}(r /r_*)$ is the position space window function, or filter, obeying the normalization condition ${\int \frac{d^3x}{r_*^3} \Tilde{W}(r/r_*)=1}$, and $W(kr_*)$ is its Fourier transform.
We emphasize that $\delta(\textbf{x})$ is the non-linear density contrast seen today, which is sourced by the gravitational dynamics of matter throughout the evolution of the universe. The magnitude of $\delta({\bf x})$ and $\bar\delta_W$ need not be small.
A special case of the filter $\tilde W$ is the TopHat window function,
\begin{equation}\label{TH}
    \tilde{W}_{\text{th}}(r/r_*)=\frac{3}{4\pi}\Theta_\text{H}\left( 1-\frac{r}{r_*}\right) \Longleftrightarrow  W_\text{th}(kr_*)=\frac{3j_1(kr_*)}{kr_*}, 
\end{equation}
where $\Theta_\text{H}(x)$ is the Heaviside function and $j_1(x)$ is the spherical Bessel function. 

We assume the initial conditions in the far past for the density perturbations to be adiabatic and Gaussian such that their two-point correlator can be written as 
\begin{equation}
    \langle \delta_i(\textbf{k}) \delta_i(\textbf{k}') \rangle = (2\pi)^3\delta_D (\textbf{k}+\textbf{k}')g^2(z_i)P(k), 
\end{equation}
where $\delta_D$ is the Dirac delta-function, $P(k)$ is the linear power spectrum at the present epoch, and $g(z)$ is the linear growth factor at redshift $z$, normalized so that $g(0)=1$. 
In what follows $g^2(z)$ will play the role of a formal expansion parameter which will be used when carrying out the saddle-point approximation. In reality, the true expansion parameter is the (redshift dependent) variance of the filtered linear density field $g^2(z)\sigma_{L,r_*}^2$, with
\begin{equation}\label{var_L}
    \sigma^2_{L,r_*}=\int_{\textbf{k}} P(k)|W(kr_*)|^2 
\end{equation}
being the variance at $z=0$. In other words,  
$g^2(z)$ merely acts as a book-keeping device to keep track of the perturbation order. 

It is convenient to rescale the density field to redshift $z$ using the linear growth factor,
\begin{equation}
    \delta_L(\textbf{k},z)=\frac{g(z)}{g(z_i)}\delta_i(\textbf{k}).
\end{equation}
This will be referred to as the `linear density field'. We will omit its redshift-dependence, as well as that of $g$, to simplify notation. 
The PDF of finding a given density contrast $\delta_*$ is given by the functional integral over the linear density fields,
\begin{equation}\label{PDFfull}
    \mathcal{P}(\delta_*)=\mathcal{N}^{-1}\int_{-i\infty}^{i\infty} \frac{d\lambda}{2\pi i g^2}\int \mathcal{D}\delta_L \exp \biggl\{ -\frac{1}{g^2}\left[ \int_{\textbf{k}} \frac{|\delta_L(\textbf{k})|^2}{2P(k)}-\lambda(\delta_*-\bar{\delta}_W\left[ \delta_L\right] ) \right]\biggr\}, 
\end{equation}
where $\mathcal{N}$ as a normalization constant and $\lambda$ is a Lagrange multiplier which, when integrated over, introduces a Dirac delta function enforcing the constraint 
\begin{equation}
\label{constr}
    \bar{\delta}_W[\delta_L]=\delta_*.
\end{equation}
This constraint ensures that the averaged density contrast of the cell, $\bar{\delta}_W$, is equal to the density contrast we are interested in, $\delta_*$. 

All information about the evolution of the density contrast from its initial value is encoded in the functional $\bar{\delta}_W[\delta_L]$. 
This evolution is carried out by the non-linear dynamics of matter in the cell, which we use to relate the linear density contrast $\delta_L(\textbf{x})$ to the non-linear density contrast $\delta(\textbf{x})$.
Evaluating the integral using the saddle-point approximation, we expect the form of the PDF to follow
\begin{equation}
\label{PDFexpansion}
    \mathcal{P}(\delta_*)=\exp \biggl\{ -\frac{1}{g^2}\left( \alpha_0 +\alpha_1g^2+\alpha_2 g^4+...\right) \biggr\}.
\end{equation}
The leading order term $\alpha_0$ corresponds to the saddle-point solution which minimizes the exponential term in (\ref{PDFfull}). The next-to-leading term $\alpha_1g^2$ comes from the Gaussian integral around the saddle-point configuration. This term corresponds to the one-loop correction in the background of saddle point solution and gives the PDF prefactor. 
The coefficient $\alpha_1$ contains a logarithmic dependence, $\alpha_1=\log g+\tilde\alpha_1$, which introduces a $1/g$ factor to the PDF. Third term $\alpha_2g^4$ represents two-loop corrections, etc. In this paper, we restrict to the leading and one-loop terms.

At this point, it is useful to impose a symmetry which will help us look for saddle-point configurations. By enforcing that the filter is spherically symmetric, and thus depends only on the magnitude of the momentum, $k$, one finds that the most probable configuration, $\hat{\delta}_L(\textbf{k})$, is also spherically symmetric. Given such a symmetry, and as long as there are no shell-crossing, we may construct a one-to-one mapping relating the linear and non-linear density fields by
\begin{equation}
\label{SCmap}
    \bar{\delta}_L(R)=F\big(\bar{\delta}(r)\big)~~ \Longleftrightarrow ~~\bar{\delta}(r)=f\big( \bar{\delta}_L(R)\big), 
\end{equation}
where the functions $f$ and $F$ come from equations governing the spherical collapse dynamics and $\bar \delta$ and $\bar\delta_L$ are the respective TopHat averages, 
\begin{equation}\label{d_avg}
    \bar{\delta}(r)=\frac{3}{r^3}\int_0^r dr' r'^2 \delta(r'), \qquad \bar{\delta}_L(R)=\frac{3}{R^3}\int_0^R dR' R'^2 \delta_L(R'). 
\end{equation} 
Here $R$ is the radius of a spherical region at some early time when perturbations inside it were linear, whereas $r$ is its radius at a later epoch when it undergoes the non-linear collapse. 
Both $R$ and $r$ are defined in the comoving Friedmann coordinates.
Conservation of mass inside the region implies the relation 
\begin{equation}
\label{Rr}
    R=r\big(1+\bar\delta(r)\big)^{1/3}\;.
\end{equation}
Note that as we go back in time to the early universe, then $\bar\delta(r)\to 0$ and $r\rightarrow R$. We may also interpret $R$ and $r$ as the Lagrangian and Eulerian radii of the region, respectively. 

The map (\ref{SCmap}) in $\Lambda$CDM cosmology almost identically coincides with that in 
Einstein-de Sitter (EdS) universe \cite{ivanov}, in which case it admits a closed-form expression,  
\begin{equation}\label{Ff}
    F(x)=\mathcal{G}\big(\mathcal{F}^{-1}(x)\big)\;, \qquad f(x)=\mathcal{F}\big(\mathcal{G}^{-1}(x)\big)\;.
\end{equation}
For overdense regions the functions $\mathcal{F}$ and $\mathcal{G}$ are 
\begin{subequations}
\label{SCs}
    \begin{equation}
    \label{SC1}
            \mathcal{F}_+(\theta)=\frac{9(\theta - \sin \theta)^2}{2(1-\cos \theta)^3}-1\;,\qquad
            \mathcal{G}_+(\theta)=\frac{3}{20}\left[ 6(\theta -\sin \theta)\right]^{2/3}\;,
    \end{equation}
whereas for underdense region they read,
    \begin{equation}
    \label{SC2}
            \mathcal{F}_-(\theta)=\frac{9(\sinh \theta - \theta)^2}{2(\cosh \theta -1 )^3}-1\;,\qquad
            \mathcal{G}_-(\theta)=-\frac{3}{20}\left[ 6(\sinh \theta-\theta )\right]^{2/3}\;.
    \end{equation}
\end{subequations}
Using this mapping function and the relation (\ref{Rr}) we can write (\ref{dw}) as
\begin{equation}\label{dw2}
    \bar{\delta}_W=\frac{4\pi}{r_*^3}\int dR \ R^2 \, \tilde{W}\left[ \frac{R}{r_*}\left(1+f\left(\bar{\delta}_L(R)\right)\right)^{-1/3}\right]-1.
\end{equation}

When evaluating the integral (\ref{PDFfull}) in the saddle-point approximation, we look for a function
$\delta_L(\textbf{k})$ and Lagrange multiplier $\lambda$ which extremize the exponent of the integrand. 
By taking variations of the latter with respect to $\delta_L(\textbf{k})$ and $\lambda$, we obtain the equations, 
\begin{subequations}\label{EL_eq}
    \begin{align}\label{EL1}
        &\frac{\hat{\delta}_L({\bf k})}{P(k)}+\hat{\lambda}\left.\frac{\partial \bar{\delta}_W}{\partial \delta_L({\bf k})}\right|_{\hat{\delta}_L({\bf k})}=0\;,\\
    \label{EL2}
        &\delta_*-\bar{\delta}_W[\hat{\delta}_L]=0\;,
    \end{align}
\end{subequations}
which must be satisfied by the saddle-point configuration 
$\hat{\delta}_L(\textbf{k})$, $\hat{\lambda}$.\footnote{Throughout the paper we use hats to denote the quantities characterizing the saddle-point of the integral (\ref{PDFfull}).} Using spherical symmetry of the saddle-point density contrast, $\hat{\delta}_L(\textbf{k})=\hat{\delta}_L(k)$, and the map (\ref{SCmap}), equation (\ref{EL1}) can be cast into the form of a one-dimensional integral equation \cite{ivanov}. 
The latter, however, is in general difficult to solve. It dramatically simplifies only for the TopHat filter,\footnote{Technically, this happens because the filtered density in this case coincides with the averaged density (\ref{d_avg}) appearing in the spherical collapse map.} in which case the saddle-point configuration can be found analytically \cite{Bernardeau:1992zw,Valageas:1998xr,Valageas:2001zr,Matarrese:2000iz}.
For other filters, one has to resort to numerical methods. Solving the integral equation  
then appears impractical.  
In Sec.~\ref{sec:numeric} we adopt an alternative route to determine $\hat{\delta}_L(k)$ 
based on a constrained minimization procedure.

The saddle-point linear density contrast directly gives the leading exponential term in the expansion (\ref{PDFexpansion}),
\begin{equation}
\label{expalpha}
 \alpha_0=\int_\k   \frac{|\hat\delta_L(k)|^2}{2P(k)}\;.
\end{equation}
Once we know $\alpha_0$ as a function of $\delta_*$, we can obtain a useful expression for the saddle-point value of the Lagrange multiplier. Indeed, taking derivative of Eq.~(\ref{PDFfull}) with respect to $\delta_*$, we find
\begin{equation}
    \frac{d\ln \mathcal{P}}{d\delta_*}=\frac{\hat{\lambda}}{g^2}+O(g^0)\;.
\end{equation}
Comparison with the expansion (\ref{PDFexpansion}) then yields
\begin{equation}\label{lam}
    \hat{\lambda}=-\frac{d\alpha_0(\delta_*)}{d\delta_*}\;. 
\end{equation}
Note that at small $\delta_*$, when the saddle-point density field is essentially linear, $\alpha_0$ is a quadratic function of $\delta_*$. This implies $\hat\lambda\big|_{\delta_*=0}=0$.   

To study the prefactor, corresponding to the $\alpha_1$-term in (\ref{PDFexpansion}), we consider perturbations around the saddle-point solution.  This involves expanding our variables into $\delta_L(\textbf{k})=\hat{\delta}_L(k)+\delta^{(1)}_L(\textbf{k})$ and $\lambda=\hat{\lambda}+\lambda^{(1)}$. Further expanding $\delta_L^{(1)}(\textbf{k})$ into spherical harmonics, substituting into (\ref{PDFfull}) and performing Gaussian integral over the amplitudes of the fluctuations,
we find that the PDF factorizes into the contributions with different multipole numbers $\ell$, 
\begin{equation}\label{pertb_pdf}
    \mathcal{P}(\delta_*) = \exp \biggl\{ -\frac{\alpha_0(\delta_*)}{g^2} \biggr\} \cdot \prod_{\ell=0}\mathcal{A}_\ell(\delta_*). 
\end{equation}
Singling out the contributions of the monopole fluctuations, we define the ``spherical part'' of the PDF as
\begin{equation}\label{SPDF}
    \mathcal{P}_{sp}(\delta_*) \equiv \mathcal{A}_0(\delta_*) \cdot \exp \biggl\{ -\frac{\alpha_0(\delta_*)}{g^2} \biggr\}. 
\end{equation}
We emphasize that this function, which we will refer to as the \textit{spherical PDF}, is not the true PDF since it lacks the contribution of perturbations with $\ell\geq 1$ into the prefactor. It does, however, make up a large fraction of the true PDF as it has the strongest dependence on $\delta_*$. 

Closed-form expression for the monopole prefactor for a general filter is derived in Appendix~\ref{app:kernal}. Here we quote the result:
\begin{equation}\label{monopre}
    \mathcal{A}_0=\frac{1}{\sqrt{2\pi g^2}} \sqrt{\frac{\det \left( \frac{\mathbbm{1}(k_1,k_2)}{4\pi P(k_1)}\right)}{-\det \mathcal{H}}}\;,
\end{equation} 
where $\mathbbm{1}(k_1,k_2)$ is the symmetric unit operator in radial $k$-space with respect to the measure (\ref{radial_measure}),
\begin{equation}\label{k_unit}
    \mathbbm{1}(k_1,k_2)=\frac{(2\pi)^3}{k_1^2}\delta_D(k_1-k_2)\;,
\end{equation}
and $\mathcal{H}$ is a block matrix,
\begin{equation}\label{Hmatrix}
    \mathcal{H}=\begin{pmatrix}
0 & S(k_2) \\
S^\intercal(k_1)&  \mathcal{O}(k_1,k_2)
\end{pmatrix}\;  ,
\end{equation}
with
\begin{equation}
\label{Okernel}
    \mathcal{O}(k_1,k_2)=\frac{\mathbbm{1}(k_1,k_2)}{4\pi P(k_1)}+\frac{\hat{\lambda} }{2\pi}Q_0(k_1,k_2)\;.
\end{equation}
Here $S(k)$ and $Q_0(k_1,k_2)$ are kernels whose rather lengthy expressions are presented in  Appendix~\ref{app:kernal}. 
Note that the determinants appearing in the monopole prefactor are functional determinants, and 
their evaluation requires regularization. In our numerical approach we use the lattice discretization, so that all kernels, such as $\mathbbm{1}$, ${\cal H}$, etc., become finite matrices.  

As long as we use the spherical collapse mapping functions (\ref{SCs}), the kernels $S(k)$, $Q_0(k_1,k_2)$ are redshift independent. Hence, all dependence of ${\cal P}_{sp}$ on $z$ is encoded in the linear growth factor $g^2(z)$, which factors out as $1/g^2$ and $1/g$ in the exponent and the monopole prefactor, respectively. Thus, once we know the spherical PDF at a single redshift, we can evaluate it at any other redshift by simply rescaling the exponent and the prefactor.

The remaining aspherical contributions to the prefactor can now be defined as 
\begin{equation}\label{ASP}
\mathcal{A}_{asp}(\delta_*)=\prod_{\ell\geq1}\mathcal{A}_\ell(\delta_*), 
\end{equation}
which we will refer to as the \textit{aspherical prefactor}. 
Its evaluation requires the study of second-order aspherical perturbations in the background of the evolving saddle-point density contrast and involves extensive numerical integration even for the TopHat filter \cite{ivanov,anton}.
In this paper we contend ourselves with two estimates for it. One uses perturbative expansion at small density contrast (see Sec.~\ref{sec:pert}), whereas another is obtained by comparing our spherical PDF with N-body simulations (see Sec.~\ref{sec:asp}). 

A few properties of $\mathcal{A}_{asp}(\delta_*)$ can be proven in general \cite{ivanov,anton} and their validation with N-body simulations will provide an important cross-check of our method. First, the aspherical prefactor only weakly depends on the redshift through the one-loop counterterms encoding the effect of non-linear short-wavelength perturbations. Thus, the dependence of the PDF on redshift comes almost exclusively from its spherical part. 

Second, the first two terms in the Taylor expansion of $\mathcal{A}_{asp}(\delta_*)$ at 
$\delta_*=0$ are fully fixed by the normalization conditions,
\begin{equation}
\label{norms}
    \int_{-1}^\infty {\cal P}(\delta_*)\,d\delta_*=1~,~~~~~~~
    \int_{-1}^\infty \delta_*{\cal P}(\delta_*)\,d\delta_*=0\;.
\end{equation}
Evaluated to leading order in the saddle-point expansion at $\delta_*=0$, they lead to (see Appendix~\ref{app:C} for details):
\begin{equation}
\label{Aaspprops}
    \mathcal{A}_{asp}(0)=1~,~~~~~~~
    \mathcal{A}'_{asp}(0)=-(\log \mathcal{A}_0)'\big|_{\delta_*=0}+\frac{1}{2}\sigma_{L,r_*}^2\alpha_0'''\big|_{\delta_*=0}\;,
\end{equation}
where prime means derivative with respect to $\delta_*$. The first of the above equalities implies that the value of the PDF at $\delta_*=0$ is determined by the monopole prefactor and can be shown to be (see Appendix~\ref{app:kernal}):  
\begin{equation}\label{PDF0}
    \mathcal{P}(0)=\mathcal{P}_{sp}(0)=\frac{1}{\sqrt{2\pi g^2 \sigma^2_{L,r_*}}}\;.
\end{equation}

\section{Spherical PDF for arbitrary filters}
\label{sec:numeric}

\subsection{Numerical saddle-point}

For the TopHat filter, Eqs.~(\ref{EL_eq}) admit a simple analytic solution \cite{ivanov}. In general, however, the saddle-point configuration must be found numerically. Solving Eqs.~(\ref{EL_eq}) then appears impractical. Instead, we develop a numerical scheme which obtains the saddle-point configuration by minimizing the Gaussian weight 
\begin{equation}\label{action}
    \mathcal{S}[\delta_L]= \int_{\textbf{k}}\frac{|\delta_L(k)|^2}{2P(k)}, 
\end{equation}
entering the integral (\ref{PDFfull}), subject to the constraint (\ref{constr}). In what follows we refer to the functional (\ref{action}) as the \textit{action}. 
Due to the symmetry of the problem, it is sufficient to perform the minimization within the class of spherically symmetric configurations $\delta_L(R)$.

We use the \texttt{SciPy} and \texttt{NumPy} libraries in \texttt{Python} and implement the following algorithm:
\begin{enumerate}
\item 
For a given $\delta_*$, discretize the Lagrangian radius over an evenly-spaced lattice with the parameters
\begin{align}\label{lattice}
\begin{split}
    R_{\text{max}}=D R_*, \quad \Delta R=R_{\text{max}}/N, \quad R_*=r_*(1+\delta_*)^{1/3}\;,
\end{split}
\end{align}
where $N$ is the number of lattice points, $D$ is a parameter controlling the size of the lattice, and $R_*$ is the initial radius of the cell in the early universe. We have found that different choices of $\delta_*$ require different values of $N$ and $D$ to reach a given level of precision. This will be discussed in more detail in section \ref{sec:results}. 

    \item Define the functions $\Delta_L(R)=R\delta_L(R)$ and its discretized version $\texttt{X}=( \Delta_L(R_1), ..., \Delta_L(R_N))$. By construction, $\Delta_L(R)$ vanishes at $R=0$. In addition, we impose a Dirichlet boundary condition on it at $R=R_{\rm max}+\Delta R$. Note that the two end-points where $\Delta_L(R)$ vanishes are not included in the array $\texttt{X}$.

\item The sine transform (\ref{FT_sin}) of $\Delta_L(R)$ gives the function $\Delta_L(k)=k\delta_L(k)$ which vanishes at $k=0$ and $k=\pi/\Delta R$. We define its discrete version $\texttt{Y}=( \Delta_L(k_1), ..., \Delta_L(k_N))$ as
\begin{equation}
    \label{DST}
        \texttt{Y}[\texttt{X}]=2\pi\Delta R \, \texttt{DST}[\texttt{X}], 
    \end{equation} 
    where
      $\texttt{DST}$ is the \texttt{SciPy} discrete sine transform (DST) function.\footnote{Note that we have a coefficient $2\pi$ in (\ref{DST}), instead of $4\pi$, because the       
      \texttt{SciPy} \texttt{DST} function is scaled by a factor of two compared to the standard definition of the sine transform (\ref{FT_sin}).}
The lattice in the $k$-space is 
    \begin{align}\label{klattice}
\begin{split}
     k_n=n\Delta k\;, \quad \Delta k= \frac{\pi}{(N+1)\Delta R}\;, \quad k_{\text{max}}=\frac{N\pi}{(N+1)\Delta R}\;.
\end{split}
\end{align}
    
    \item Discretize the action over the lattice as 
    \begin{equation}\label{dis_action}
        \mathcal{S}[\texttt{X}]=\sum_{\text{int.}} \frac{1}{(2\pi)^2} \frac{\texttt{Y}_n^2[\texttt{X}]}{P(k_n)},
    \end{equation}
    where $\sum_{\text{int.}}$ denotes numerical integration.\footnote{We found that for  monotonic filters (\ref{filters}) considered below the trapezoid integration scheme suffices and leads to a more stable convergence of the optimization procedure than the higher order Simpson method. To ensure convergence for the non-monotonic filter (\ref{DWmain}), we had to add a spline interpolation of the array before applying the \texttt{quad} integration function in \texttt{SciPy}.}

\item Construct a function $\texttt{Avg}[\texttt{X}]$ which, given the array $\texttt{X}$, computes an array of averaged density contrast (\ref{d_avg}). Its output has the form $(\bar{\delta}_L(R_1), ..., \bar{\delta}_L(R_N))$. 

\item Define numerically the spherical collapse mapping function $f(x)$ from Eq.~(\ref{Ff}). To construct the inverse function $\mathcal{G}_{\pm}^{-1}(x)$, we use a root-finding function \texttt{fsolve}. 

    \item Discretize the constraint function~\eqref{dw2} as
    \begin{equation}
    \label{dw2disc}
        \bar{\delta}_W[\texttt{X}]=-1+\frac{4\pi}{r_*^3}\sum_{\text{int.}}R_n^2 \Tilde{\texttt{W}}\left[\frac{R_n}{r_*}\left( 1+\texttt{f}[\texttt{Avg}[\texttt{X}]]\right)^{-1/3}\right]
    \end{equation}
    where $\Tilde{\texttt{W}}$ and \texttt{f} are numerical implementations of the window function and the spherical collapse map.

    \item Select a value for $\delta_*$ and construct an initial array, $\texttt{X}_n=F(\delta_*) R_n$ representing the initial guess for optimization scheme. Note that the continuous version of the array, corresponding to $\Delta_L(R)=F(\delta_*)R$, satisfies the constraint given in Eq. (\ref{EL2}).
    
    \item Use the sequential least squares programming (SLSQP) solver to minimize (\ref{dis_action}) subject to the non-linear constraint $\bar{\delta}_W=\delta_*$. This optimizer is implemented in the \texttt{minimize} function from the \texttt{scipy.optimize} library.
\end{enumerate}
This algorithm returns the optimized array, $\hat{\texttt{X}}$, representing the discretized function for $R\hat{\delta}_L(R)$, at a given $\delta_*$ for any specified window function. The algorithm takes, on average, $\sim 3$ hours to complete on a single core. We haven't attempted to accelerate the code in this work, limiting ourselves to a proof of principle. Optimization of the code for possible Monte Carlo applications is left for future.  

The minimal value of the action $\mathcal{S}[\hat{\texttt{X}}]$ gives the exponential part of the PDF, $\alpha_0(\delta_*)$ in Eq.~(\ref{SPDF}),
\begin{equation}
\label{expPDF}
    \alpha_0(\delta_*)=\mathcal{S}[\hat{\texttt{X}}].
\end{equation}
Given a sufficiently fine grid of points over $\delta_*$, the saddle-point value of the Lagrange multiplier $\hat\lambda$ is computed via numerical differentiation of $\alpha_0(\delta_*)$, according to Eq.~(\ref{lam}). 
In our applications, we sample 100 values of $1+\delta_*$ in the range $[0.1, 10]$, evenly-spaced on a log-scale, and use cubic spline interpolation to obtain a smooth function $\alpha_0(\delta_*)$. 

\subsection{Computation of the monopole prefactor}
\label{ssec:pref}

Given the optimized profile, $\hat{\texttt{X}}=(\hat{\Delta}_L(R_1), ...,\hat{\Delta}_L(R_N))$, where $\hat{\Delta}(R_n)=R_n\hat{\delta}_L(R_n)$, we evaluate the monopole prefactor 
(\ref{monopre}) numerically. 
We proceed as follows:
\begin{enumerate}
    \item Interpolate $\hat{\texttt{X}}$ over the interval $[0,R_{\rm max}]$ using a cubic spline function. Note that we interpolate to $R$=0 by setting $\hat{\Delta}_L(0)$=0. 

    \item Using this interpolating function, construct the averaged linear density $\bar\delta_L(R)$. The integration is performed with the \texttt{quad} function of \texttt{SciPy}.  

    \item Construct $S(k)$ and $Q_0(k_1,k_2)$ by numerically implementing (\ref{SQint}) over the points $k_n, k_m$, for $n,m=1,2,...,N$.

    \item Discretize $\mathbbm{1}(k_1,k_2)$ over the lattice into a diagonal ${N}\times {N}$ matrix given by 
    \begin{equation}
        \mathbbm{1}_{nm}=\frac{(2\pi)^3}{k_n^2}\frac{\delta_{nm}}{\Delta k};.
    \end{equation}
    
    \item  Construct the $({N}+1)\times ({N}+1)$ discretized $\mathcal{H}$ matrix as, 
    \begin{equation}
        \mathcal{H}_{nm}=\begin{pmatrix}
        0 & S (k_m) \\
        S^\intercal(k_n)& \mathcal{O}_{nm}
    \end{pmatrix}, 
    \end{equation}
    with
    \begin{equation}
        \mathcal{O}_{nm}=  \frac{2 \pi^2}{\Delta k} \frac{\delta_{ij}}{k_n^2 P(k_n) } + \frac{\hat{\lambda}}{2\pi}Q_0(k_n,k_m).
    \end{equation}

    \item Compute the determinants of $\mathcal{H}_{nm}$ and $\mathbbm{1}_{nm}$ using the \texttt{scipy.linalg} library in \texttt{Python} and substitute into (\ref{monopre}). 
    
\end{enumerate}
This gives us the monopole prefactor ${\cal A}_0(\delta_*)$.
The average runtime of this algorithm is 1 minute on a single core.

\section{Application to selected filters}\label{sec:results}
In this section, we illustrate our numerical procedure on several examples. The linear power spectrum is generated using the Boltzmann solver \texttt{CLASS} \cite{CLASS} with cosmological parameters matching that of the \texttt{Quijote} N-body simulations \cite{quijote}. While the power spectrum and N-body simulations used in this paper are based on the $\Lambda$CDM cosmology, we approximate the map between linear and non-linear densities in 
spherical cells with the EdS expressions (\ref{Ff}), (\ref{SCs}).
This is known to be a very good approximation (see e.g. \cite{ivanov}) and we adopt it for simplicity.
We start with the TopHat filter and examine the accuracy of the code by comparing the numerical results with the analytical solutions \cite{ivanov}. This allows us to identify suitable values of the lattice parameters $N, D$ introduced in Sec.~\ref{sec:numeric}. We then apply the code to several different filters and compare the results with the TopHat case.

\subsection{Validation of the numerical procedure with TopHat}
\label{ssec:TopHat}
In order to test the accuracy of the numerical procedure, we apply it to 
the TopHat filter (\ref{TH}). We choose the radius of the cell to be $r_*=$10 Mpc/$h$. 
The analytic solution for the saddle-point profile reads \cite{ivanov}, 
\begin{equation}
\label{TH_analytic}
    \hat{\delta}_L(k) = \frac{F(\delta_*)}{\sigma^2_{L,R_*}}P(k)W_{\text{th}}(kR_*), 
\end{equation}
where $F(\delta_*)$ is the spherical collapse map (\ref{SCmap}) and
$\sigma^2_{L,R_*}$ is the linear density variance defined similar to (\ref{var_L}), with $r_*$ replaced by the $\delta_*$-dependent Lagrangian radius $R_*$ (see Eqs.~(\ref{lattice})). It leads to the spherical PDF \cite{anton}
\begin{equation}
\label{TH_PDF}
    {\cal P}_{sp}(\delta_*)=\frac{1}{\sqrt{2\pi g^2}}\frac{d\nu(\delta_*)}{d\delta_*}\exp{-\frac{\nu^2(\delta_*)}{2g^2}}~,\qquad \nu(\delta_*)\equiv\frac{F(\delta_*)}{\sigma_{L,R_*}}\;.
\end{equation}
To implement the filter (\ref{TH}) in the numerical code, we smear the discontinuity of 
$\Theta_H(x)$ replacing it by a sigmoid function (see Eq.~(\ref{FTanh}) below, where we take $a=0.03$). We use the lattice parameters $(N,D)=(450,30)$ for $\delta_*\leq-0.4$ and (300,20) for $\delta_*>-0.4$. Different parameter choices for different $\delta_*$ come from the need to ensure the numerical convergence, see Appendix \ref{app:figs} for further discussion of the convergence tests. 

\begin{figure}[t]
\centering
\hspace*{-1cm}                                                           
\includegraphics[width=0.4\paperwidth]{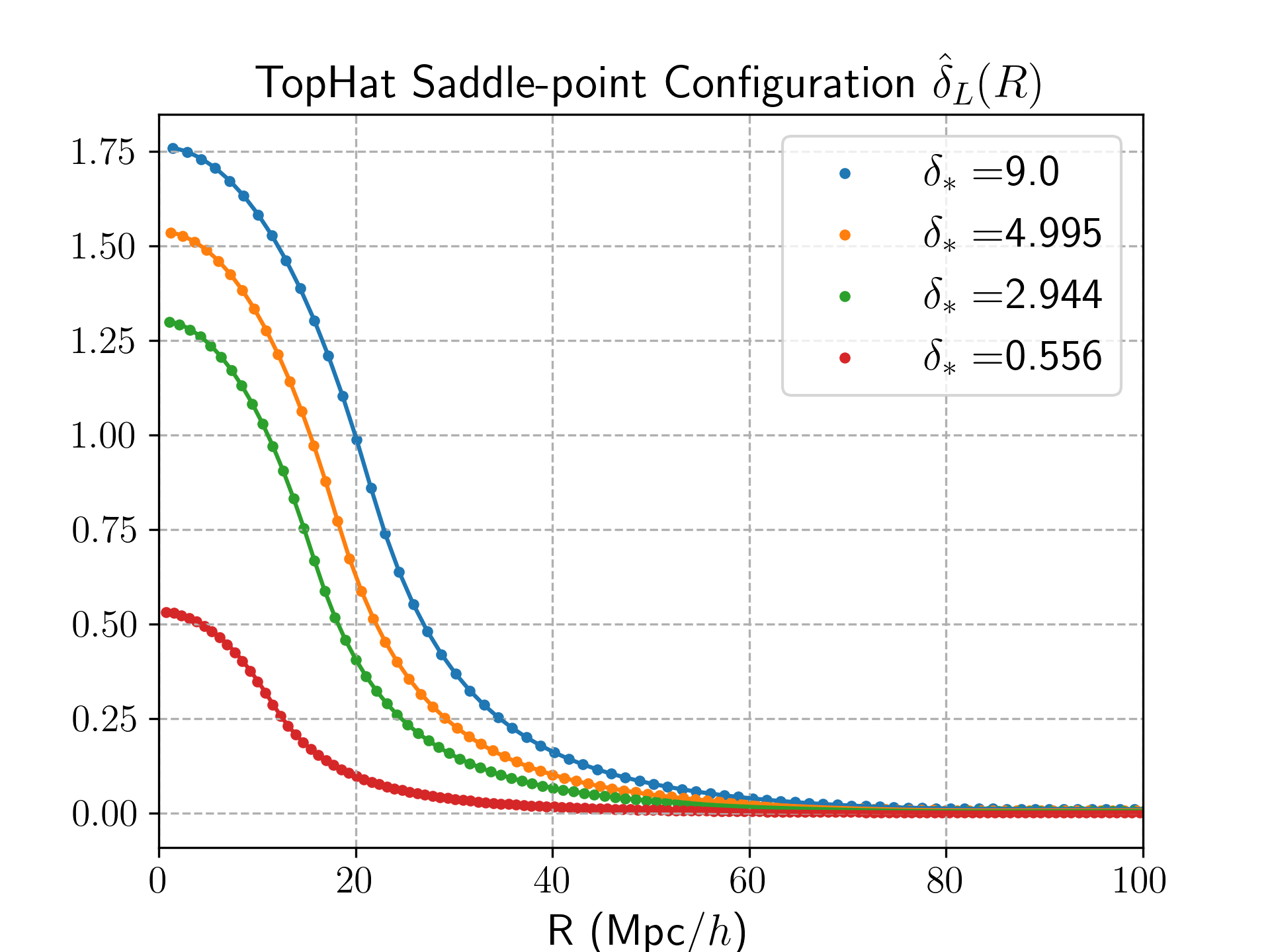}
\includegraphics[width=0.4\paperwidth]{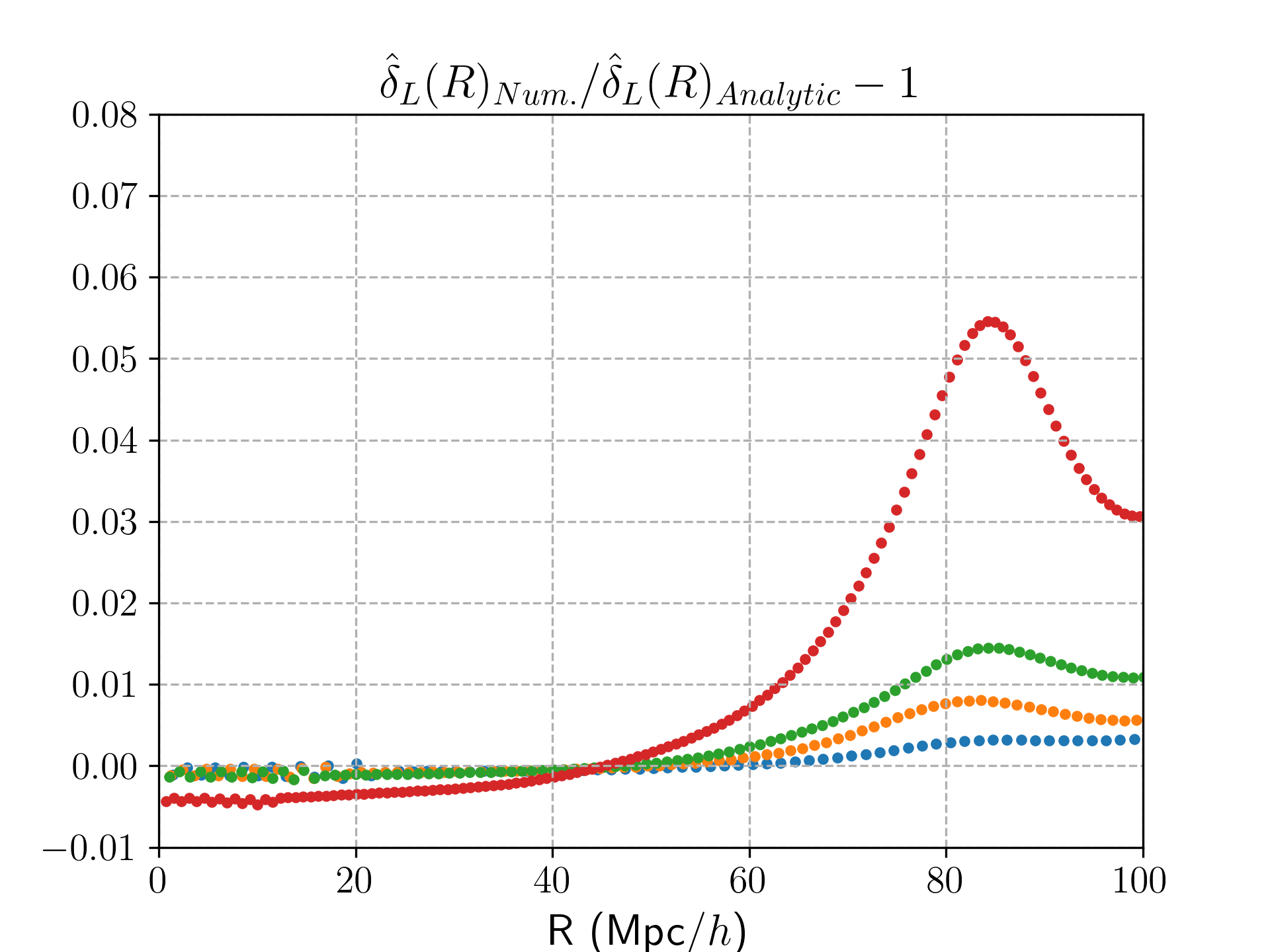}
\hspace*{-1cm}                                                           
\includegraphics[width=0.4\paperwidth]{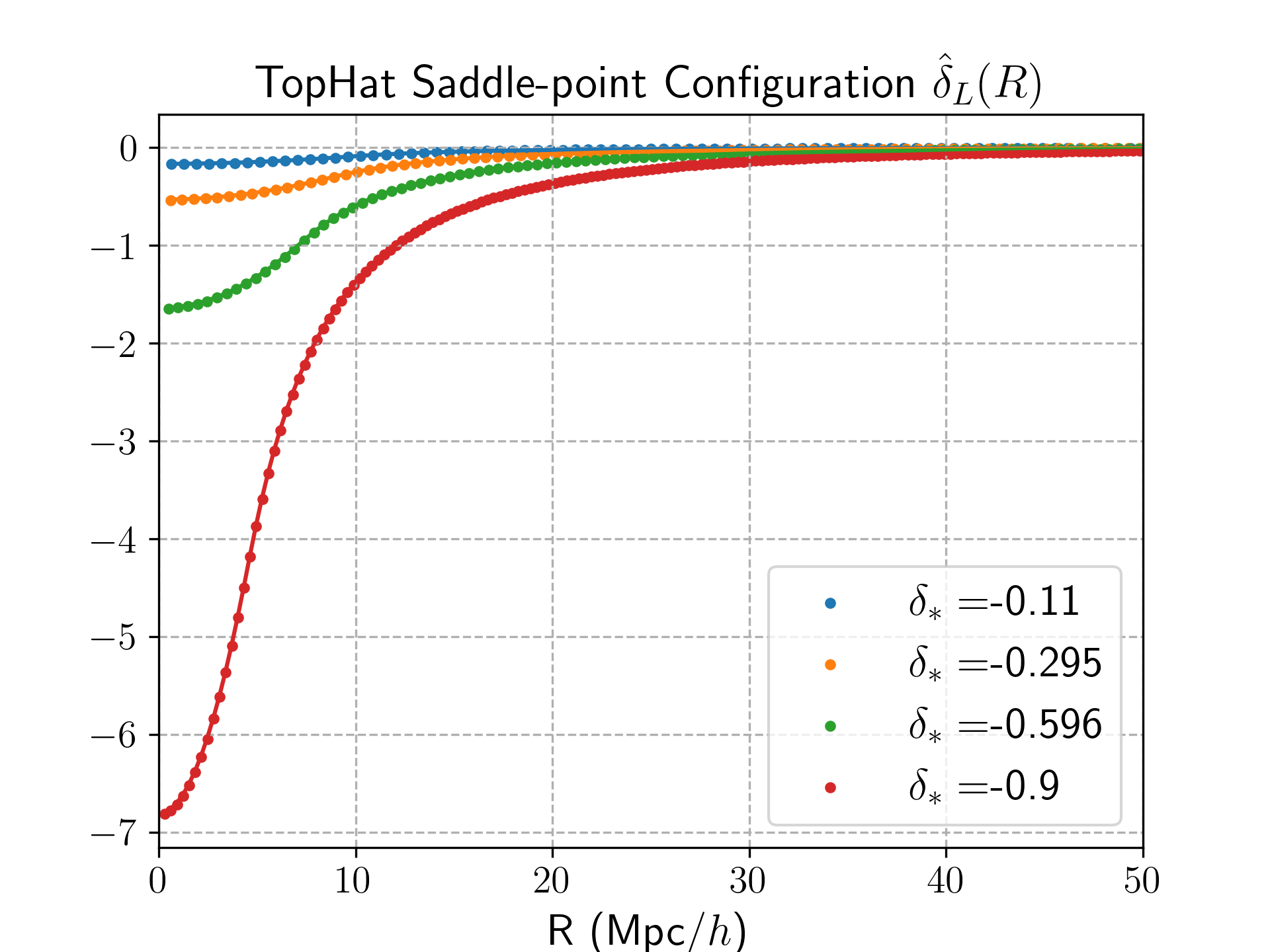}
\includegraphics[width=0.4\paperwidth]{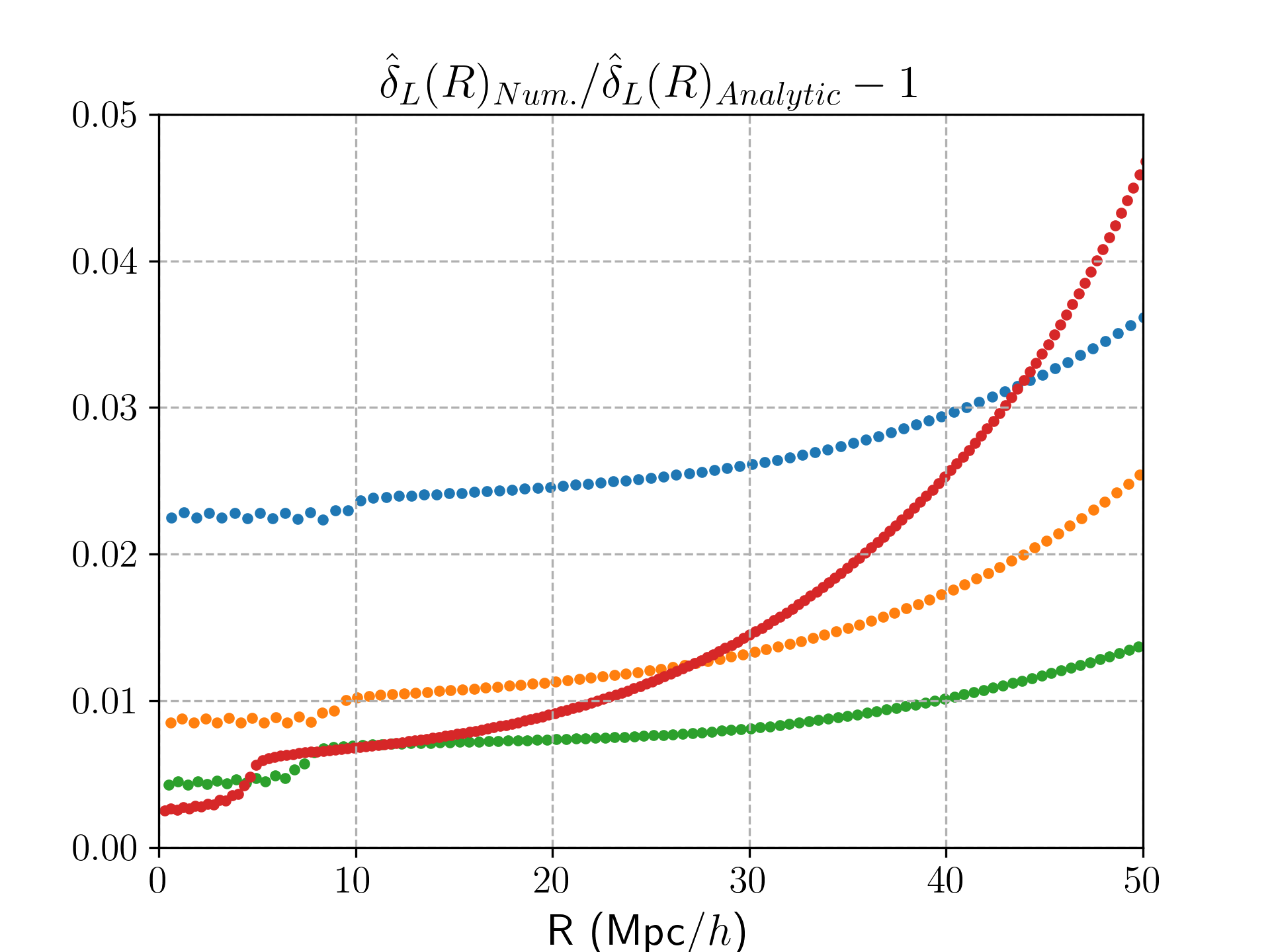}
\caption{{\it Left panels}: Numerical (dots) and analytical (solid line) saddle-point linear density profiles for several values of~$\delta_*$ for TopHat filter with radius $r_*=10\,{\rm Mpc}/h$. The profiles are functions of the Lagrangian radius $R$. The upper panel shows overdensities, $\delta_*>0$, while the lower panel shows underdensities, $\delta_*<0$. {\it Right panels}: Residuals between the numerical and analytical profiles.}
\label{fig:TH_prof}
\end{figure}

In Fig.~\ref{fig:TH_prof} we present the results for the linear saddle-point profiles $\hat{\delta}_L(R)$ along with residuals comparing the numerical data with the exact analytical expressions. Both methods are in good agreement with one another. The residuals are at the level $1-2\%$ within the center of the profiles and only modestly increase towards the tails, where the profiles themselves are small. 
This confirms that our numerical minimization captures the correct saddle-point configurations. 

When the spherically averaged linear profile $\bar\delta_L(R)$ exceeds a critical density $\delta_c=1.686$,\footnote{The critical density $\delta_c=1.686$ corresponds to the EdS universe and its value slightly changes in the $\Lambda$CDM cosmology. However, the change is very small since $\delta_c$ only weakly depends on the background cosmology. } 
shell-crossing occurs at the center of the cell. We find that it happens at $\delta_*\gtrsim 7$, in agreement with \cite{ivanov}. The spherical collapse map $f(\bar\delta_L)$ diverges at $\bar\delta_L \to \delta_c$ and looses physical meaning at $\bar\delta_L > \delta_c$. Note, however, that the expression (\ref{dw2}) for the filtered matter overdensity and its discrete version (\ref{dw2disc}) used in the code remains finite. The code thus produces smooth profiles matching the analytical solution (\ref{TH_analytic}) at least up to $\delta_*=9$. To understand whether these profiles are physically meaningful, we observe that $f(\bar\delta_L)$ is positive and large at $\bar\delta_L > \delta_c$. Inspection of Eq.~(\ref{dw2}) then tells us that the mass from a finite inner region of the cell in the Lagrangian coordinates gets effectively concentrated in the neighborhood of the origin $r=0$ in the Eulerian space. In reality, the inner region will get virialized leading to redistribution of mass across the cell --- a process not taken into account by the spherical collapse map (\ref{SCmap}). However, for the TopHat filter, the PDF is sensitive only to the total mass inside the cell (cf. discussion of shell crossing in \cite{ivanov}). Thus the results of our minimization procedure remain trustworthy as long as the the size of the virialized region is smaller than the size of the cell.
The situation is different for other filters which are sensitive to the distribution of mass across the cell. In this case, the virialization of the inner region introduces a poorly controlled uncertainty into calculation of the PDF. 
In the analysis of general filters below we will restrict the PDF to the range of density contrasts without the shell-crossing.

\begin{figure}[t]
\centering
\hspace*{-1cm}                                                           
 \includegraphics[width=0.4\paperwidth]{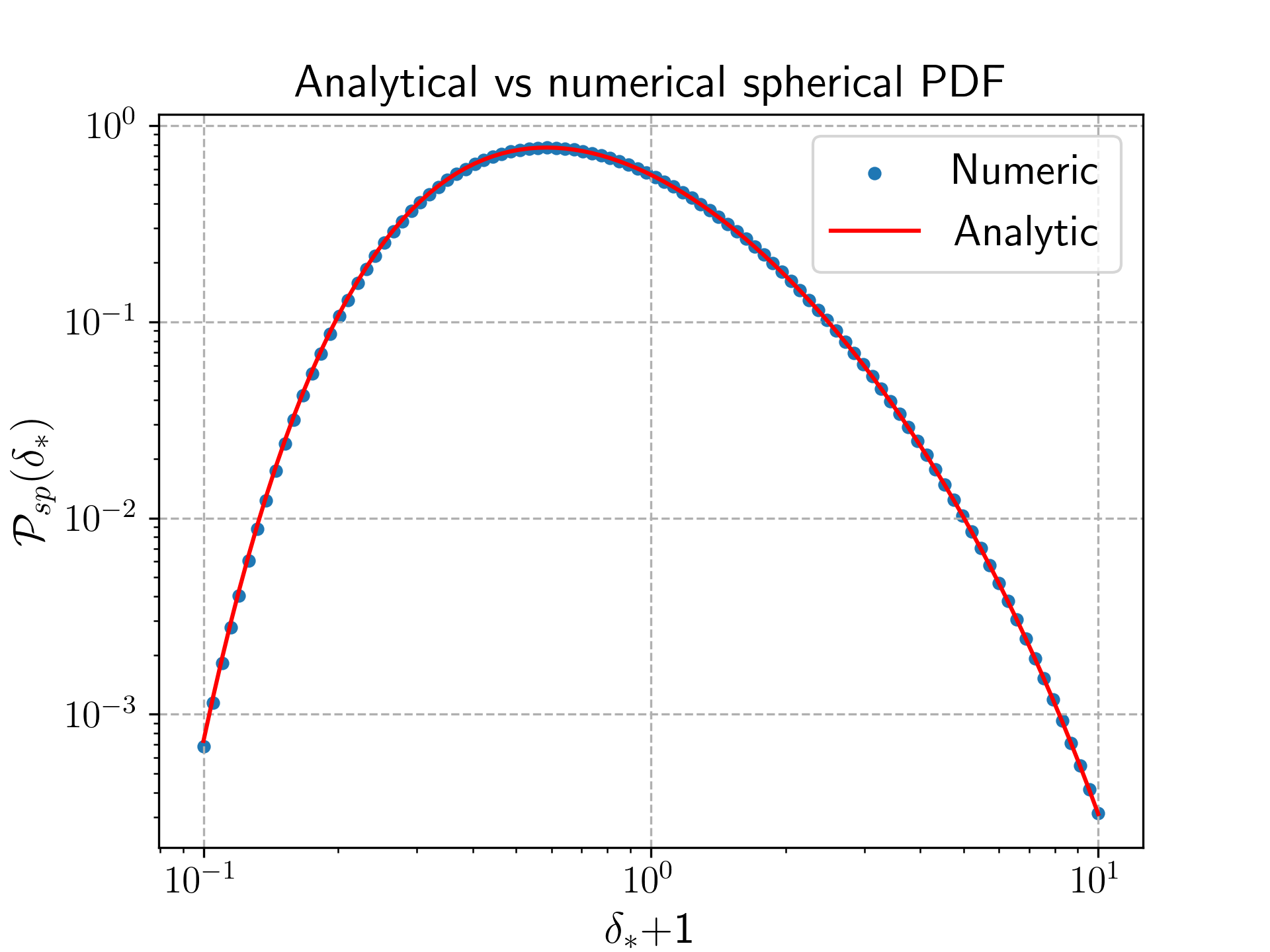}
\includegraphics[width=0.4\paperwidth]{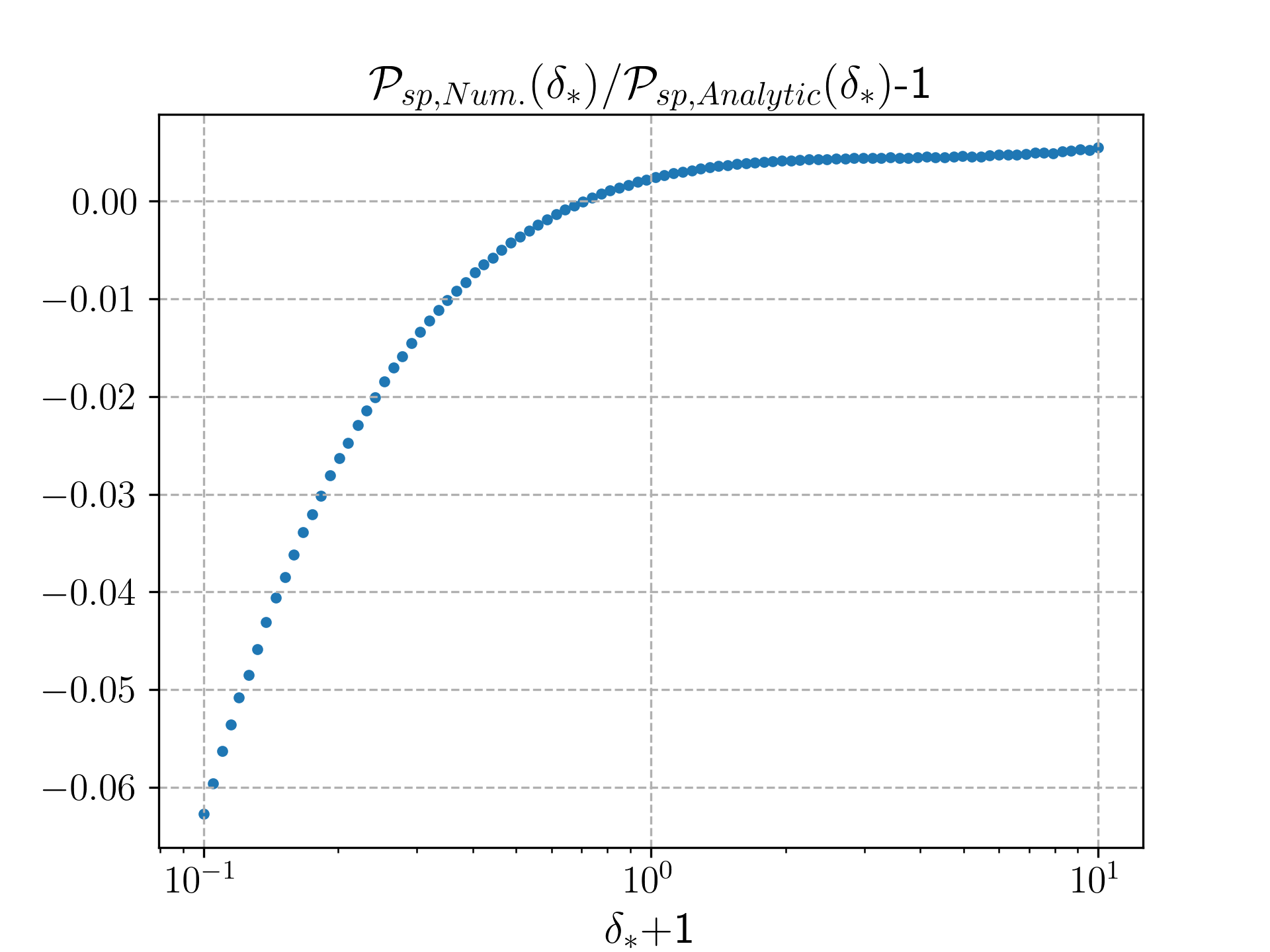} 
\caption{\textit{Left panel}: Spherical PDF for TopHat filter of radius $r_*=10\,{\rm Mpc}/h$ at redshift $z=0$. Numerical results (dots) are compared against the analytic expression (solid line). \textit{Right panel}: Residuals between the numerical and analytic results. 
}
\label{fig:TH_Psp}
\end{figure}

The numerical code directly gives us the saddle-point value of the action $\mathcal{S}[\hat{\delta}_L]$ and hence the exponential part of the PDF, see Eq.~(\ref{expPDF}). Further, numerically differentiating it with respect to the density contrast $\delta_*$ we obtain the saddle-point value of the Lagrange multiplier $\hat\lambda$.  
The spherical prefactor is computed using the algorithm of Sec.~\ref{ssec:pref}. 
We find that the latter reproduces 
the analytic expression for the prefactor to within $0.3\%$ across the considered range of densities.

Combining the exponential part and the prefactor, we obtain the spherical PDF shown in 
Fig.~\ref{fig:TH_Psp}, where it is compared to the analytic expression (\ref{TH_PDF}). Since the redshift dependence of the spherical PDF is uniquely determined by that of the growth factor $g(z)$, it suffices to perform the comparison for $z=0$.
We see that the numerical $\mathcal{P}_{sp}(\delta_*)$ is within a $1\%$ from the exact result in
the central part of the PDF and its overdensity tail. The extreme underdense regions exhibit a higher error up to $6\%$. 
We expect that this error can be reduced by using larger and finer lattices, which, however, would increase the computational cost. 
From practical viewpoint, a moderate degradation of the accuracy of the PDF model at the tails appears acceptable. Indeed, the tails correspond to rare events and, when measured from the data, suffer from large statistical uncertainties dominating over the model errors.

The results on the accuracy of the code obtained in this subsection provide us with the guidelines in the case of non-TopHat filters, when no analytic results are available.

\subsection{Smooth monotonic filters}
\label{ssec:smooth}

\begin{figure}[ht]
\centering
\hspace*{-1cm}                                                           
\includegraphics[width=0.4\paperwidth]{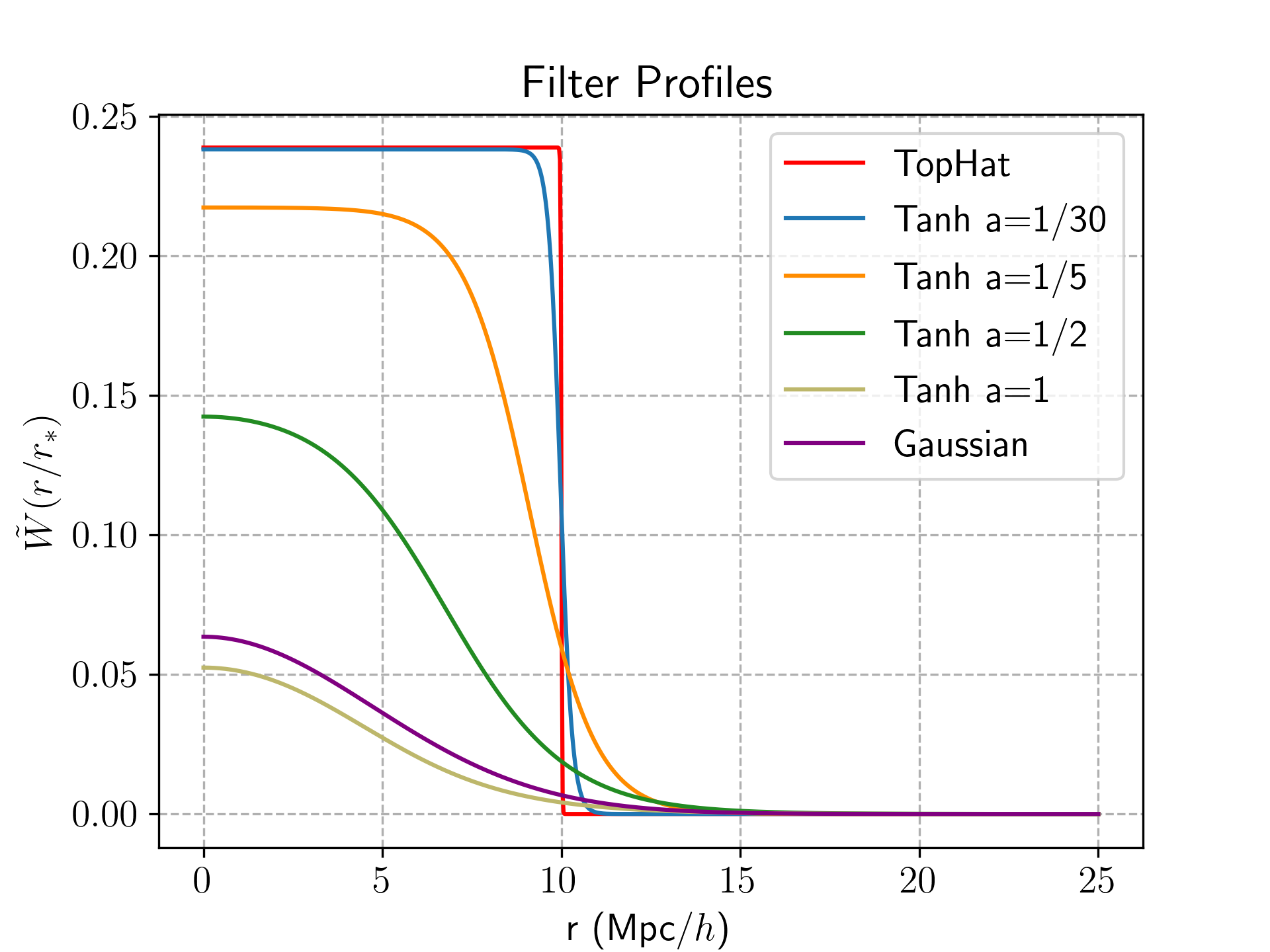}
\includegraphics[width=0.4\paperwidth]{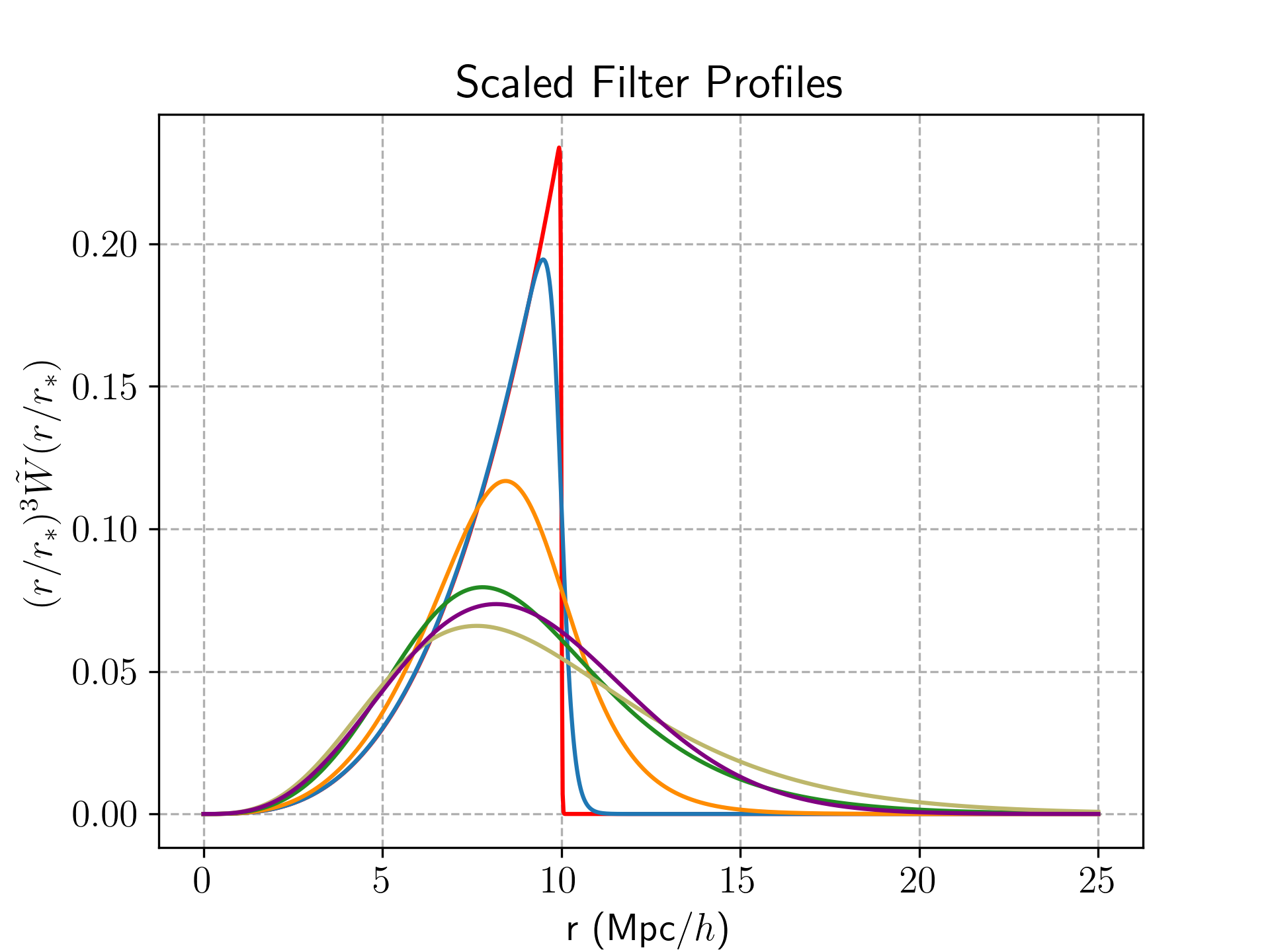}
\caption{\textit{Left panel:} Tanh and Gaussian filters considered in this subsection. The parameters $a$ and $r_*$ are taken from Table~\ref{table:ar} to ensure that all filters give the same linear variance. TopHat filter corresponding to the same linear variance is also shown. \textit{Right panel:} Filter profiles multiplied by $(r/r_*)^3$. The area under each curve is equal to $1$. }
\label{fig:filters}
\end{figure}

Having validated the code, we now apply it to a variety of different window functions. In this subsection, we examine the following radial filters, 
\begin{subequations}\label{filters}
    \begin{gather}\label{FGauss}
        \text{Gaussian}: \quad \Tilde{W}_{\text{Gauss}}(r/r_*)=\frac{1}{(2\pi)^{3/2}} e^{-\frac{1}{2} \left(\frac{r}{r_*} \right)^2} \quad \Longleftrightarrow \quad W_\text{Gauss}(kr_*)= e^{-\frac{(kr_*)^2}{2}}\;,\\  
    \label{FTanh}
        \text{Hyperbolic Tangent (Tanh):} \quad \Tilde{W}_{\text{Tanh}}(r/r_*)=\frac{3}{2\pi (4+a^2\pi^2)} \left[ \tanh{\left(\frac{1-\frac{r}{r_*}}{a}\right)} + \tanh{\left(\frac{1+\frac{r}{r_*}}{a}\right)} \right]. 
    \end{gather}
\end{subequations}
Note that the Fourier transform of the Tanh filter does not admit a simple analytic expression. 
The Tanh filter has, in addition to the overall size, a second free parameter $a$ which characterizes the smearing of its boundary. The limit $a\to 0$ recovers the TopHat filter, while $a\sim 1$ corresponds to a smooth curve similar to a Gaussian,\footnote{It is worth stressing that despite the similarity, the Tanh filter with $a=1$ and the Gaussian filter are still different.} see Fig.~\ref{fig:filters}.

For a Gaussian random field, the PDFs obtained using different filters are always Gaussian and depend on a single parameter, the filtered density variance (\ref{var_L}). On the other hand, we are interested in the sensitivity of the PDFs to the non-linear (non-Gaussian) properties of the density field. In order to remove the difference between different filters at the Gaussian level and directly compare their non-Gaussian properties, we choose their parameters in the way that they all lead to the same linear density variance $\sigma^2_{L,r_*}=0.5056$. In the TopHat case, it corresponds to the cell radius $r_*=10\,{\rm Mpc}/h$. For other filters, the scale $r_*$ is different; its values are listed in Table~\ref{table:ar}.

\begin{table}[ht]
\setlength{\tabcolsep}{18pt}
\centering
\begin{tabular}{|c|c|c|}
\hline
                & smearing parameter $a$        & $r_*$ (Mpc/$h$)         \\ \hline
TopHat          & -          & 10                    \\ \hline
Tanh            & 1/30       & 9.9695                      \\ \hline
Tanh            & 1/5        & 9.1121                      \\ \hline
Tanh            & 1/2        & 6.7547                      \\ \hline
Tanh            & 1          & 4.2267                      \\ \hline
Gaussian        & -          & 4.7186                      \\ \hline
\end{tabular}
\caption{Parameters of the filters considered in this subsection. They scale $r_*$ is adjusted to ensure that all filters lead to the same linear variance $\sigma^2_{L,r_*}=0.5056$. }
\label{table:ar}
\end{table}

\begin{figure}[t]
\centering
\hspace*{-1cm}                                              
\includegraphics[width=0.4\paperwidth]{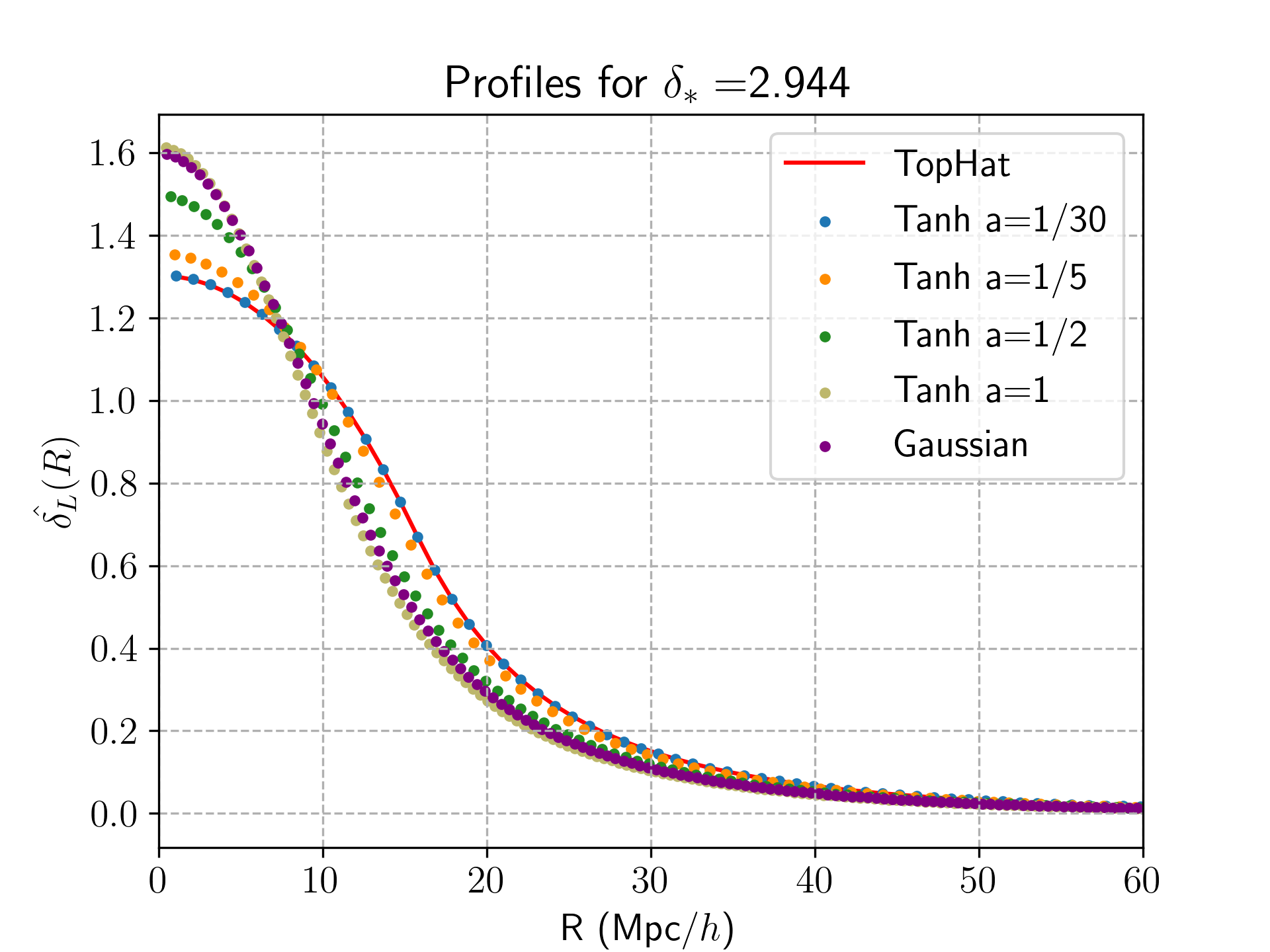}
\hspace*{-0.5cm}                                              
\includegraphics[width=0.4\paperwidth]{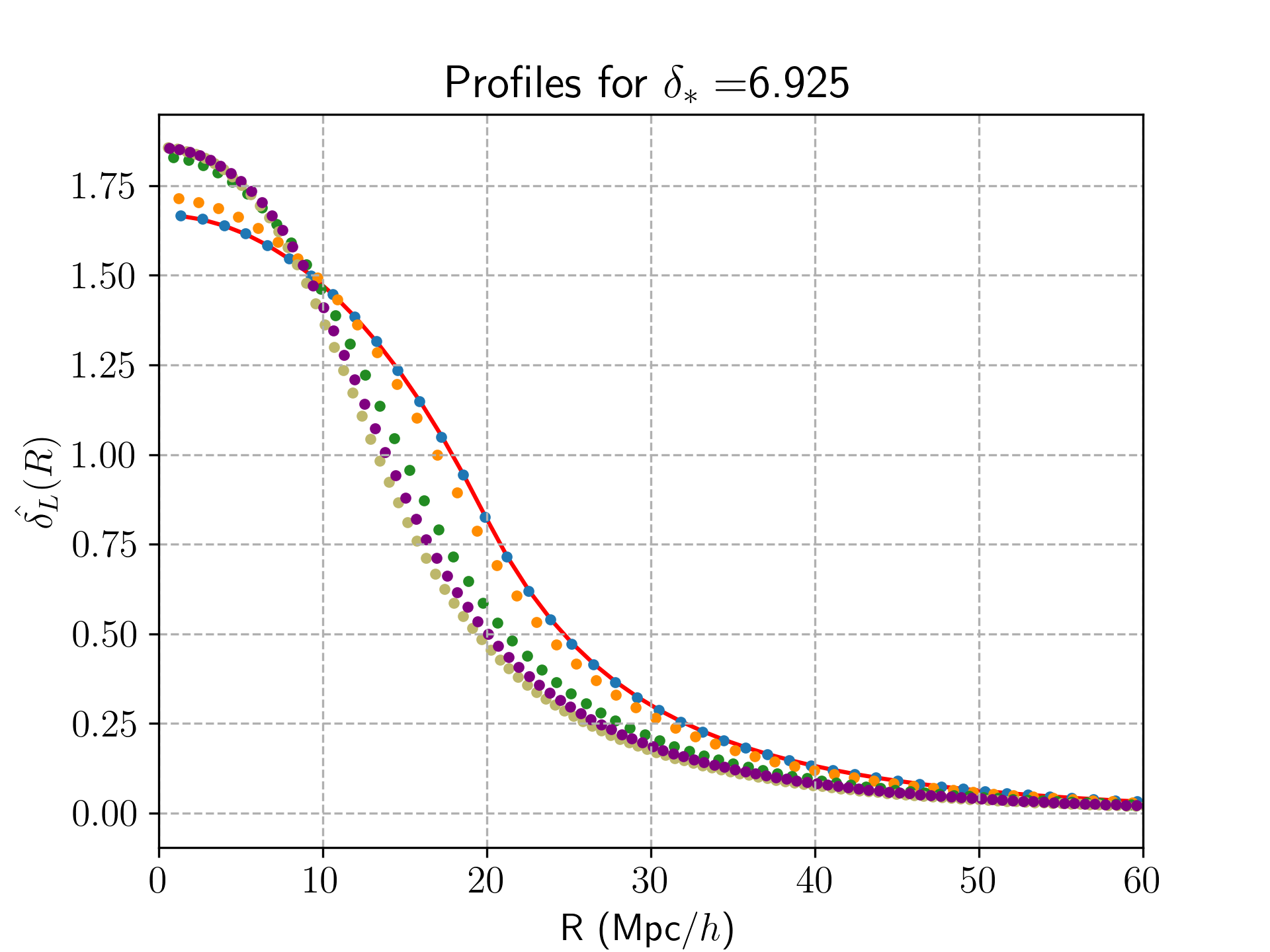}
\hspace*{-1cm}                                              
\includegraphics[width=0.4\paperwidth]{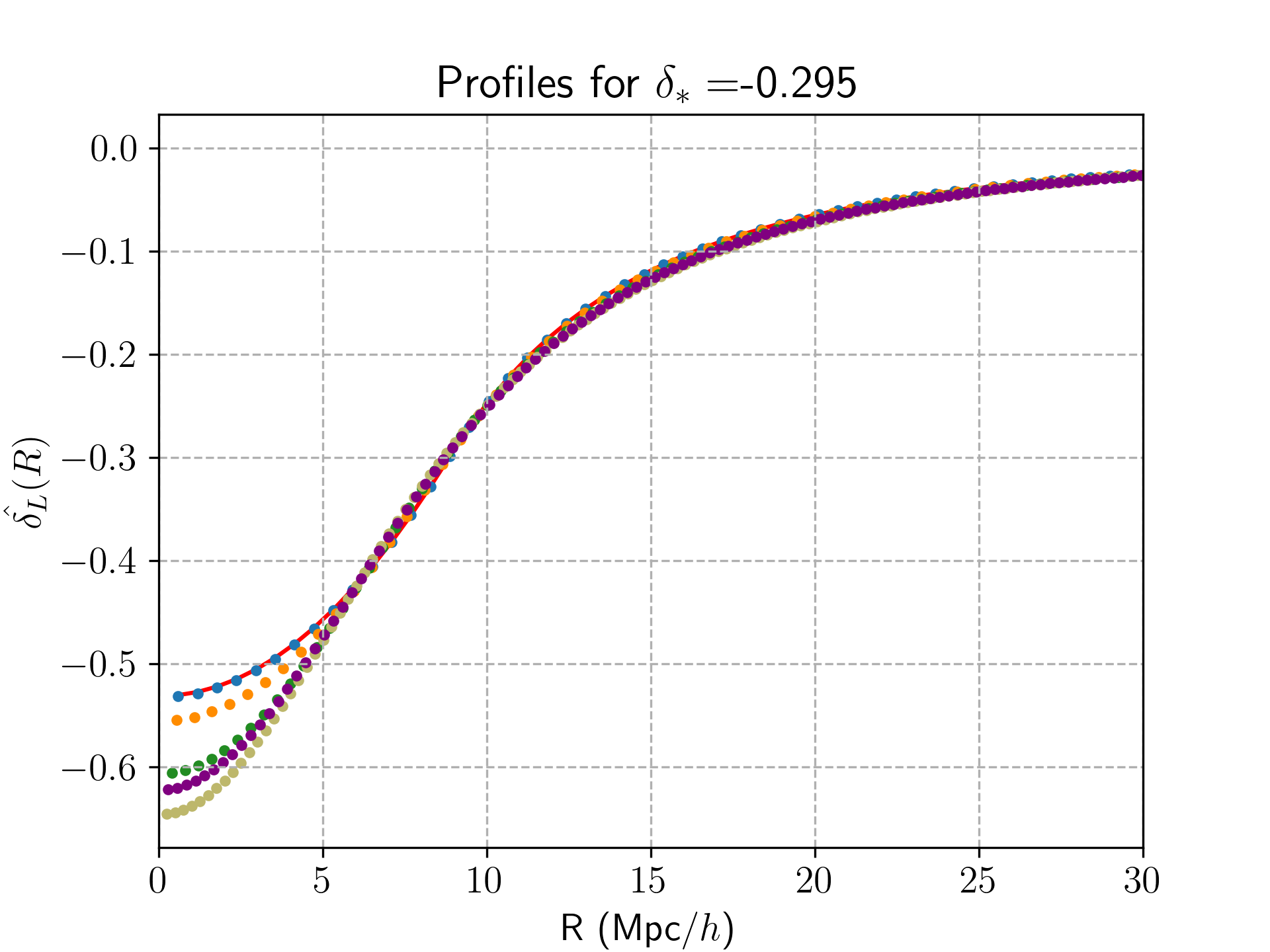}
\hspace*{-0.5cm}                                              
\includegraphics[width=0.4\paperwidth]{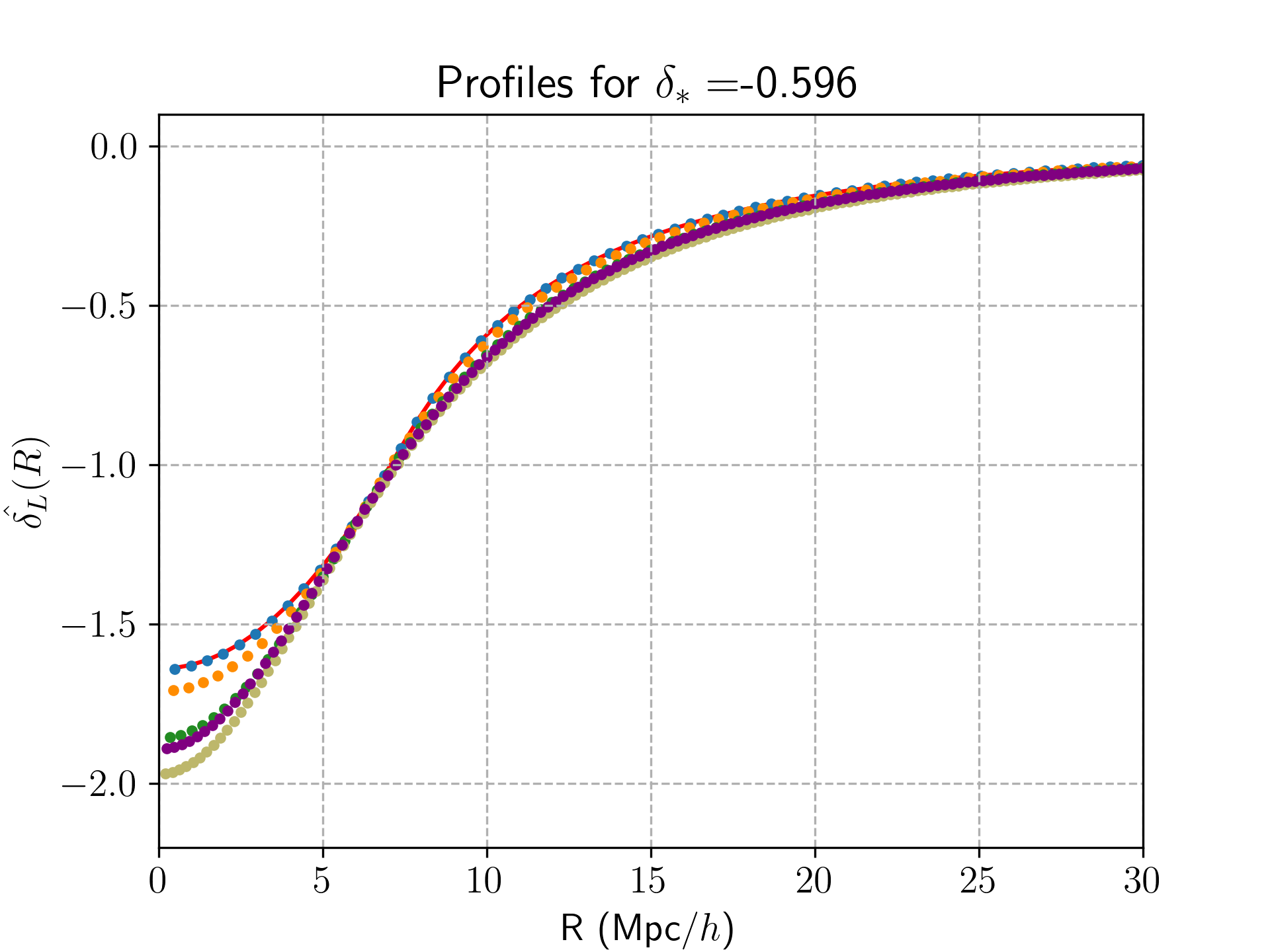}
\caption{Saddle-point linear density profiles for several density contrasts filtered with the Tanh and Gaussian window functions (dots). Analytic profile for the TopHat filter is shown for comparison (solid line). Top panels show the overdense profiles, with underdense profiles in the bottom. }
\label{fig:prof_comp}
\end{figure}

Figure~\ref{fig:prof_comp} shows examples of the saddle-point linear density profiles obtained numerically for different filters and several values of the density contrast\footnote{Note that some density profiles exceed the critical overdensity $\delta_c=1.686$ in the center and thus lead to shell-crossing. As discussed above, this does not imply a breakdown of the numerical code since the filtered nonlinear density (\ref{dw2}) remains finite.} $\delta_*$. We use the lattice with the same parameters as in Sec.~\ref{ssec:TopHat}: $(N,D)=(450,30)$ for $\delta_*\leq-0.4$ and (300,20) for $\delta_*>-0.4$. 
In Appendix~\ref{app:figs} we verify that these parameters ensure the proper convergence of the action minimization code for the considered filters by varying $N$ and $D$ away from the fiducial values and checking that the variation of the saddle-point action does not exceed $\sim 2\%$. 
We emphasize that this test is important since the action for the smooth filters typically exhibits slower convergence than in the TopHat case.
The variation of the action is generally stronger at underdensities leading to the maximal systematic uncertainty of the PDF of order $20\%$ at the extreme underdense tail, $\delta_*<-0.8$.
This appears acceptable in view of extreme rarity of such underdensities implying a large statistical uncertainty of their PDF.  
In principle, further increase of the accuracy of the spherical PDF appears possible, but would require significant computational resources. 

The profiles in Fig.~\ref{fig:prof_comp} are compared to the profiles for the TopHat filter obtained from the analytic expression (\ref{TH_analytic}). We observe that the Tanh profiles with $a=1/30$ closely reproduce the TopHat results, both for overdensities and underdensities. As $a$ increases, the overdense profiles become higher and narrower implying that the smooth filters pick up matter distributions concentrated near the center of the cell. On the other hand, the underdense profiles only get deeper, with their width essentially unchanged. The profiles corresponding to the Gaussian filter lie in between the Tanh profiles with $a=1/2$ and $a=1$. 

An inspection of the linear density profiles yields the overdensity $\delta_*$, at which the shell crossing in the center of the cell occurs for a given filter. For the Tanh filters with $a=1/30,\;1/5,\; 1/2, \; 1$ and the Gaussian filter we find that the critical overdensities are $\delta_* \approx 7,\;6,\;4,\;3$ and $3$, respectively. Note that for $a=1/30$ the value is the same as for the TopHat filter. We see that smoother filters lead to shell crossing at lower $\delta_*$, consistently with our previous observation that they prefer more concentrated density configurations. To avoid uncontrollable errors associated with the shell-crossing, we restrict the PDF model for smooth filters to the overdensities below the critical values.

\begin{figure}[t]
\centering
\hspace*{-1cm}
\includegraphics[width=0.4\paperwidth]{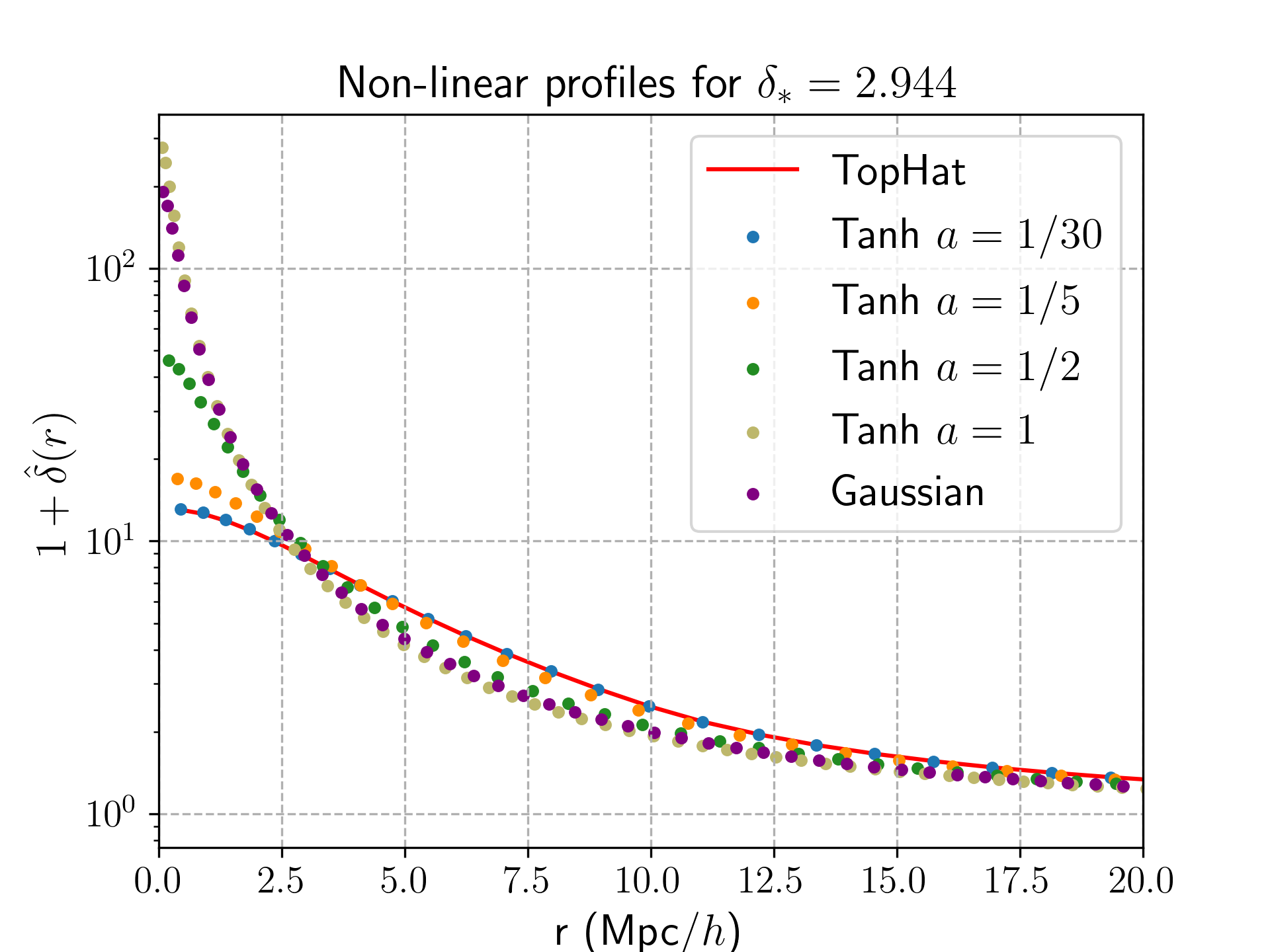}
\hspace*{-0.5cm}
\includegraphics[width=0.4\paperwidth]{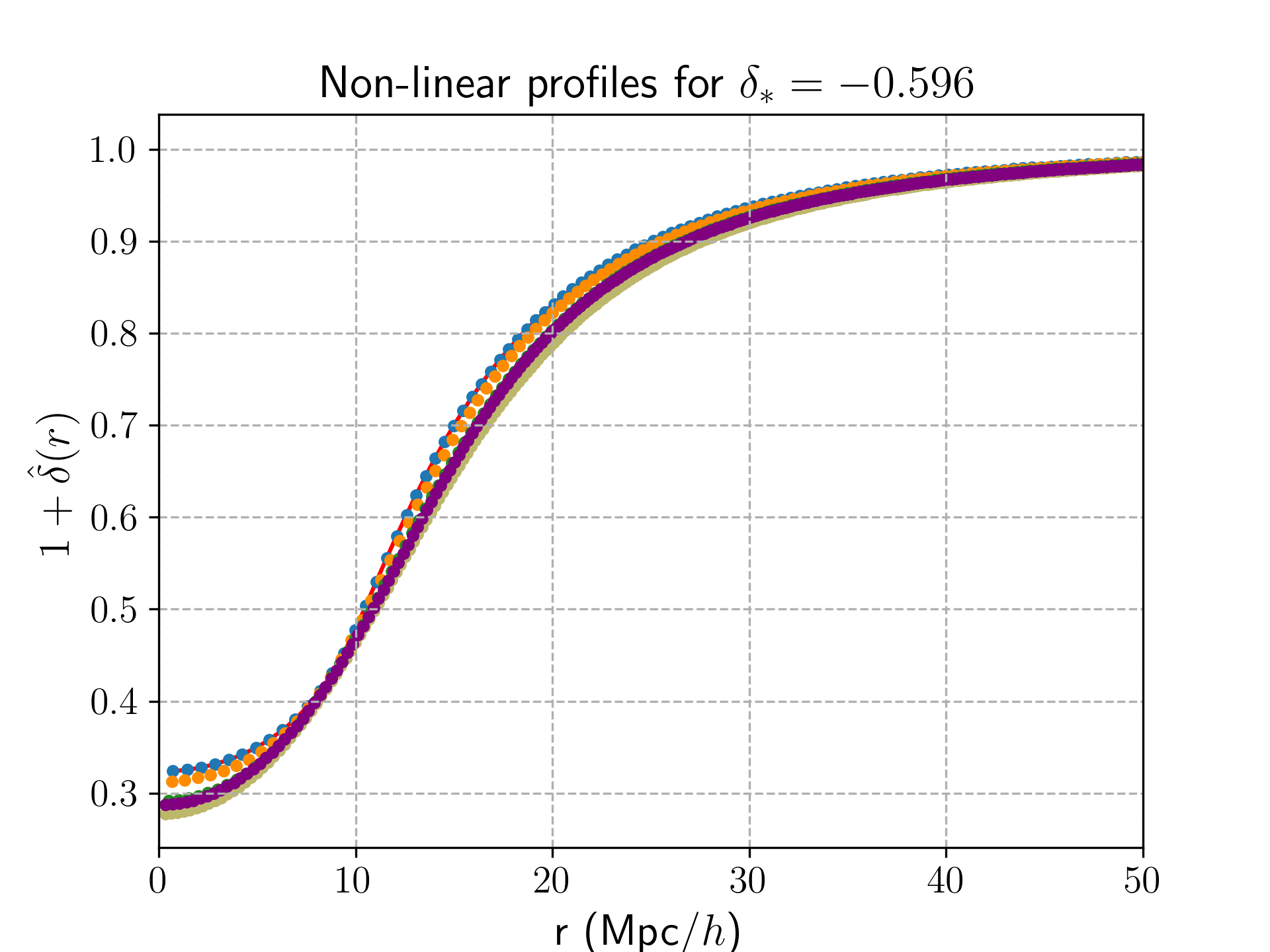}
\caption{Non-linear saddle-point density profiles for different filters and same values of overdensity ({\it left panel}) and underdensity ({\it right panel}). Note the logarithmic scale on the vertical axis in the left panel. }
\label{fig:nonlin_multi}
\end{figure}

To gain further insight into the density distributions contributing to PDFs for different filters, we show in 
Fig.~\ref{fig:nonlin_multi} the non-linear saddle-point density profiles for a representative overdensity and underdensity. In the overdense case we see that the central part of the profiles varies 
drastically, depending on the filter. In particular, the Gaussian filter pick up a central density more than two orders of magnitude larger than the TopHat filter. By contrast, in the underdense case the non-linear profiles exhibit remarkable universality and closely match each other at all radii, including the center.  
This suggests that the void profiles depend only on the overall underdensity $\delta_*$ and are insensitive to the shape of the filter used to define it, as long as the filter provides a given linear density variance.

\begin{figure}[t]
\centering
\hspace*{-1cm}                                                           
\includegraphics[width=0.75\paperwidth]{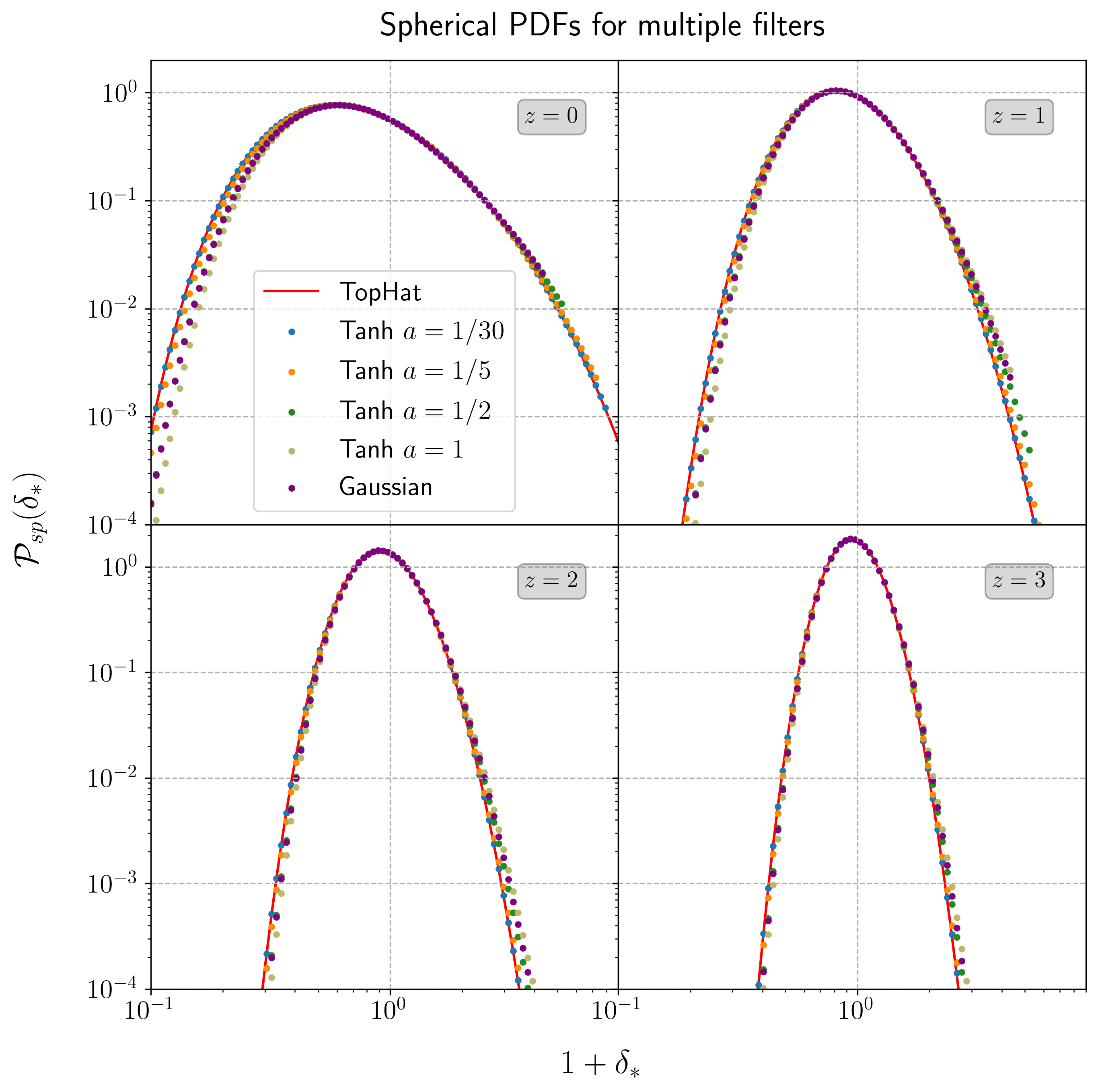}
\caption{Spherical PDFs for the Tanh and Gaussian filters compared to the TopHat PDF at various redshifts ($z=0,1,2,3$). The PDFs are restricted to the values of $\delta_*$ below the critical overdensity leading to shell-crossing. }
\label{fig:PDF_multi}
\end{figure}

Figure~\ref{fig:PDF_multi} presents the spherical PDFs ${\cal P}_{sp}(\delta_*)$ for different filters and for several redshifts. The PDFs are built from the saddle-point action and the monopole prefactor computed with our code. It is worth stressing that these quantities are redshift-independent, so computing them once allows us to obtain the PDF at arbitrary redshift by simply rescaling the action and 
 prefactor with the linear growth factor $g(z)$ according to Eqs.~(\ref{SPDF}), (\ref{monopre}). The data are restricted to the densities below the critical overdensity leading to shell-crossing and the ensuing break-down of the spherical collapse map (\ref{SCmap}).
For comparison, we also show the analytic TopHat PDF (\ref{TH_PDF}). 
  
  All PDFs coincide in the central region $\delta_*\sim 0$. This is expected since we ensured that all filters correspond to the same linear density variance and thus are equivalent at the Gaussian level. By contrast, the close similarity of all spherical PDFs in the whole overdense region $\delta_*>0$ seen in Fig.~\ref{fig:PDF_multi} is surprising. The coincidence becomes almost perfect at $z=0$, which is striking, given  
  the vast difference between the density profiles dominating PDFs for different filters shown in the left panel of Fig.~\ref{fig:nonlin_multi}.

The difference between different PDFs in Fig.~\ref{fig:PDF_multi}
is more pronounced for underdensities, reaching up to an order of magnitude at $z=0$ and $\delta_*=-0.9$. This may appear to contradict our earlier observation that the non-linear underdensity profiles weakly depend on the choice of the filter. The contradiction is resolved by recalling that the leading exponential part of the PDF (\ref{expalpha}) is sensitive to the linear density profile. Due to the properties of the spherical collapse map (\ref{SCmap}), similar non-linear underdensities can correspond to significantly different linear profiles, as seen by comparing the right panel of Fig.~\ref{fig:nonlin_multi} with the lower right panel of Fig.~\ref{fig:prof_comp}. Note that extreme underdensities correspond to voids and 
measuring even a factor of a few difference in their PDF would be challenging due to the difficulty with the precise observational charaterization of the mass distribution in them, see \cite{Contarini:2026yfv} and references therein. 

Figure~\ref{fig:PDF_multi} shows that within $\Lambda$CDM various monotonic filters chosen to yield the same linear matter density variance lead to very similar spherical PDFs, essentially coinciding with the analytic result (\ref{TH_PDF}). As we discuss in Sec.~\ref{sec:asp} below, this universality extends also to the full PDFs including the aspherical prefactors. It appears to be a robust property of $\Lambda$CDM and can provide a 
null test of the consistency of the data with the standard cosmology. 
It will be interesting to understand if the degeneracy between different filters gets broken in extended cosmological models and can be used to discriminate between them.

\subsection{Hockey-Stick filter}\label{subsec:res_DWell}

\begin{figure}[t]
    \centering
\includegraphics[width=0.6\linewidth]{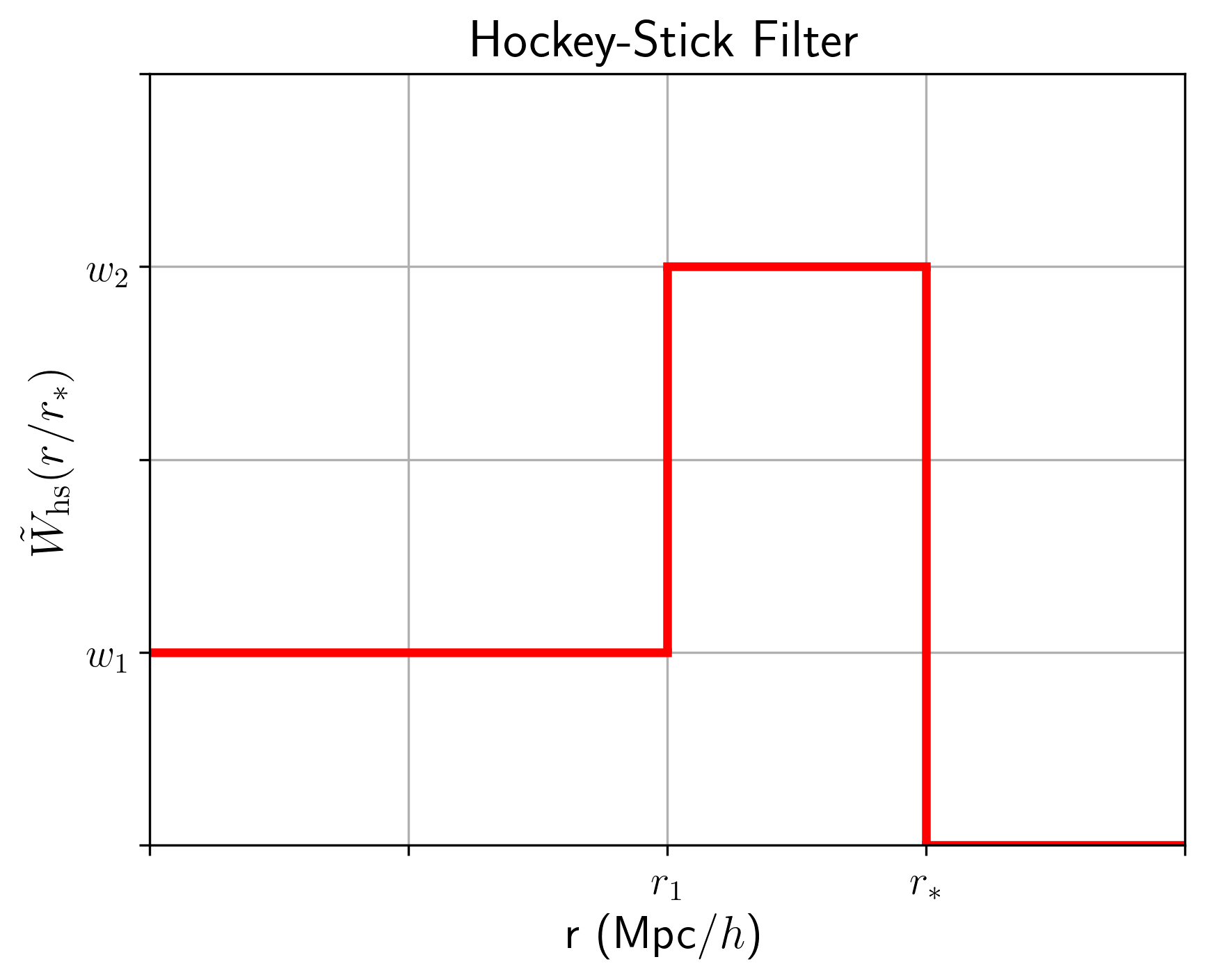}
    \caption{Non-monotonic Hockey-Stick filter.}
    \label{fig:DW}
\end{figure}

\begin{figure}[t]
\centering
\hspace*{-1cm}                                              
\includegraphics[width=0.34\paperwidth]{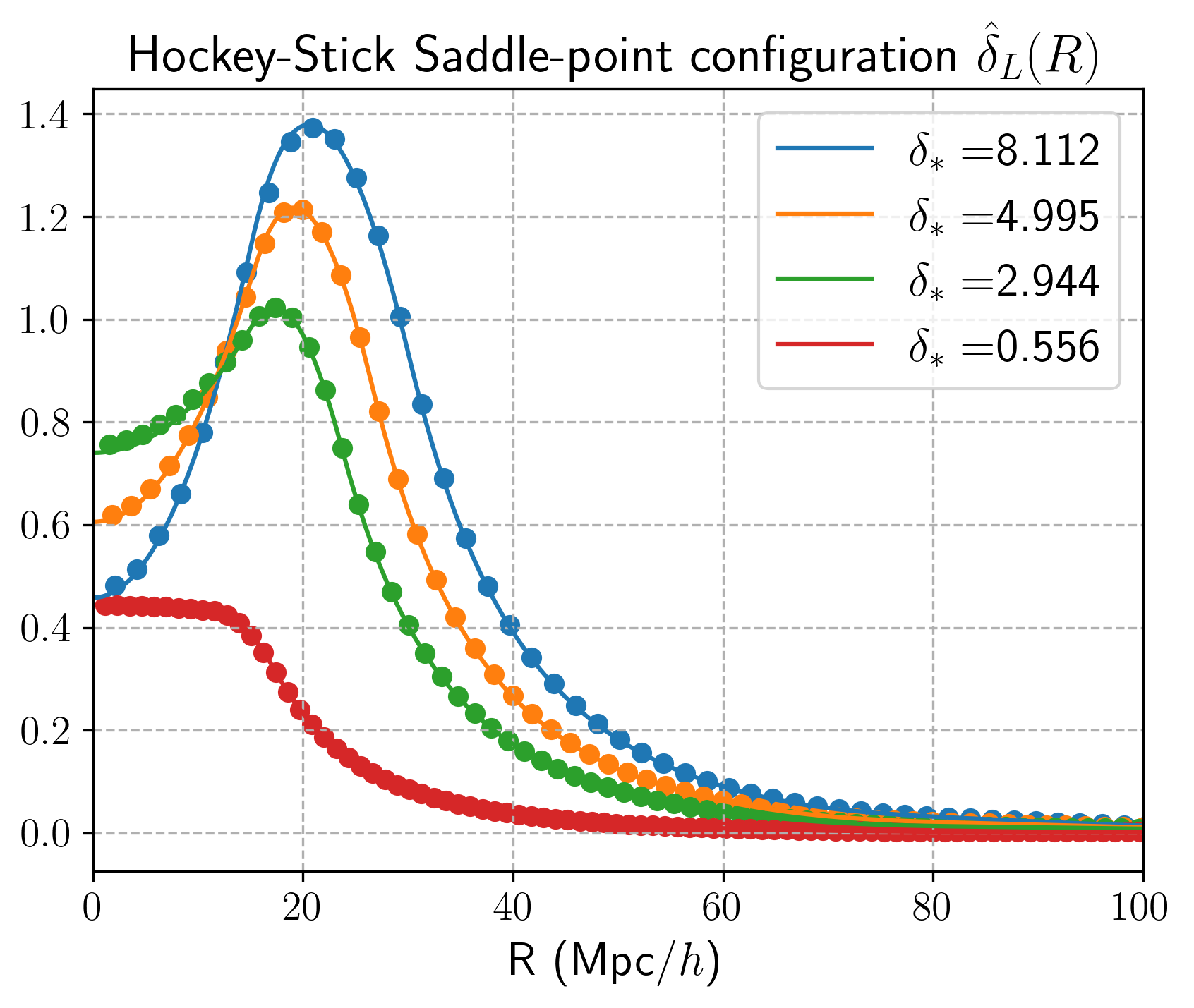}
\hspace*{0.5cm}                                            
\includegraphics[width=0.35\paperwidth]{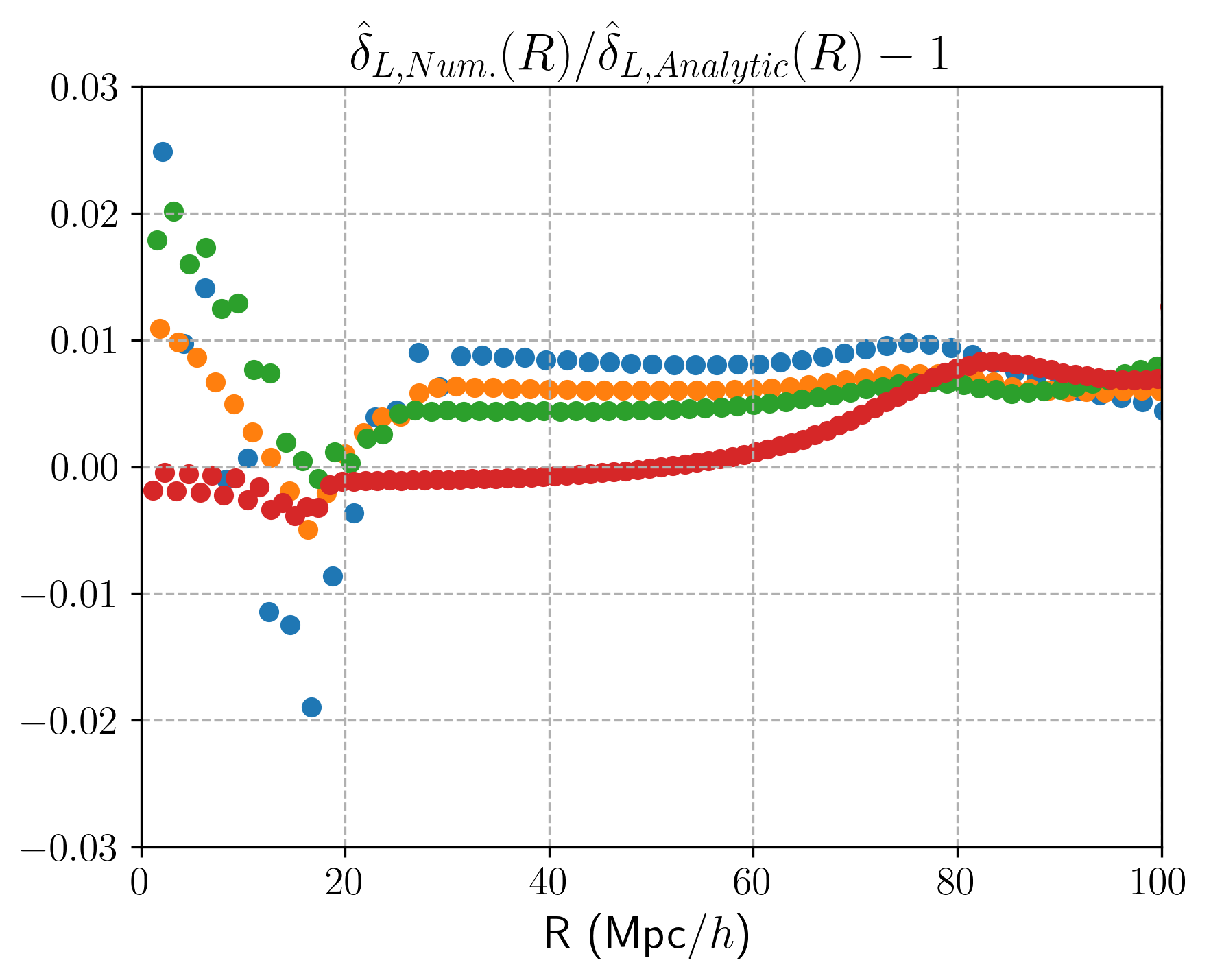}
\hspace*{-1cm}                                              
\includegraphics[width=0.34\paperwidth]{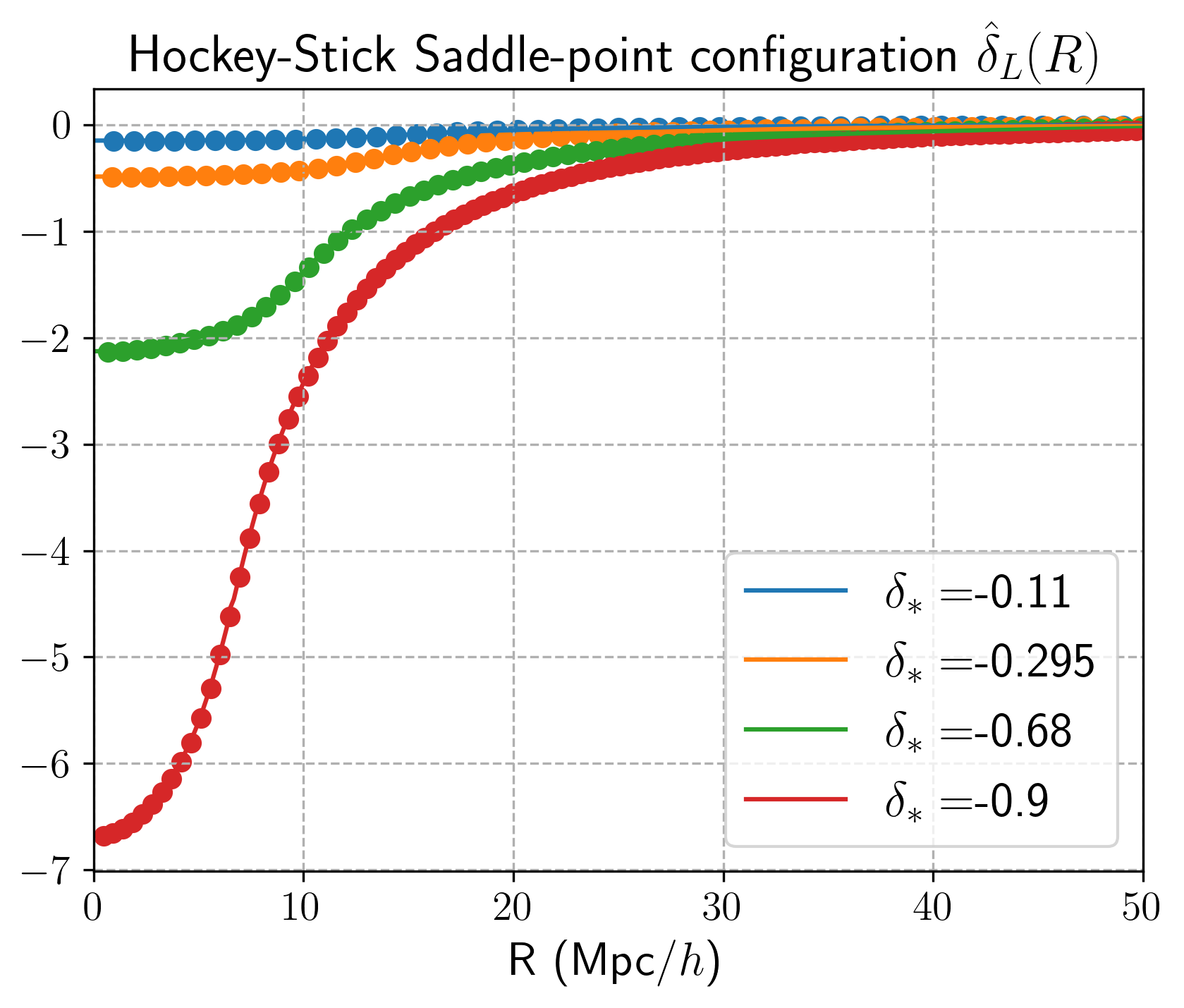}
\hspace*{0.5cm}                                            
\includegraphics[width=0.35\paperwidth]{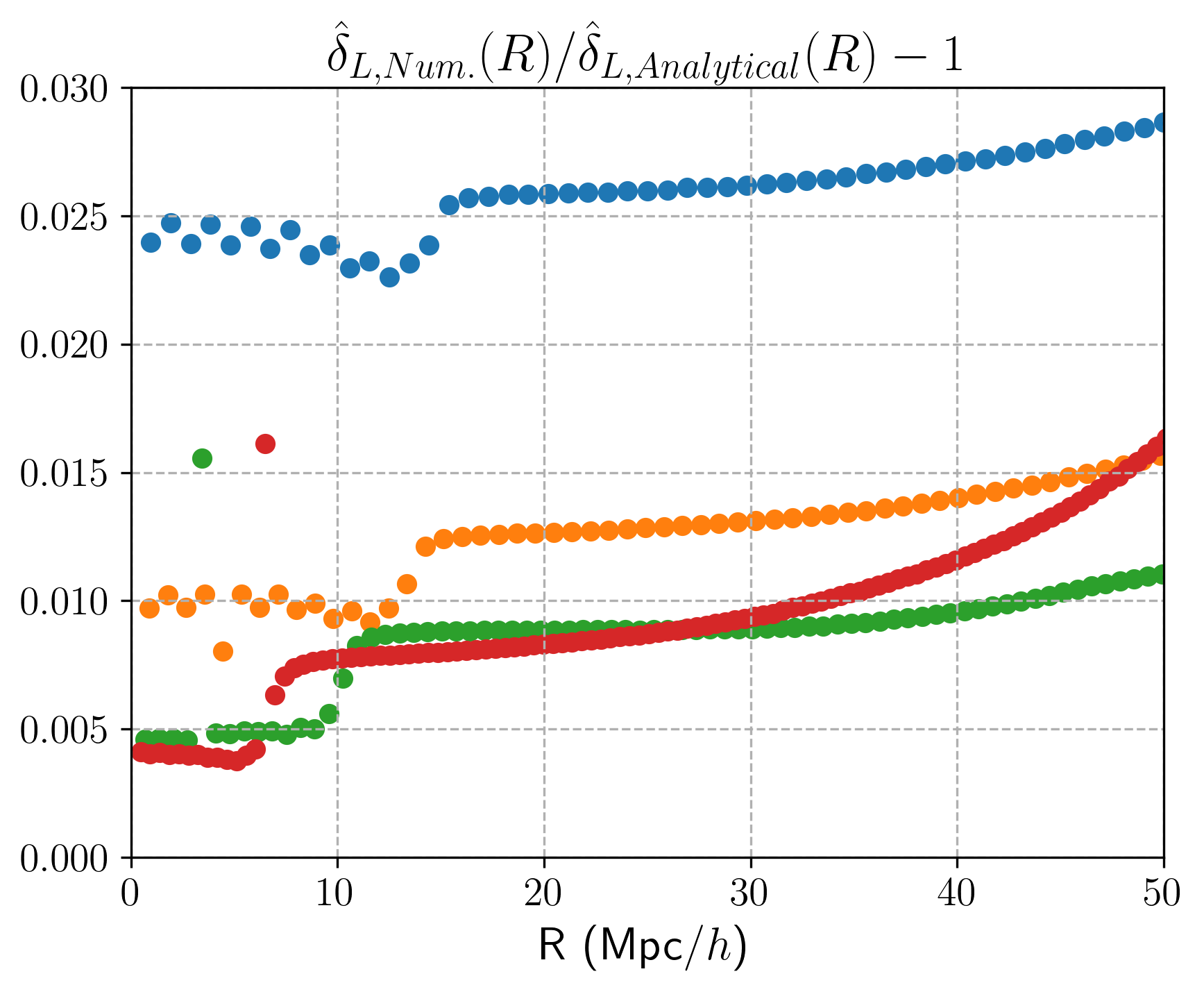}
\caption{{\it Left panels}: Numerical (dots) and analytical (solid line) saddle-point linear density profiles for several values of~$\delta_*$ for the Hockey-Stick filter (\ref{DWmain}). The filter parameters are $r_*=15\,{\rm Mpc}/h$, $\alpha=1/3$, $\beta=2/3$.
The upper panel shows overdensities, $\delta_*>0$, while the lower panel shows underdensities, $\delta_*<0$. {\it Right panels}: Residuals between the numerical and analytical profiles.}
\label{fig:dw_prof}
\end{figure}

So far, we have studied monotonic filters that qualitatively resemble TopHat. We now consider a filter of different type shown in Fig.~\ref{fig:DW}. It peaks at a finite distance from the center implying that it is more sensitive to the amount of matter in a spherical shell around the cell, rather than in its inner part. We will refer to this filter as  `Hockey-Stick'. It is given by a linear combination of two TopHat window functions, parameterized, apart from $r_*$, by two dimensionless ratios, $\alpha=w_1/w_2$ and $\beta=r_1/r_*$:
\begin{subequations}
\label{DWmain}
    \begin{gather}
    \label{DWmain1}
        \Tilde{W}_{\text{hs}}(r/r_*)= \frac{1}{(\alpha -1)\beta^3 +1} \left[ (\alpha -1) \Tilde{W}_{\text{th}}\left(\frac{r}{\beta r_*}\right) + \Tilde{W}_{\text{th}} \left( \frac{r}{r_*}\right) \right]  \\
        \label{DWmain2}
        \Longleftrightarrow \quad
        W_{\text{hs}}(kr_*)= \frac{1}{(\alpha -1)\beta^3 +1} \big[ (\alpha -1)\beta^3\, W_{\text{th}}(\beta kr_*) + W_{\text{th}}(kr_*) \big].
    \end{gather}
\end{subequations}
The spherical PDF with this filter admits analytic treatment which is presented in Appendix~\ref{app:DWell}. Here we apply to it our numerical pipeline. Our purpose is twofold. First, we subject the pipeline to a stress-test, confirming its ability to converge to the right solution for filters beyond monotonic examples. Second, we further explore the universality of spherical PDFs for different filters found in Sec.~\ref{ssec:smooth}. For numerical calculations we use the parameters $r_*=15$ Mpc/$h$, $\alpha=1/3$, $\beta=2/3$ (corresponding to $r_1=10\;{\rm Mpc}/h$). The linear variance (\ref{var_L}) for this choice is $\sigma_{L,r_*}^2=0.251$; it coincides with the variance for a TopHat filter with radius $\tilde r_*=15.627\;{\rm Mpc}/h$.

The linear density profiles minimizing the action for various $\delta_*$ are shown in  
Fig.~\ref{fig:dw_prof}. We observe that they are non-monotonic for high overdensities, reflecting the shape of the Hockey-Stick filter. On the other hand, at underdensities and low overdensities, the monotonicity of the profiles is preserved. The figure also compares the numerical profiles to the analytic results from Appendix~\ref{app:DWell}. The overall agreement is good, the residuals remaining within $\sim 2\%$ throughout the considered range of density contrasts. 

\begin{figure}[t]
\hspace*{-1cm}                                              
\includegraphics[width=0.45\paperwidth]{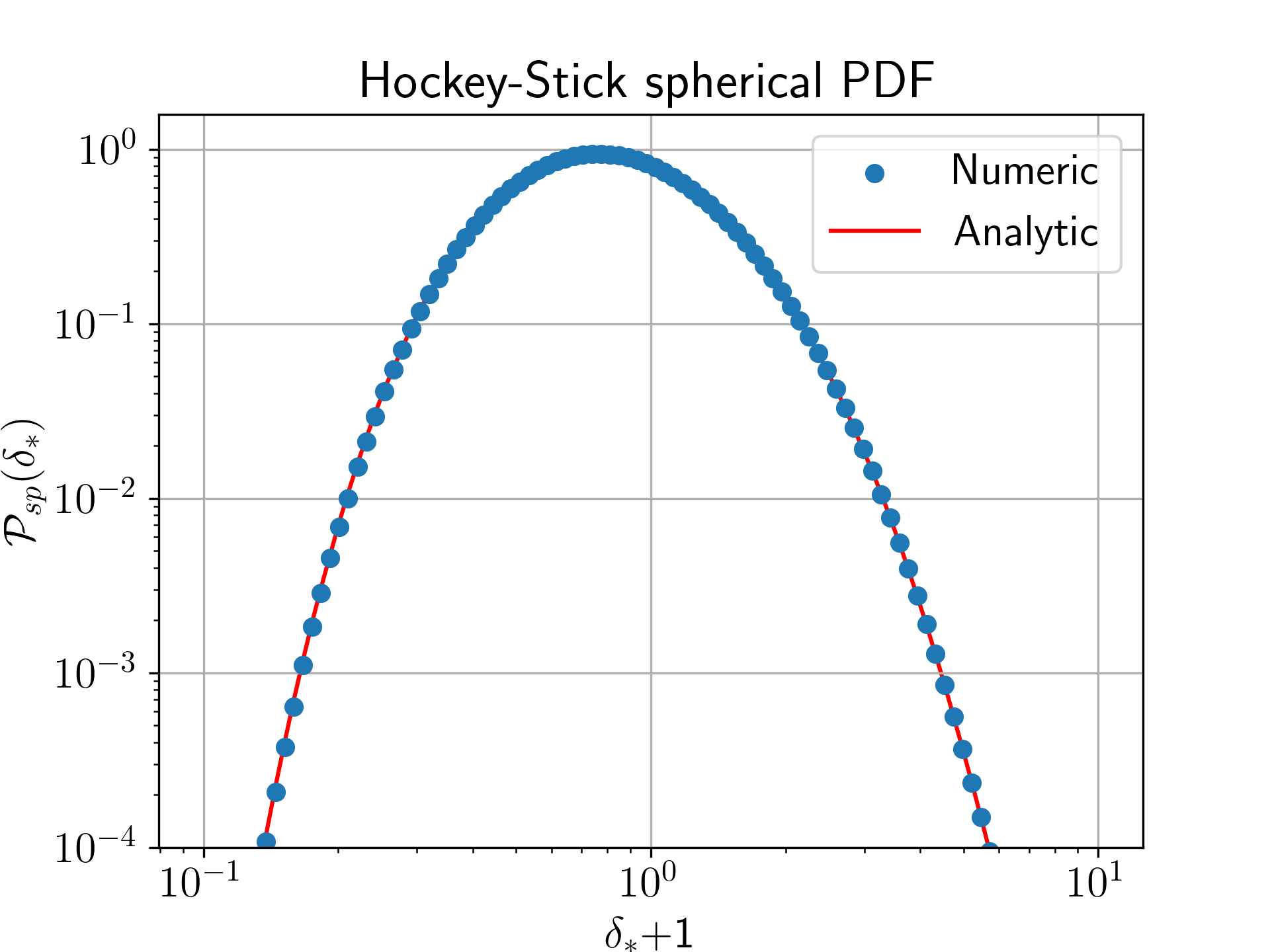}
\hspace*{-1cm}                                            
\includegraphics[width=0.45\paperwidth]{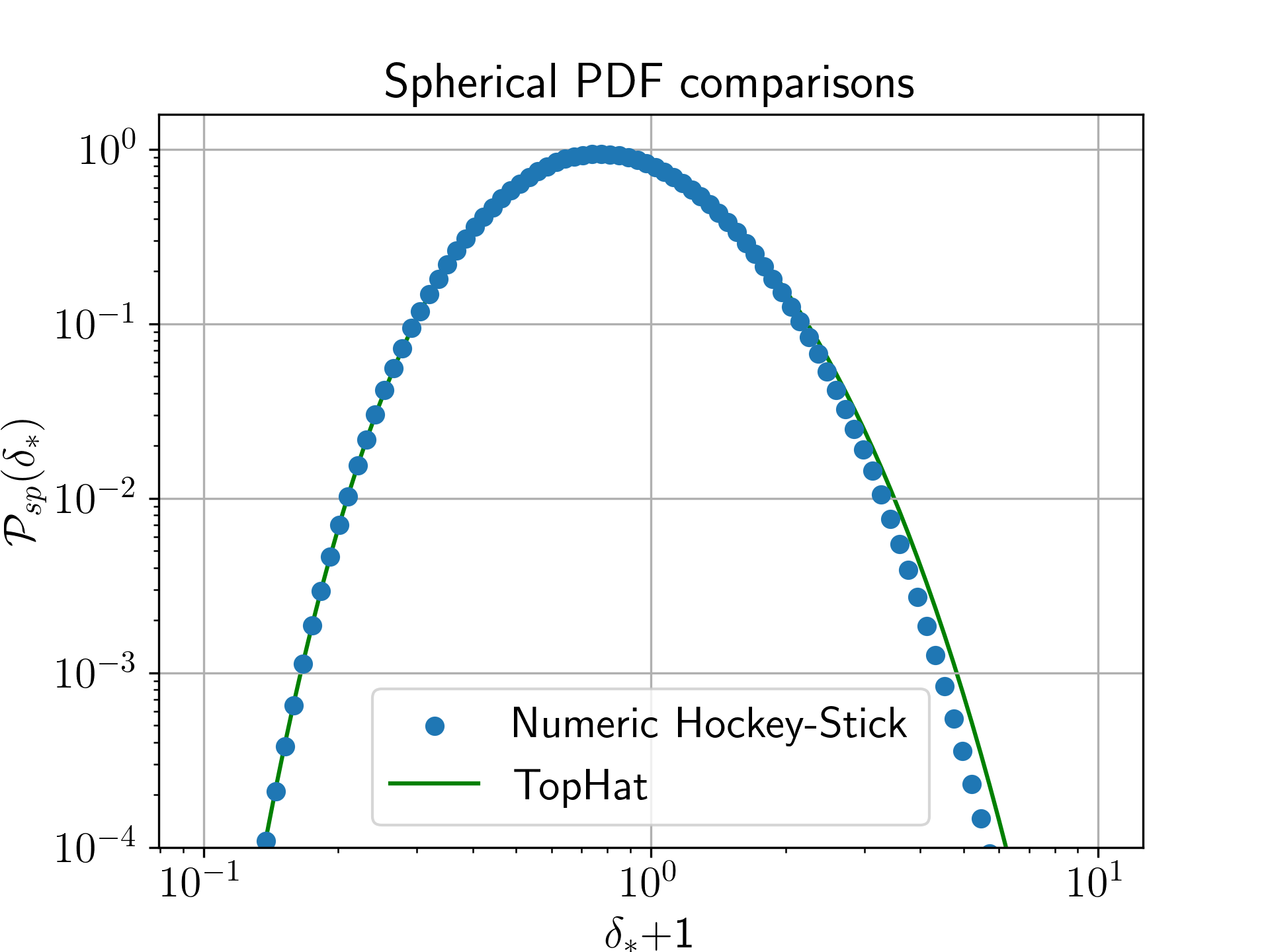}
\caption{\textit{Left panel}: Spherical PDF for the Hockey-stick filter coming from the numerical pipeline (blue dots) and analytic results (red line). \textit{Right panel}: Plots of the spherical PDF for the Hockey-Stick and the TopHat filter (green solid) with $\tilde{r}_*=15.627$ Mpc/$h$. }
\label{fig:dw_pdfs}
\end{figure}

 The spherical PDF for the Hockey-Stick filter is shown in Fig.~\ref{fig:dw_pdfs}. We see that numerical and analytical results are practically indistinguishable. An analysis of the residuals (not shown) reveals that they are below $2\%$ in the central part ($0.3<1+\delta_*<6$) and increase to $\sim 8\%$ at the tails. In the right panel we compare the Hockey-Stick spherical PDF to the spherical PDF for the TopHat filter with radius $\tilde r_{*}=15.627\;{\rm Mpc}/h$ which shares the same linear variance. We observe that the two PDFs practically coincide, except a minor difference at large overdensities. This is striking, given the marked difference between the two filters and the respective saddle-point density profiles. We have here one more manifestation of the remarkable universality of the PDF in the $\Lambda$CDM cosmology.

\section{Perturbative examination of the aspherical prefactor}
\label{sec:pert}

The spherical PDF captures the leading dependence of the PDF on the density contrast. However, one still needs to multiply it by the aspherical prefactor (\ref{ASP}) to get the actual PDF. While ${\cal A}_{asp}(\delta_*)$ exhibits relatively mild dependence on $\delta_*$, it is conceptually important for ensuring the vanishing of the mean density contrast; see the second condition in (\ref{norms}) \cite{ivanov}. Besides, it bears information about the sensitivity of the PDF to short-distance physics and connects to renormalization of the correlation functions within the EFTofLSS \cite{anton}.

An algorithm for precise evaluation of ${\cal A}_{asp}(\delta_*)$ was developed in \cite{ivanov,anton} for the case of the TopHat filter. It requires a significant numerical effort and we do not attempt to generalize it here for arbitrary filters. Instead, we obtain a perturbative expression for ${\cal A}_{asp}(\delta_*)$ at small values of $\delta_*$. Comparison with the results of N-body simulations in the next section shows that this expression is surprisingly accurate in the whole range $|\delta_*|<1$, in particular it provides a good description of the aspherical prefactor in the whole range of considered underdensities $-0.9<\delta_*<0$.

The first two terms in the Taylor expansion of ${\cal A}_{asp}(\delta_*)$ at $\delta_*=0$ are given in Eqs.~(\ref{Aaspprops}) which are straightforward consequence of the consistency conditions (\ref{norms}); see the derivation in Appendix~\ref{app:C1}. Here we want to go a step further and obtain the contribution ${\cal O}(\delta_*^2)$. To this end, we use the exact relation 
\cite{ivanov}, 
\begin{equation}\label{asp_AtotD0}
    \mathcal{A}_{asp}= \sqrt{\frac{D_0}{D_{tot}}}, 
\end{equation}
where $D_0$ and $D_{tot}$ are determinants of the operators acting on spherical and general density fluctuations, respectively:
\begin{subequations}
\label{DDS}
    \begin{align}
\label{D0}
    &D_{0}= \det \Big[ \mathbbm{1}(k_1,k_2)+2\hat{\lambda}\sqrt{P(k_1)}\,Q_0(k_1,k_2)\sqrt{P(k_2)}\Big]\;,\\
\label{Dtot}
    &D_{tot}=\det\Big[ \mathbbm{1}({\bf k}_1,{\bf k}_2) +2\hat{\lambda}\sqrt{P(k_1)}\,
    Q_{tot}({\bf k}_1,{\bf k}_2)\sqrt{P(k_2)}\Big]\;. 
 \end{align}
\end{subequations}
Here $\mathbbm{1}(k_1,k_2)$ is the monopole unit operator (\ref{k_unit}), 
$\mathbbm{1}({\bf k}_1,{\bf k}_2)=(2\pi)^3\delta_D({\bf k}_1-{\bf k}_2)$ is its 3-dimensional analog, $\hat\lambda$ is the saddle-point value of the Lagrange multiplier, and the kernels $Q_{tot}({\bf k}_1,{\bf k}_2)$ and $Q_0(k_1,k_2)$ are defined in Appendix~\ref{app:kernal}, Eqs.~(\ref{Qtotapp}), (\ref{Q0}). 

From Eqs.~(\ref{expalpha}), (\ref{lam}) we infer that $\hat\lambda$ is of order ${\cal O}(\delta_*)$. In more detail, we have
\begin{equation}
\label{lam_exp}
    \hat{\lambda}=-\frac{\delta_*}{\sigma^2_{L,r_*}}+a_2\delta_*^2+{\cal O}(\delta_*^3)\;, 
\end{equation}
where we have used Eq.~(\ref{alphaTexp}) to fix the linear term. The coefficient $a_2$ can also be connected to the power spectrum and the window function. The calculation is rather tedious and is relegated to Appendix~\ref{asp_pert}; the final expression is given in Eq.~(\ref{a2app}). 

To capture the terms quadratic in $\delta_*$, it is sufficient to expand the determinants (\ref{DDS}) up to the second order in $\hat\lambda$. We write,
\begin{align}
\label{D0mass}
    \sqrt{D_0}=\exp \left[ \frac{1}{2}\Tr \ln \left( \mathbbm{1} + 2\hat{\lambda} \sqrt{P}\, Q_{0} \sqrt{P} \right)\right] \approx \exp \left[\hat{\lambda}\mathcal{T}_1  -  \hat{\lambda}^2\mathcal{T}_2\right]\;,
\end{align}
where we have defined
\begin{subequations}
\label{Taus}
\begin{align} 
\label{Tau1}
\mathcal{T}_1& \equiv \Tr(Q_0P)= \int \left[dk\right] \, P(k)\,Q_0(k,k)\;,\\
\label{Tau2}
\mathcal{T}_2&\equiv \Tr(Q_0PQ_0P) 
=  \int \left[dk\right] \, \left[dq\right] \, P(k)P(q)\big(Q_0(k,q)\big)^2\;.
\end{align}
\end{subequations}
Note that these quantities are functions of $\delta_*$. Taylor expanding 
\begin{equation}
   {\cal T}_1= {\cal T}_{10}+{\cal T}_{11}\delta_*+\ldots\;,\qquad
   {\cal T}_2= {\cal T}_{20}+\ldots\,
\end{equation}
and substituting into (\ref{D0mass}) we arrive at
\begin{equation}
\label{D0_exp}
    \sqrt{D_0}\approx\exp\biggl\{-\delta_*\frac{\mathcal{T}_{10}}{\sigma^2_{L,r_*}}+ \delta_*^2\biggl(-\frac{\mathcal{T}_{11}}{\sigma^2_{L,r_*}}+a_2 \mathcal{T}_{10}
    -\frac{\mathcal{T}_{20}}{\sigma^4_{L,r_*}}\biggr) \biggr\}. 
\end{equation}
Explicit expressions for ${\cal T}_{10}$, ${\cal T}_{11}$ and ${\cal T}_{20}$ are given in Eqs.~(\ref{t_eqs}). 

The determinant $D_{tot}$ is calculated in a similar way, with the results (see Appendix~\ref{asp_pert} for the details of the calculation),
\begin{equation}\label{Dtot_pert}
    \sqrt{D_{tot}}\approx \exp \left[- \delta_*^2\,\frac{\sigma^2_{\text{1-loop}}}{2\sigma_{L,r_*}^4}\right]\;, 
\end{equation}
where $\sigma_{\text{1-loop}}^2$ is the 1-loop correction to the filtered density variance, calculated within the EFTofLSS.\footnote{We define the 1-loop correction to the variance and the power spectrum as quantities that would be redshift-independent in SPT.
The full time dependent power spectrum including this correction then reads, 
\[
    P(k,z)=g^2(z)P(k)+g^4(z)P_{\text{1-loop}}(k,z)\;.
\]} 
It is expressed through the 1-loop correction to the power spectrum,
 \begin{equation}
 \label{sig-1loop_main}
        \sigma^2_{\text{1-loop}}=\int_{\textbf{k}} P_{\text{1-loop}}(k)\,|W(kr_*)|^2\;.
\end{equation}
Note that the latter must be properly renormalized, to remove the spurious contributions of the short-wavelength modes. This amounts to subtracting the EFT counterterm from the result obtained using the Standard Perturbation Theory \cite{EFT_ivanov}:
\begin{equation}
\label{Pcntr}
    P_{\text{1-loop}}(k,z)=P_{\text{1-loop}}^{\rm SPT}(k)-\frac{2\gamma(z)}{g^2(z)}\,k^2P(k)\;. 
\end{equation}
The counterterm coefficient $\gamma(z)$ is sometime called `effective sound speed'. Its redshift dependence is approximately $\gamma(z)\propto [g(z)]^{2.3}$ (see Eq.~(\ref{cntrgamma}) below), which introduces a weak redshift dependence into $\sqrt{D_{tot}}$, and hence into ${\cal A}_{asp}$. 
If not for the counterterm, ${\cal A}_{asp}(\delta_*)$ would be completely redshift independent.\footnote{As long as we neglect the 2-loop correction represented by the $\alpha_2 g^4$ term in Eq.~(\ref{PDFexpansion}).}

\begin{figure}[t!]
\centering
\hspace*{-0cm}
\includegraphics[width=0.5\paperwidth]{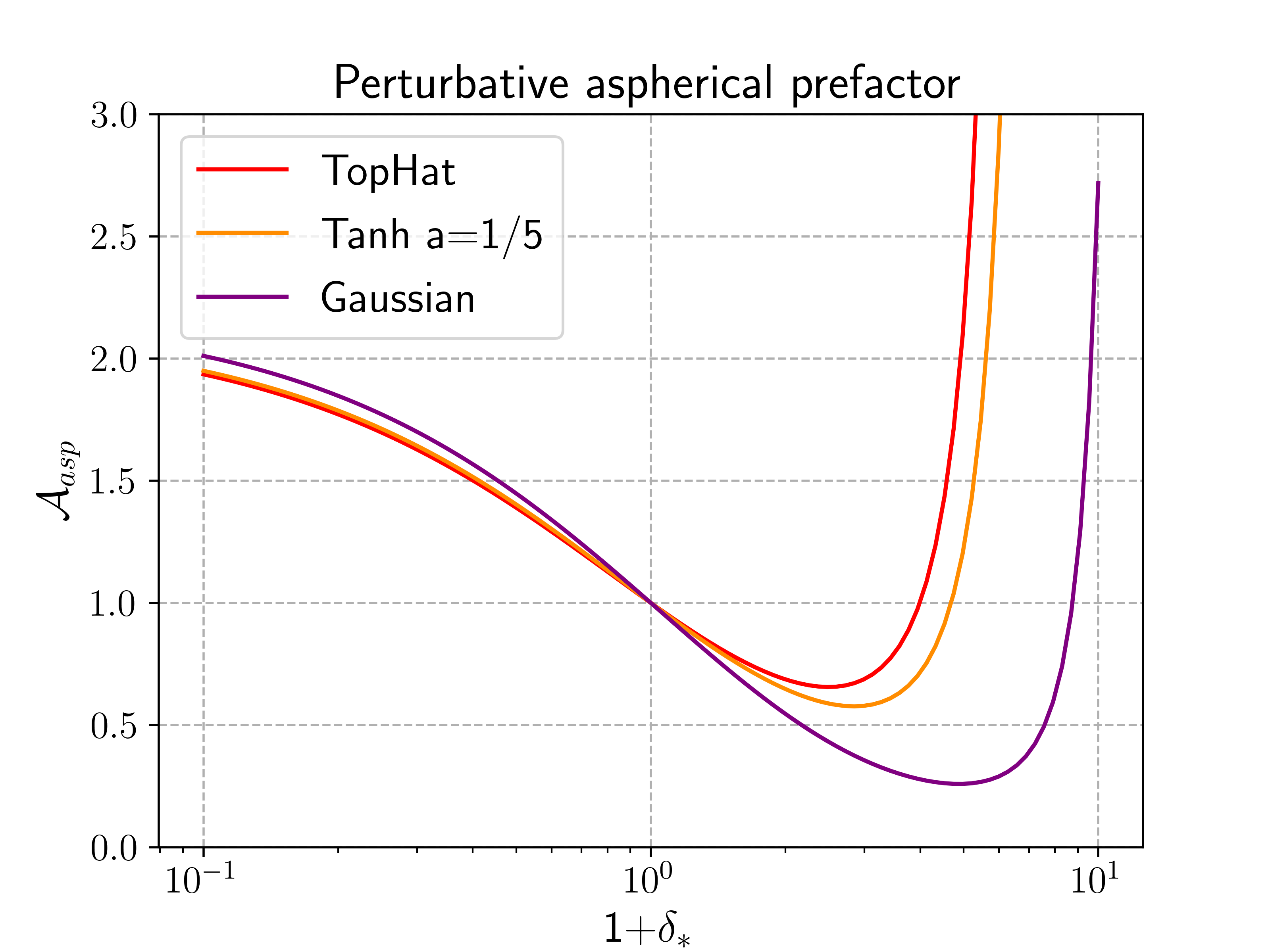}
\caption{ Perturbative approximation to the aspherical prefactor for various filters. 
The EFT counterterm is taken to be $\gamma=1.39$ (Mpc/$h)^2$, consistent with the N-body simulations used in Sec.~\ref{sec:asp} at $z=0$.  }
\label{fig:asp_analytic}
\end{figure}

Combining Eqs.~(\ref{asp_AtotD0}), (\ref{D0_exp}) and (\ref{Dtot_pert}) we obtain our final expression, 
\begin{equation}\label{asp_ren1}
    \mathcal{A}_{asp}(\delta_*)=\exp\biggl\{-\delta_*\frac{\mathcal{T}_{10}}{\sigma^2_{L,r_*}} + \delta_*^2\bigg(\frac{\sigma^2_{\text{1-loop}}}{2\sigma^4_{L,r_*}}-\frac{\mathcal{T}_{11}}{\sigma^2_{L,r_*}}+a_2 \mathcal{T}_{10}-\frac{\mathcal{T}_{20}}{\sigma^4_{L,r_*}}\bigg) 
    +{\cal O}(\delta_*^3)\biggr\}\,, 
\end{equation}
where we recall that $a_2$, ${\cal T}_{10}$, ${\cal T}_{11}$, ${\cal T}_{20}$ are given in Eqs.~(\ref{a2app}), 
(\ref{t_eqs}). Results of evaluation of this expression dropping the ${\cal O}(\delta_*^3)$ term are presented in Fig.~\ref{fig:asp_analytic} for the TopHat, Tanh and Gaussian filters. The abrupt upward turn of the curves at overdensities $\delta_*\gtrsim 1$ signals a breakdown of the Taylor expansion in $\delta_*$. On the other hand, at $\delta_*\lesssim 1$, including the whole range of underdensities, the curves are well-behaved and remarkably similar. We are going to see in Sec.~\ref{sec:asp} that in this range of density contrasts they provide a fair approximation to the exact ${\cal A}_{asp}(\delta_*)$.

The above analysis allows us to address and important aspect of the PDF --- its sensitivity to the short-distance (`ultraviolet, UV') physics. Since this physics is non-perturbative, it cannot be captured exactly by the saddle-point expansion, and must be instead modeled, in the spirit of the EFTofLSS, by a counterterm in the aspherical prefactor \cite{anton}. This `counterterm prefactor' is quite significant: for the TopHat filter, it rescales the PDF by a factor $\sim 0.7$ ($\sim 0.5$) at the underdense (overdense) tail. The ambiguities in its modeling introduce a systematic uncertainty in the PDF. 

One might think that the numerical importance of the counterterm prefactor is due to the sharp boundary of the TopHat filter, and for smooth filters it would be reduced. If this were true, PDF for smooth filters would enjoy smaller modeling error. Unfortunately, this does not seem to happen. We see it by comparing the counterterm prefactors for different filters at small values of $\delta_*$, where they are under perturbative control. Here the counterterm prefactor is directly related to the counterterm in the 1-loop density variance $\sigma_{\text{1-loop}}^2$. According to Eqs.~(\ref{sig-1loop_main}), (\ref{Pcntr}), the latter reads,
\begin{equation}
\label{var_ren}
    \sigma^2_{\text{1-loop}}=\sigma^2_{\text{1-loop}}\big|^{\rm SPT} - \frac{2\gamma(z)}{g^2(z)}\,\Sigma_{L,r_*}^2\;, \qquad \Sigma_{L,r_*}^2=\int_{\textbf{k}} k^2 P(k)\, |W(kr_*)|^2\;.
\end{equation}
The parameter $\gamma(z)$ is determined from the counterterm in the power spectrum and is the same for all filters. Thus, the sensitivity of the PDF to the counterterm is controlled by the quantity $\Sigma_{L,r_*}^2$ which we call `counterterm variance'. 
In Table.~\ref{table:renom}, we list the values of the counterterm variance for various smooth filters, relative to its value for the TopHat filter with radius $10\;{\rm Mpc}/h$. As before, all filters share the same linear variance $\sigma_{L,r_*}^2$. 
We see that $\Sigma_{L,r_*}^2$ differs at most by $10\%$, whence we conclude that the PDFs have comparable UV sensitivity for all filters.

\begin{table}[ht]
    \renewcommand{\arraystretch}{1.25}
    \centering
    \begin{tabular}{|c|c|c|c|c|c|}
    \hline
    & Gaussian & Tanh $a=1$ & Tanh $a=1/2$ & Tanh $a=1/5$  & Tanh $a=1/30$ \\ \hline
    $ \Sigma^2_{L,r_*}/\Sigma^2_{L,r_*| {\rm th}}$ &  0.9369 & 1.0009 &  0.9263 & 0.8967 & 0.9794  \\ \hline
    \end{tabular}
    \caption{Ratios of the counterterm variance for different filters compared to the TopHat.  }
\label{table:renom}
\end{table}

\section{Comparison with N-body simulations}
\label{sec:asp}

In this section, we compare our PDF model to the results of pure dark matter N-body simulations. 
We use the high-resolution (HR) \texttt{Quijote} dataset \cite{quijote}. The HR dataset consists of 100 realizations of a given $\Lambda$CDM cosmology which evolve $1024^3$ cold dark matter (CDM) particles in a cosmological volume of 1 ($h^{-1}$Gpc$)^3$ with a  gravitational softening length of 25 $h^{-1}$ kpc. Initial conditions for all simulations are generated at $z=127$ using second-order Lagrangian perturbation theory (2LPT). The simulation uses the following cosmological parameters, which closely follow the Planck 2018 best-fit values: $\Omega_\text{m}$=0.3175, $\Omega_\text{b}$=0.049, $h=$0.6711, $n_s$=0.9624, $\sigma_8$=0.834, $w=-1$, and massless neutrinos. Simulation snapshots are saved at redshifts $z=0$, $0.5$, $1$, $2$, and $3$. 

\begin{figure}[t!]
\centering
\hspace*{-0cm}
\includegraphics[width=0.5\paperwidth]{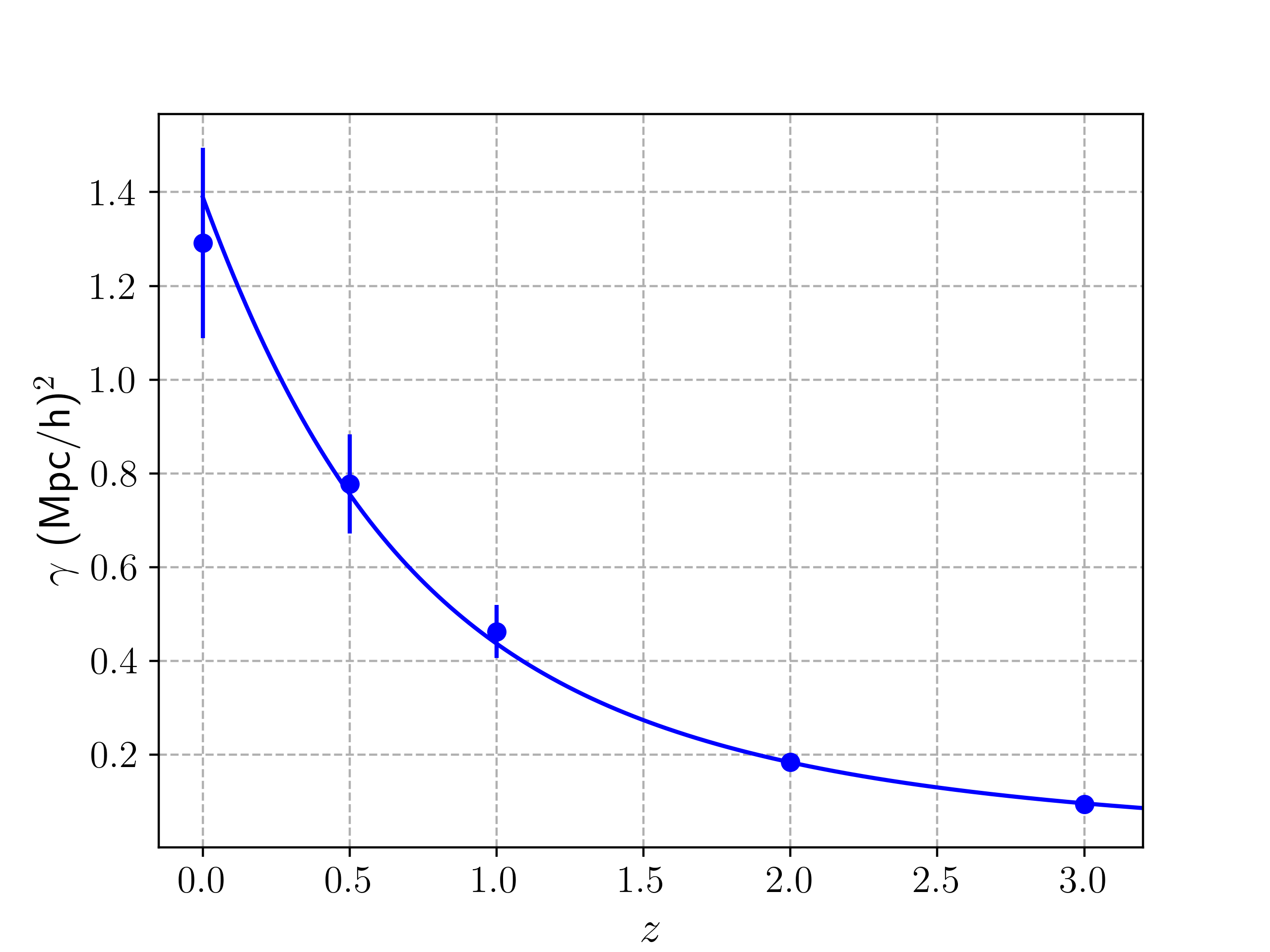}
\caption{ The EFT counterterm coefficient $\gamma(z)$ over various redshifts as measured from the \texttt{Quijote} simulations (dots), along with the fitted power-law (line). The errorbars include the statistical and theoretical uncertainty accounting for the omitted two-loop contribution.  }
\label{fig:soundspeed}
\end{figure}

For what follows, we need to determine the EFT counterterm coefficient 
$\gamma(z)$. We fit the power spectrum measured in the simulations with the theoretical prediction (\ref{Pcntr}). The 1-loop power spectrum is computed using \texttt{CLASS-PT} \cite{Chudaykin:2020aoj}.  
We then model the $\gamma(z)$ using the power-law template,
\begin{equation}
\label{cntrgamma}
    \gamma(z)=\gamma_0[g(z)]^m
\end{equation}
with $\gamma_0=1.39\pm0.14$ (Mpc/$h)^2$ and $m=2.30\pm0.12$. Figure~\ref{fig:soundspeed} shows the results of the fitted power-law, as well as the measured values at given redshift.

\subsection{PDF and aspherical prefactor from N-body}

\begin{figure}[t!]
\centering
\hspace*{-1cm}
\includegraphics[width=0.75\paperwidth]{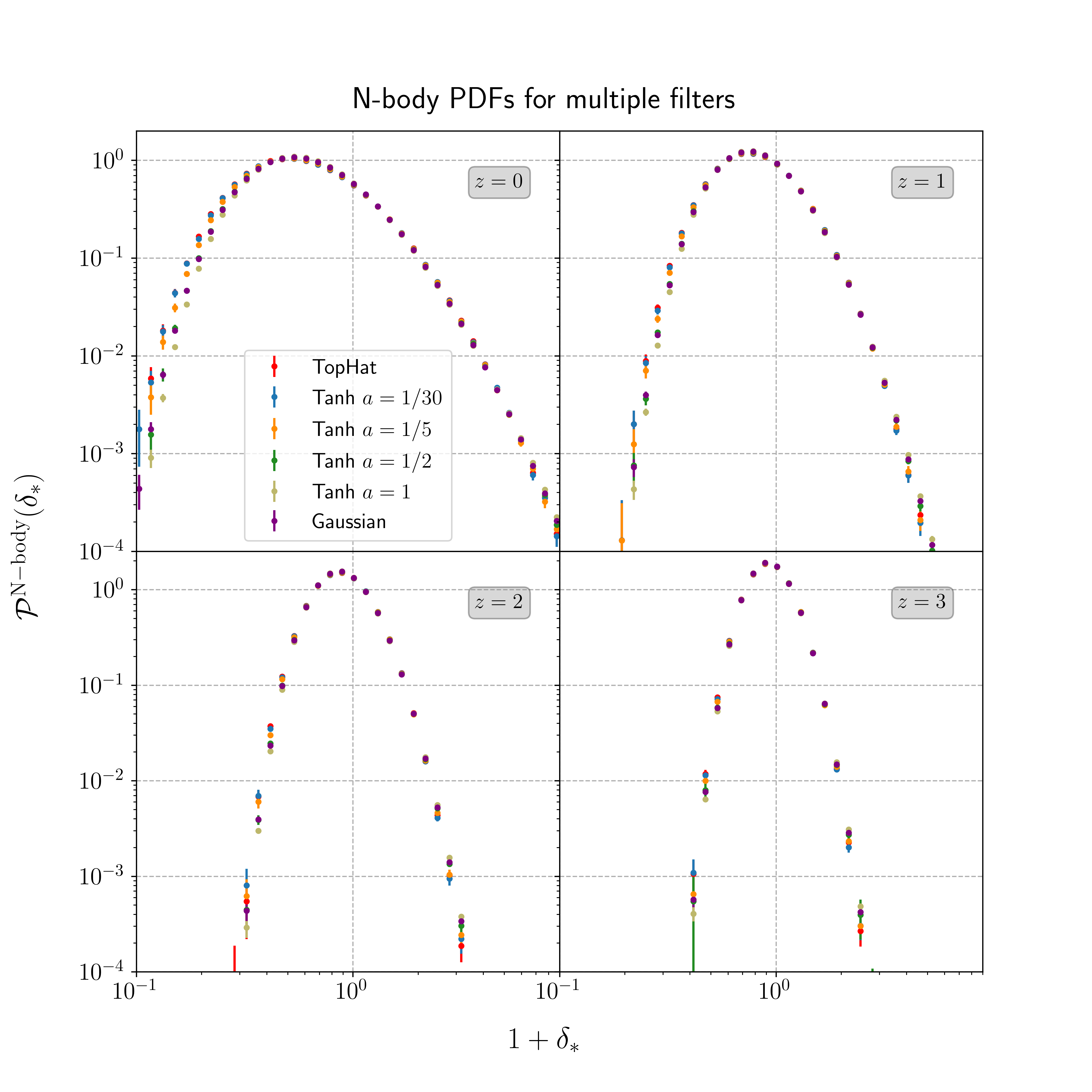} 
\caption{ Numerical PDFs constructed from the \texttt{Quijote} N-body simulations for the TopHat, Gaussian, and Tanh filters with the parameters listed in Table~\ref{table:ar},
at various redshifts ($z=0,1,2,3$).}
\label{fig:Nbody_pdf}
\end{figure}

\begin{figure}[t!]
\centering
\hspace*{-1cm}
\includegraphics[width=0.75\paperwidth]{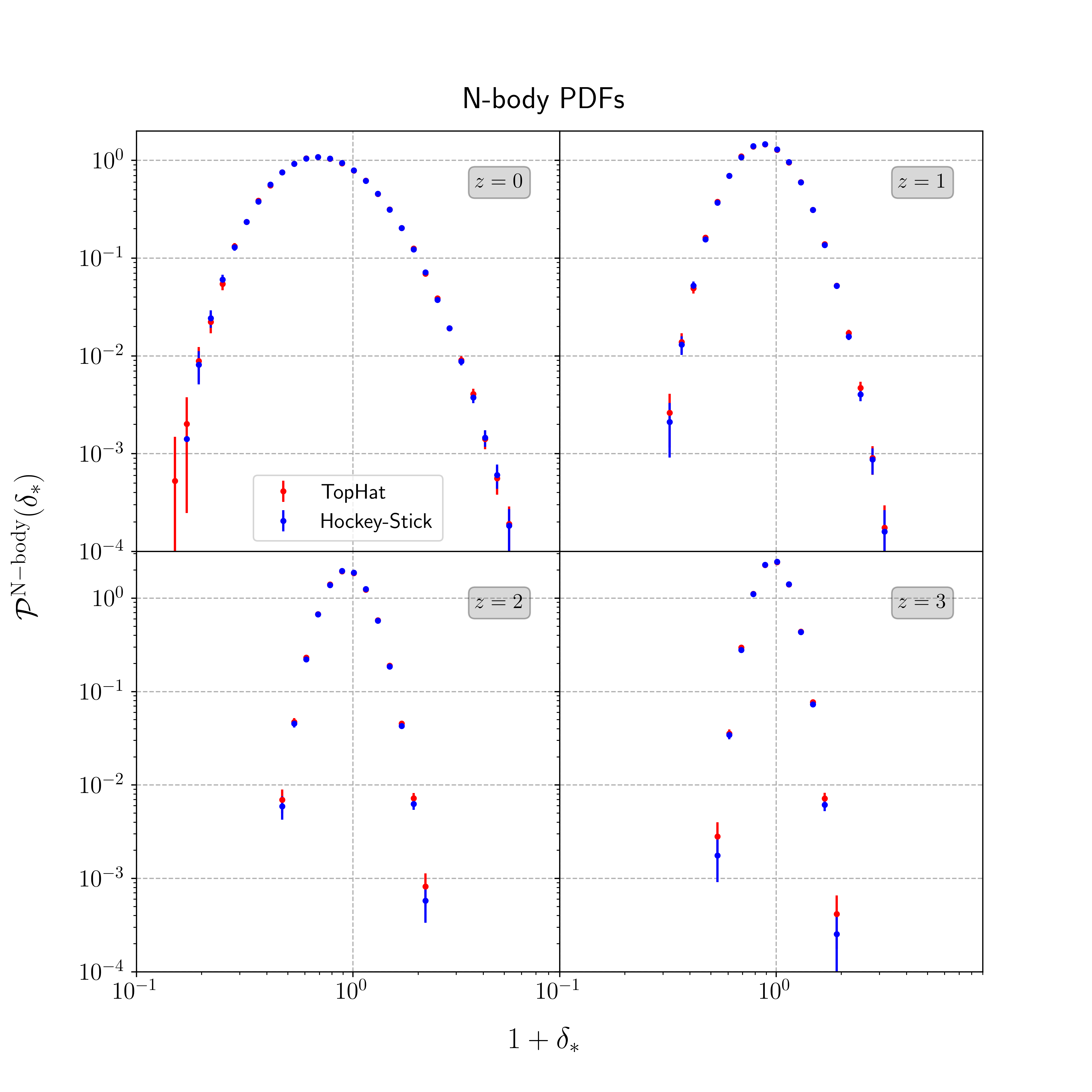} 
\caption{ Numerical PDFs constructed from the \texttt{Quijote} N-body simulations for the Hockey-Stick and TopHat (with $\tilde{r}_*=15.627$ Mpc/$h$) filters at various redshifts ($z=0,1,2,3$). Only density bins with non-vanishing number of cells are shown, leading to a cutoff of the PDFs at high redshifts at the edges.
}
\label{fig:Nbody2_pdf}
\end{figure}

\begin{figure}[t]
\centering
\hspace*{-1cm}                                                     
\includegraphics[width=0.4\paperwidth]{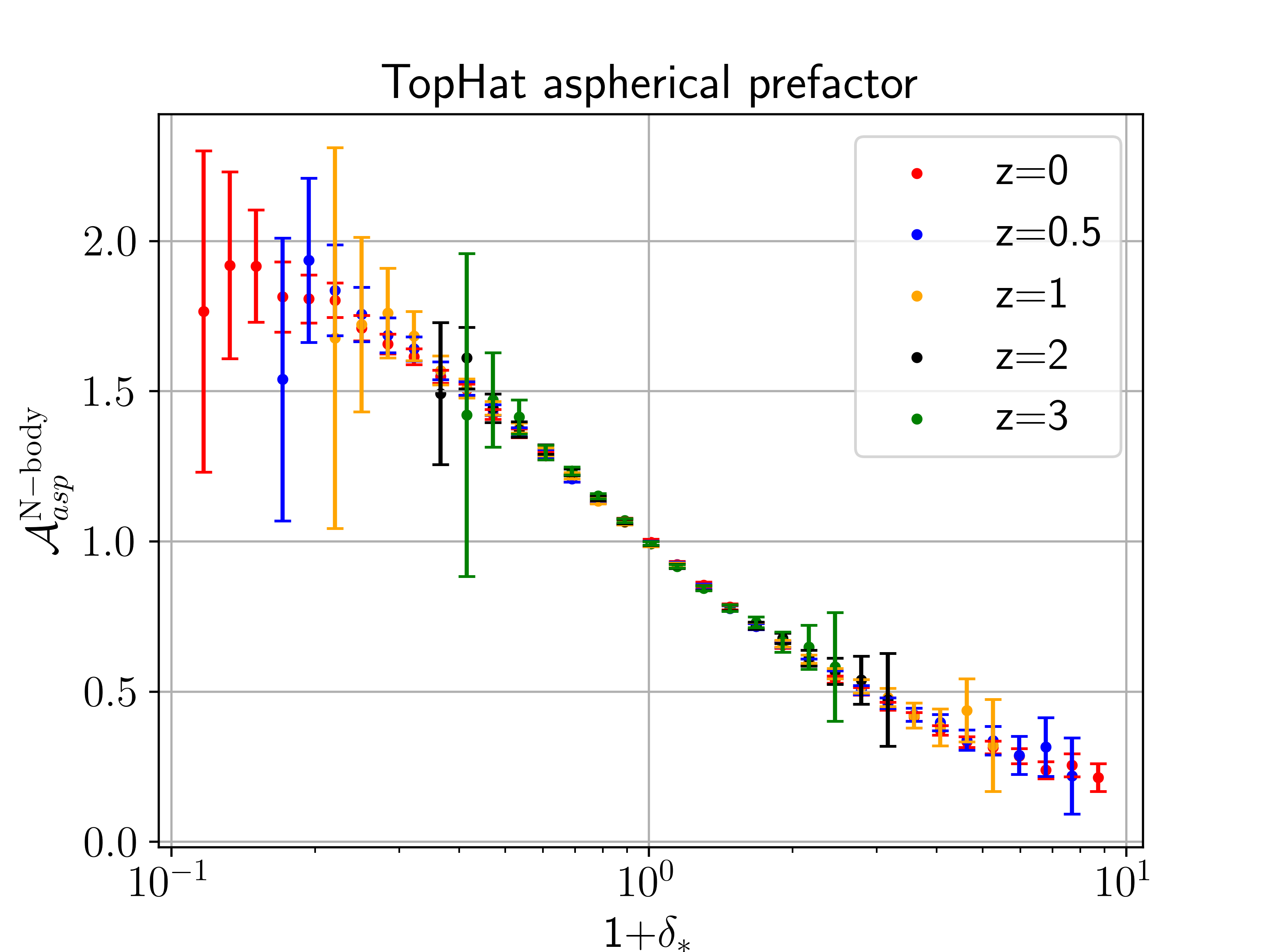}
\includegraphics[width=0.4\paperwidth]{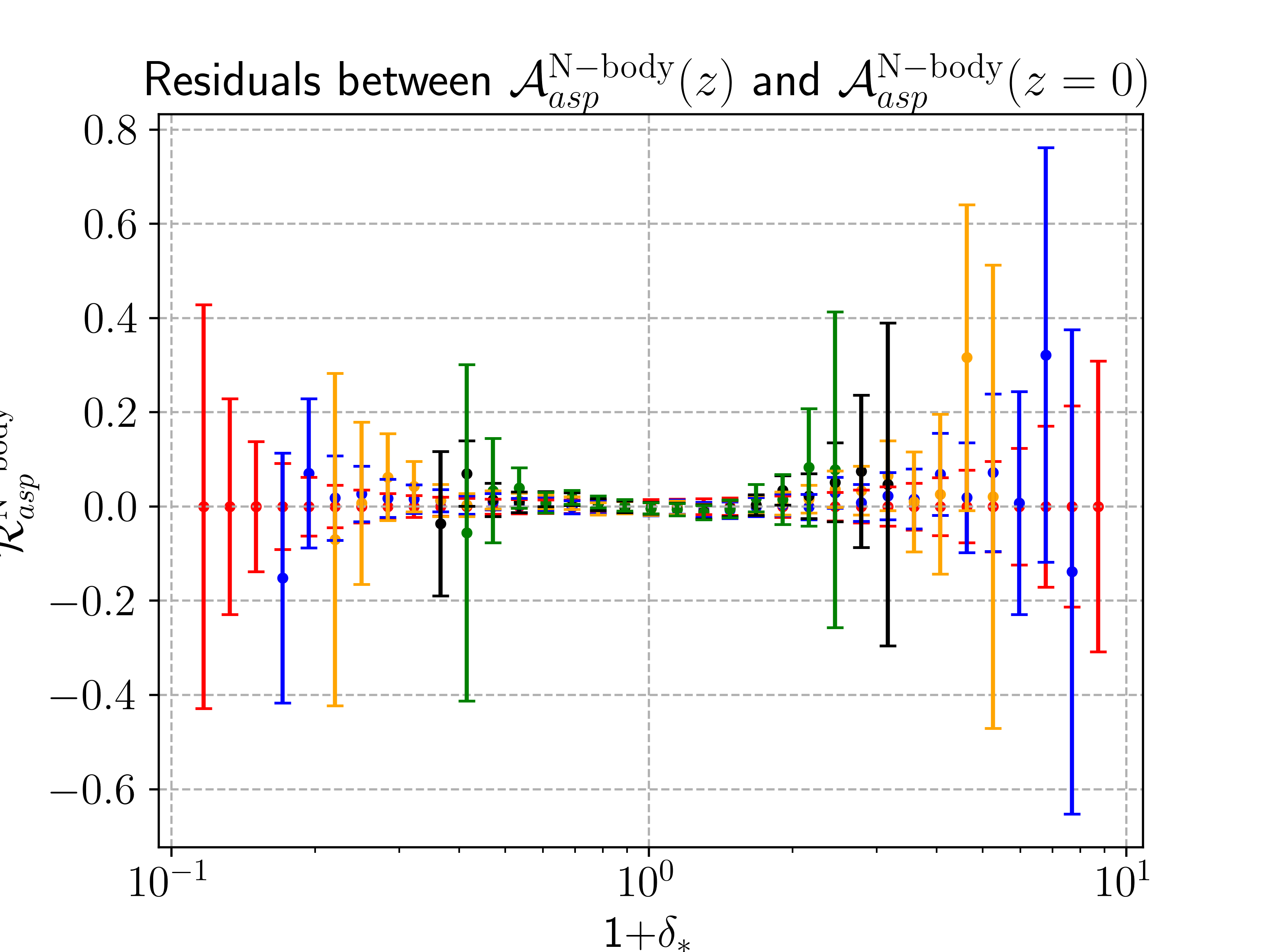}
\caption{\textit{Left panel:} Estimator of the aspherical prefactor, Eq.~(\ref{Asp}), constructed from the snapshots of the \texttt{Quijote} simulation at several redshifts. The PDFs are obtained using the TopHat filter with $r_*=10\,{\rm Mpc}/h$. 
The errorbars correspond to the statistical uncertainty  of the PDF obtained from the N-body simulations. 
\textit{Right panel:}  Residuals between aspherical prefactor at different redshifts compared to the $z=0$ case, Eq.~(\ref{RAsp}). 
}
\label{fig:TH_ASP}
\end{figure}

\begin{figure}[t!]
\centering
\hspace*{-1cm} 
\includegraphics[width=0.385\paperwidth]{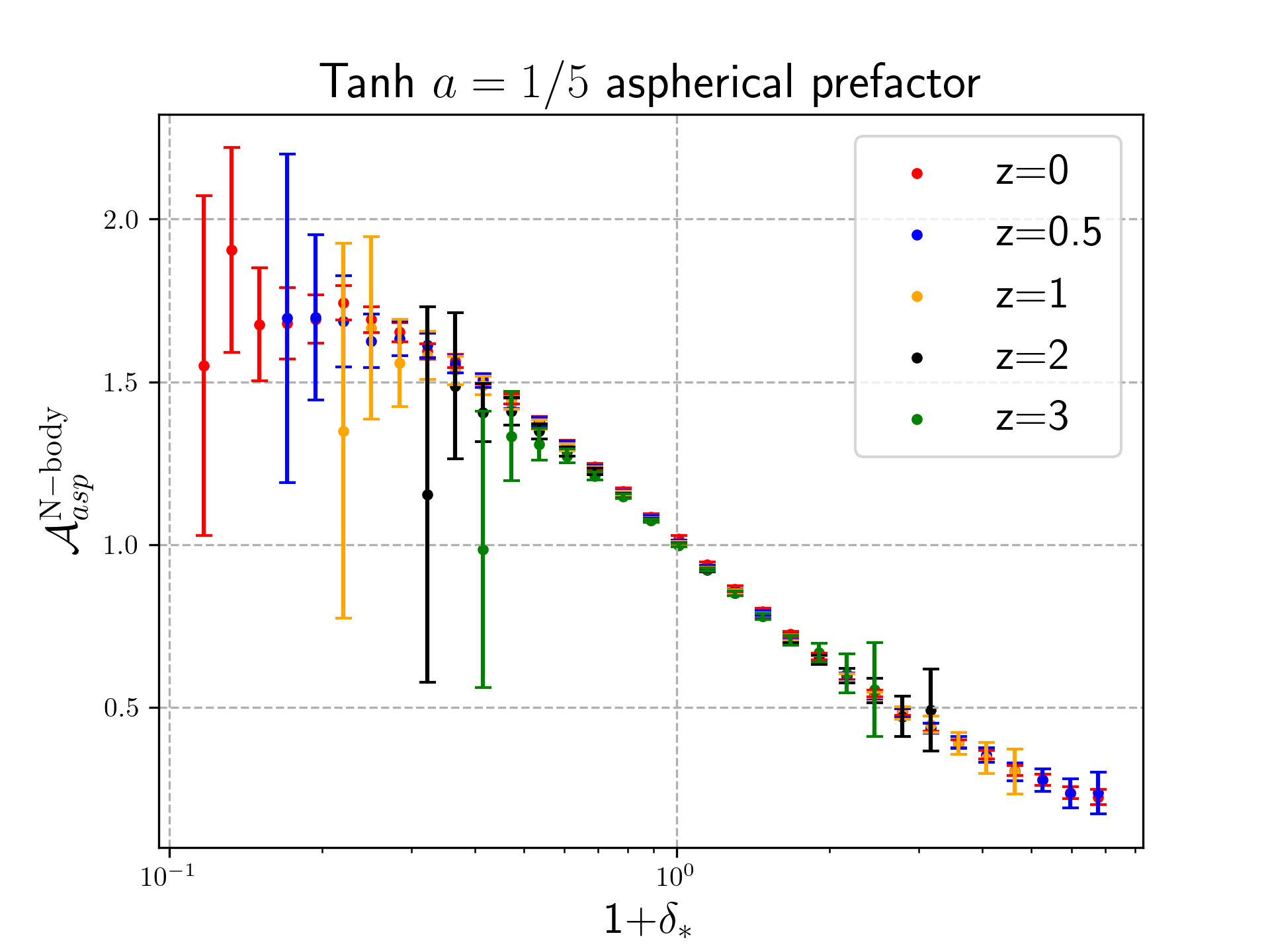}
\hspace*{-0cm}                                                     
\includegraphics[width=0.385\paperwidth]{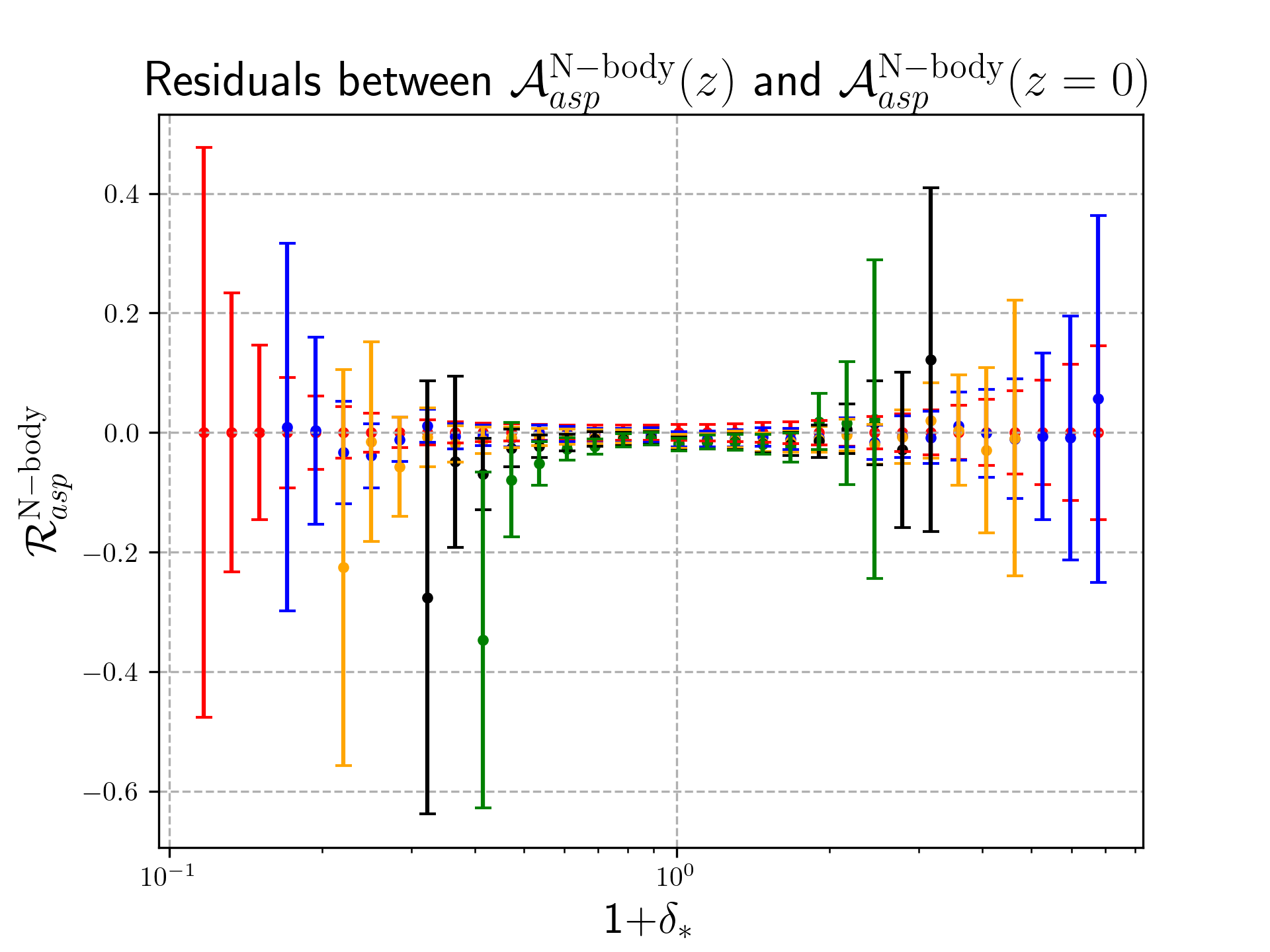}
\hspace*{-1cm}                                                     
\includegraphics[width=0.385\paperwidth]{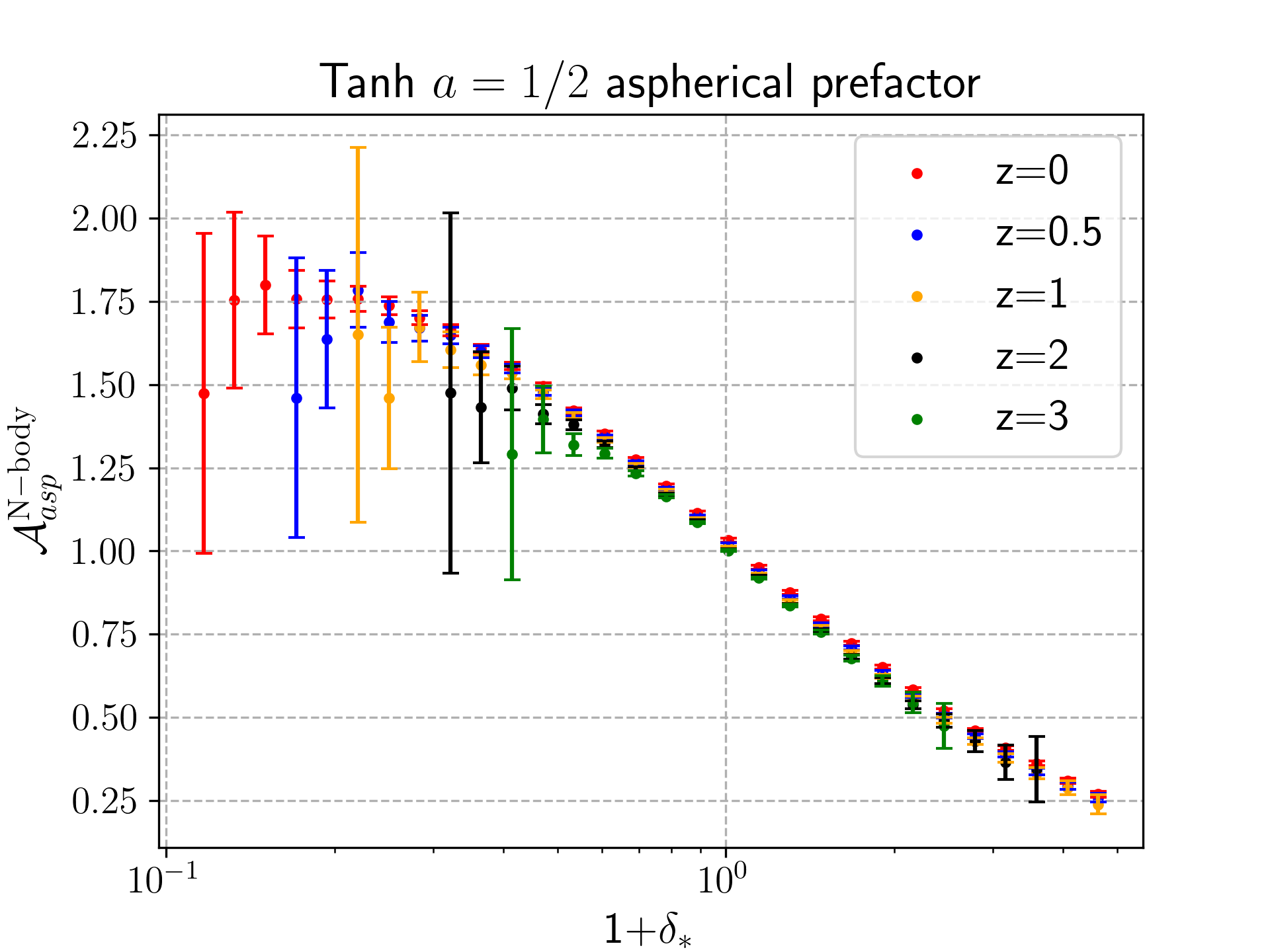}
\hspace*{-0cm}                                                     
\includegraphics[width=0.385\paperwidth]{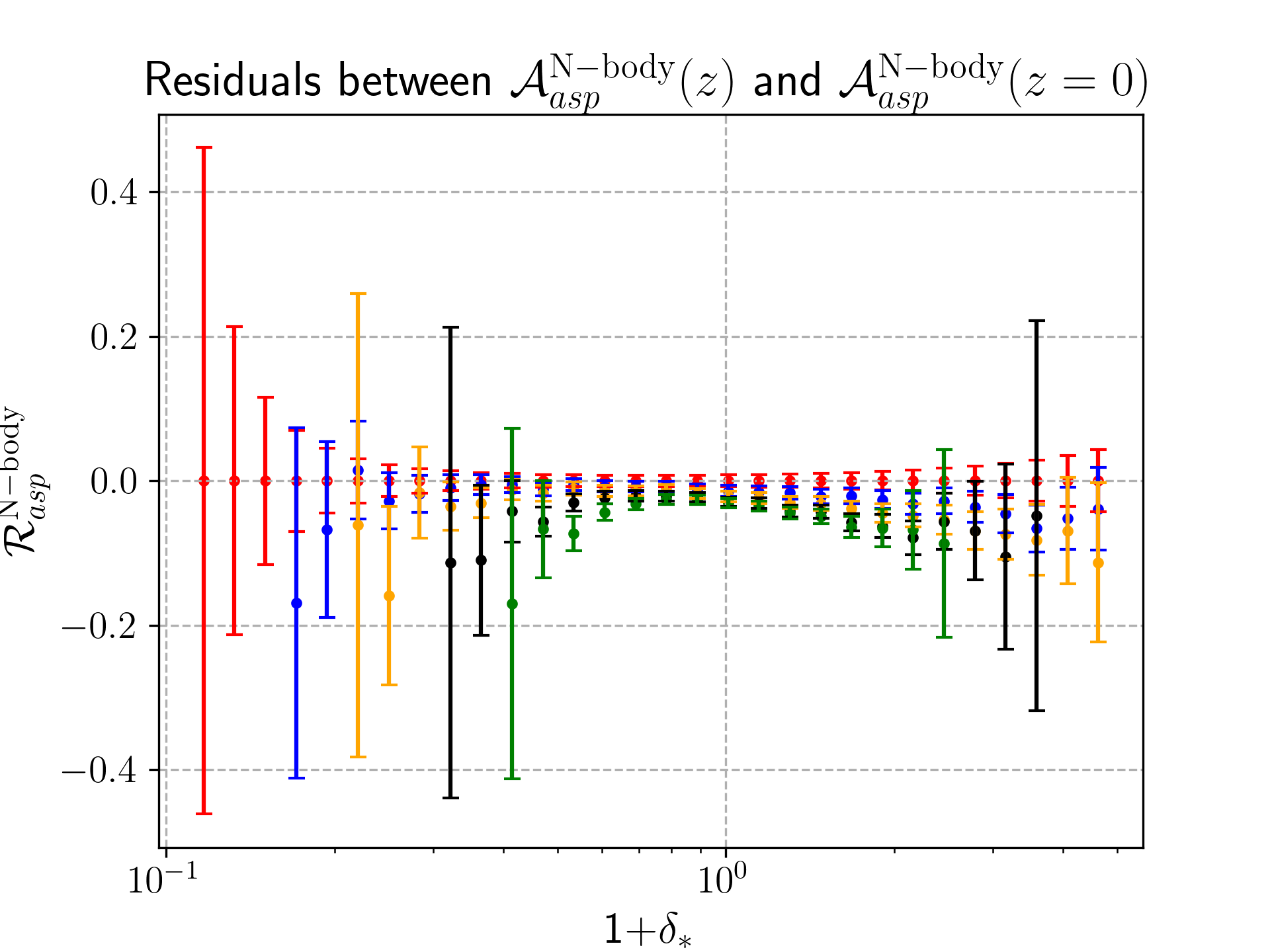}                \hspace*{-1cm} 
\includegraphics[width=0.385\paperwidth]{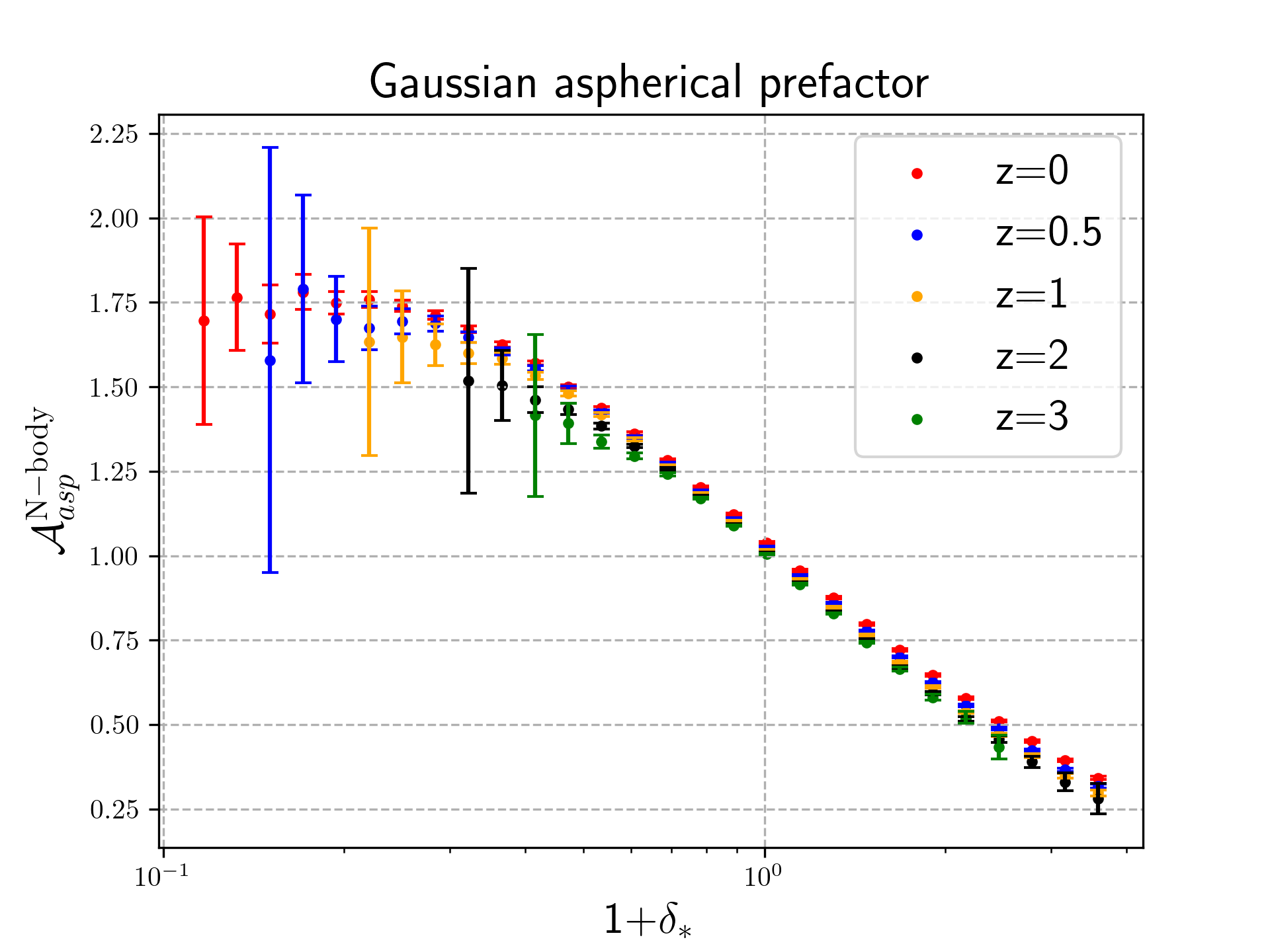}
\hspace*{-0cm}                                                     
\includegraphics[width=0.385\paperwidth]{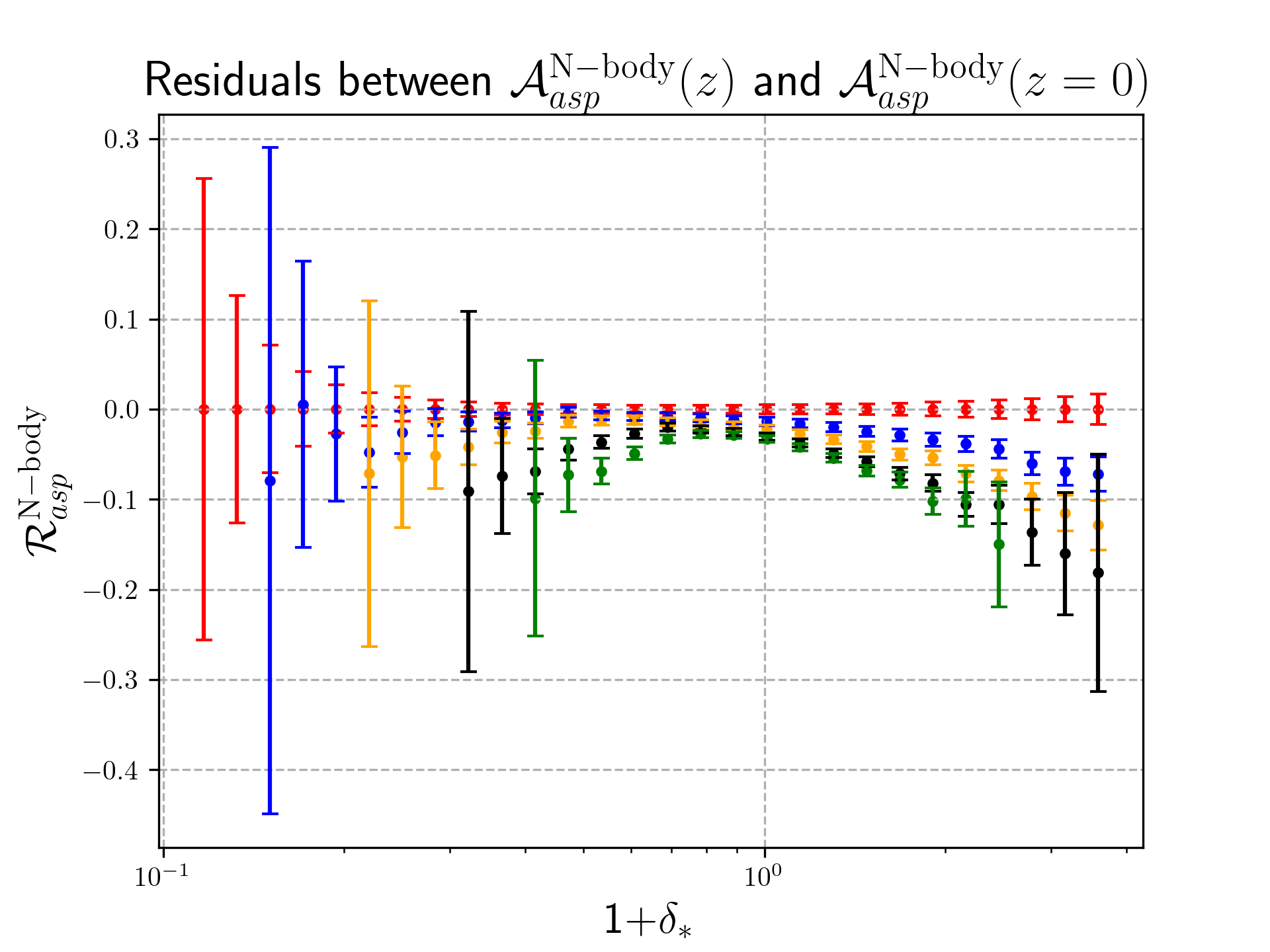}
\caption{Same as Fig.~\ref{fig:TH_ASP}, for Tanh (upper and middle rows) 
and Gaussian (bottom row) filters. The parameters of the filters are listed in Table~\ref{table:ar}.
}
\label{fig:filter_ASP}
\end{figure}

In order to construct the N-body PDF, we use the \texttt{Pylians} library \cite{Pylians}. First, the algorithm distributes particle masses over the grid space using the Cloud-in-Cell mass assignment scheme. 

Then, by initializing our filter choice and sphere radius $r_*$, we construct a set of cells with the centers separated by the distance $2r_*$ and compute their filtered density contrasts. Note that, unlike the TopHat case, the Gaussian and TopHat filters extend beyond their cells, causing the tails of neighboring filters to overlap; still, our choice of the cells ensures that the overlap remains small. The cell densities are distributed into 127 bins, evenly spaced in $\log(1+\delta_*)$ in the range $1+\delta_* \in [10^{-3}, 10^4]$. This gives us a 
histogram of the smoothed cell densities which we translate into the PDF by a proper normalization.
The procedure is repeated using 10 simulation snapshots at each redshift\footnote{We find that 10 independent realizations are sufficient to obtain an accurate measurement of the PDF and using more realizations does not significantly change the mean.} and averaged over, giving us $\mathcal{P}^{\text{N-body}}(\delta_*)$. The uncertainty of the PDF constructed in this way is estimated using the 
Poisson statistics within each $\delta_*$ bin.\footnote{The assumption of Poisson errors is 
expected to be less accurate for Gaussian-like filters compared to the TopHat filter, as in the former case the nearby cells are more correlated. Covariance can be extracted from large suits of N-body simulations, such as \texttt{Quijote}. 
We leave a detailed analysis incorporating accurate covariance for future.} 
Figure~\ref{fig:Nbody_pdf} shows the resulting N-body PDFs for various filters listed in Table~\ref{table:ar}, at various redshifts. 
An immediate observation is that all PDFs closely resemble each other. We could already expect this from our previous study of the spherical PDFs. Statistically significant  differences are observed only at extreme underdensities ($\delta_*<-0.5$).

In Fig.~\ref{fig:Nbody2_pdf} we further compare the PDF obtained with the Hockey-Stick filter (\ref{DWmain}) to the PDF for the TopHat filter with the same linear density variance. The agreement between the two PDFs is almost perfect, and stronger than for their spherical counterparts (cf. right panel of Fig.~\ref{fig:dw_pdfs}).

We now 
compare the N-body PDF to the spherical PDF 
${\cal P}_{sp}(\delta_*)$ constructed using the method of Sec.~\ref{sec:numeric}. A convenient way to do this is by considering an estimator for the aspherical prefactor,
\begin{equation}
\label{Asp}
    \mathcal{A}_{asp}^\text{N-body}(\delta_*,z)=\frac{\mathcal{P}^{\text{N-body}}(\delta_*,z)}{\mathcal{P}_{\text{sp}}(\delta_*,z)}\;.
\end{equation}
As discussed in Secs.~\ref{sec:review} and \ref{sec:pert}, the aspherical prefactor is expected to be essentially independent of the redshift, irrespective of the shape of the filter. A weak redshift dependence can arise only through the EFT counterterm and 2-loop corrections. 

In the left panel of Fig.~\ref{fig:TH_ASP} we show the estimator (\ref{Asp}) for the TopHat filter with $r_*=10\,{\rm Mpc}/h$ at several redshifts and on the right panel of the same figure the residuals
\begin{equation}
\label{RAsp}
    {\cal R}_{asp}^\text{N-body}(\delta_*,z)=\frac{\mathcal{A}_{asp}^\text{N-body}(\delta_*,z)}{\mathcal{A}_{asp}^\text{N-body}(\delta_*,z=0)}-1\;.
\end{equation}
The errorbars represent the statistical uncertainties inherited from the N-body PDF ${\cal P}^\text{N-body}(\delta_*,z)$. We note that the estimator (\ref{Asp}) also possesses a systematic uncertainty coming from the numerical error in the determination of the spherical PDF ${\cal P}_{sp}(\delta_*,z)$; the results of Sec.~\ref{ssec:TopHat} show that, at least for the TopHat filter, this is subdominant, so we do not include it.
We see that the aspherical prefactor is indeed redshift independent, within the errorbars, which validates our numerical pipeline. The prefactor changes between $\sim 1.8$ at extreme underdensities and $\sim 0.2$ at extreme overdensities. Though significant, this change is much smaller than several orders of magnitude variation of the spherical PDF in the same density range. The results on the plot are consistent with those of Refs.~\cite{ivanov,anton}.

\begin{figure}[t]
\centering
\hspace*{-1cm}                                              
\includegraphics[width=0.38\paperwidth]{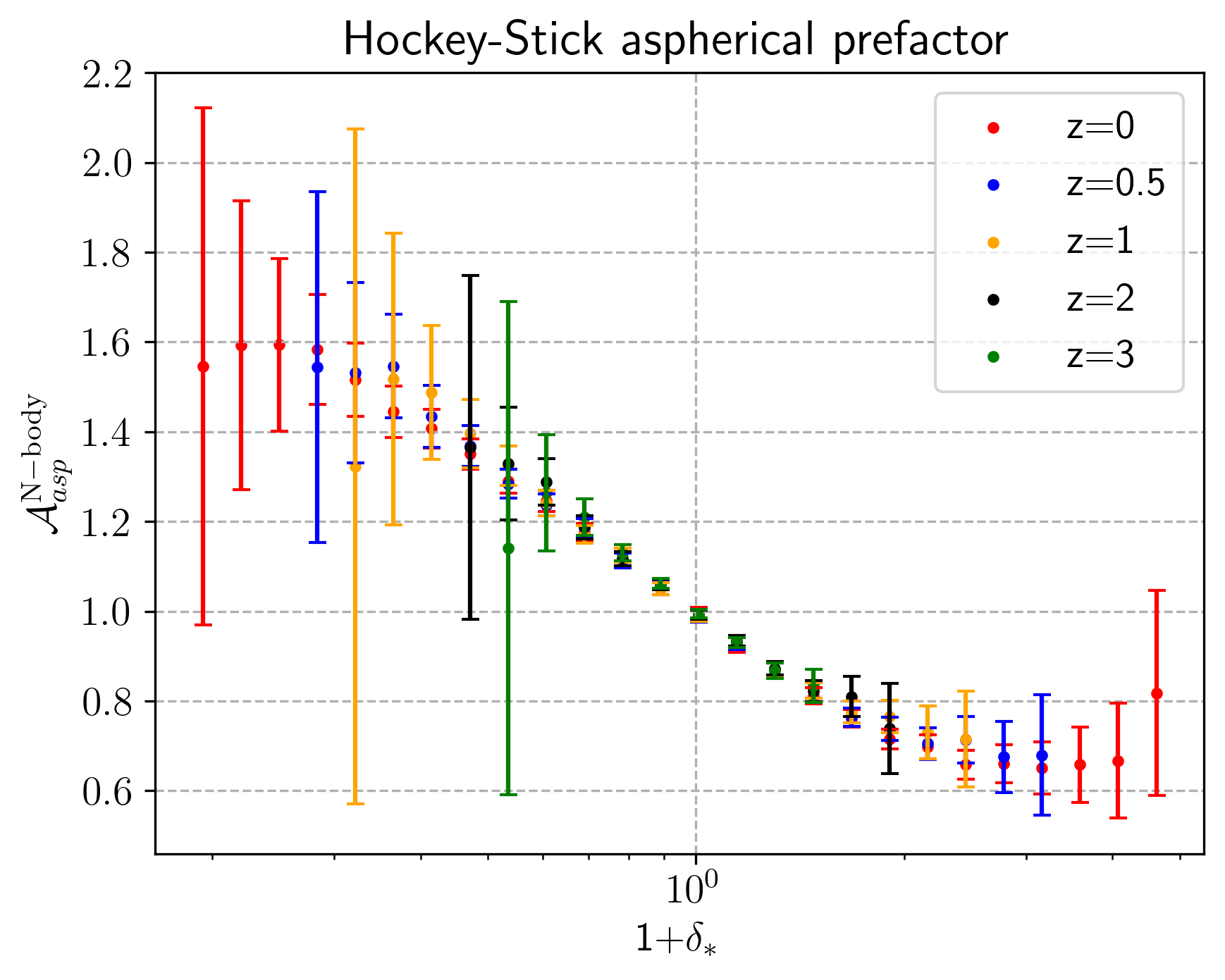}
\hspace*{0cm}                                            
\includegraphics[width=0.395\paperwidth]{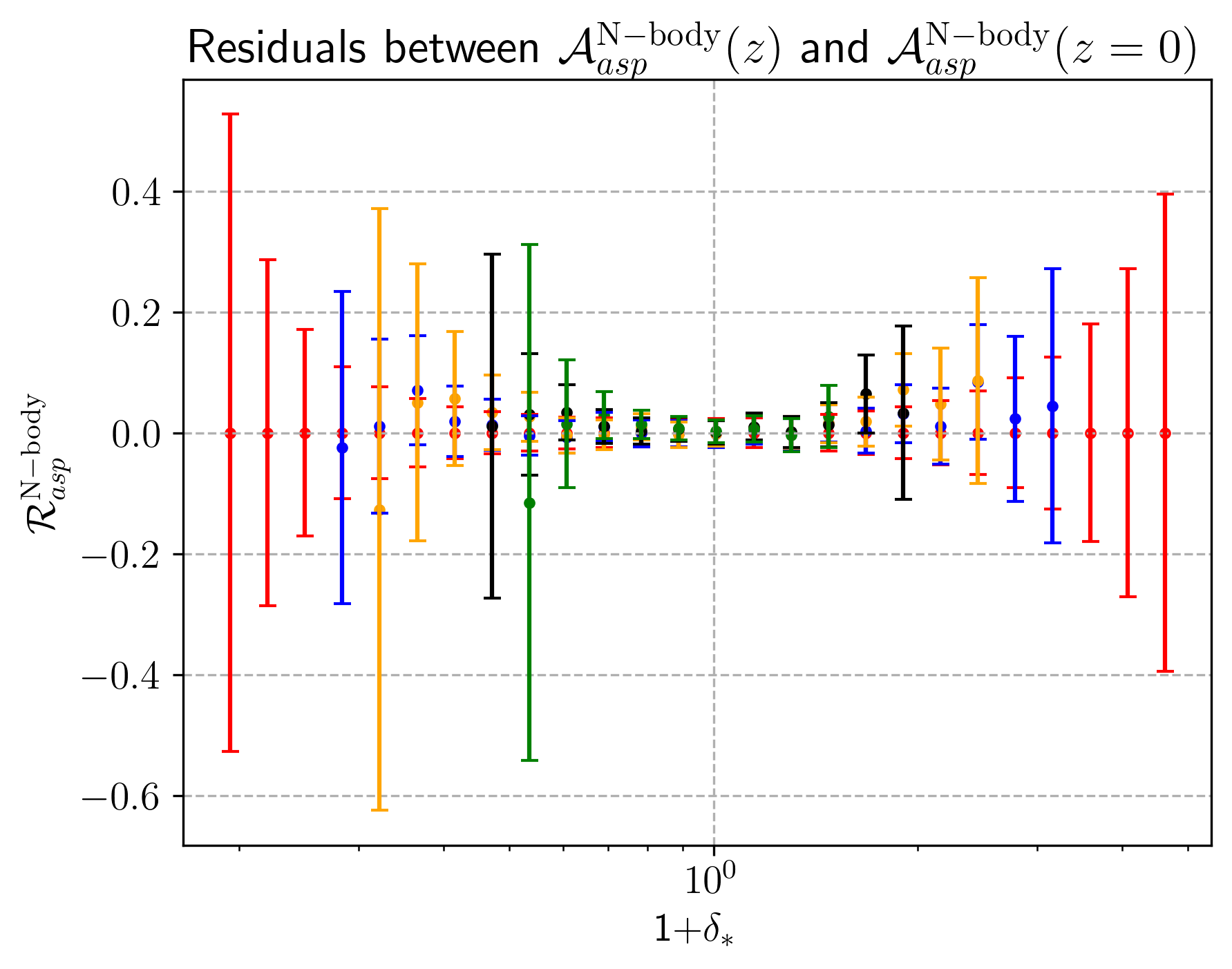}
\caption{ Same as Fig.~\ref{fig:TH_ASP}, for the Hockey-Stick filter (\ref{DWmain}). 
}
\label{fig:DW_ASP}
\end{figure}

Figure~\ref{fig:filter_ASP} shows the estimator (\ref{Asp}) for the smooth filters studied in Sec.~\ref{ssec:smooth}. We  
 do not present plots for the Tanh filters with $a=1/30$ and $a=1$ since they essentially coincide with the TopHat and Gaussian cases. 
To a good accuracy, all aspherical prefactors shown in the left panels are redshift independent, as expected. They also closely resemble the TopHat aspherical prefactor of Fig.~\ref{fig:TH_ASP}. 
However, a closer inspection of the residuals between the estimator (\ref{Asp}) at zero and non-zero redshifts reveals a mild redshift dependence, which becomes stronger as we move from TopHat to Gaussian filter. 
The residuals follow a quadratic curve near $\delta_*=0$. Partially, this can be explained by the redshift dependence of the EFT counterterm: as shown in Sec.~\ref{sec:pert}, the latter enters into the $\delta_*^2$ term in the aspherical prefactor. In addition, there are 
2-loop corrections, which we now discuss.

We observe that the residuals at $\delta_*=0$ drift away from zero as the redshift grows. This implies a non-trivial dependence of ${\cal A}_{asp}^{\text{N-body}}(\delta_*=0,z)$ on the redshift, apparently violating the first consistency condition in (\ref{Aaspprops}). We recall, however, that this condition has been obtained in the approximation neglecting the 2-loop correction to the prefactor ($\alpha_2g^4$ term in Eq.~(\ref{PDFexpansion})). Inclusion of the 2-loop contribution produces a deviation of ${\cal A}_{asp}(\delta_*=0,z)$ from unity, which is proportional to $g^2(z)$. Thus, the 2-loop correction becomes stronger at smaller redshifts. It is also expected to be larger for smaller cells, since these correspond to stronger non-linearities. 
The trend observed in Fig.~\ref{fig:filter_ASP} is consistent with this expectation: the smoother filters exhibiting larger residuals correspond to smaller cell radii; see Table~\ref{table:ar}. In particular, the Gaussian filter has $r_*\sim 5\,{\rm Mpc}/h$. Note that statistically significant deviations of the aspherical prefactor from unity were previously observed for TopHat cells of radius $r_*=5\,{\rm Mpc}/h$ \cite{anton}. 
The attribution of the redshift dependence of aspherical prefactors in Fig.~\ref{fig:filter_ASP} to the 2-loop corrections thus appears plausible. We stress, however, that a thorough investigation of the PDF systematics is needed to establish this connection firmly, and we leave it for future study.

Finally, in Fig.~\ref{fig:DW_ASP} we present the aspherical prefactor for the Hockey-Stick filter. Since it corresponds to a larger cell radius, $r_*=15\,{\rm Mpc}/h$, the 2-loop correction is expected to be negligible. Indeed, we observe that the residuals of the aspherical prefactor at different redshifts are consistent with zero. Additionally, the overall variation of ${\cal A}_{asp}^\text{N-body}$ over the considered density range is reduced, going from $\sim 1.5$ at underdensities to $\sim 0.7$ at overdensities.

\subsection{Comparison with perturbative aspherical prefactor}

\begin{figure}[t!]
\centering
\hspace*{-1cm}                                              
\includegraphics[width=0.4\paperwidth]{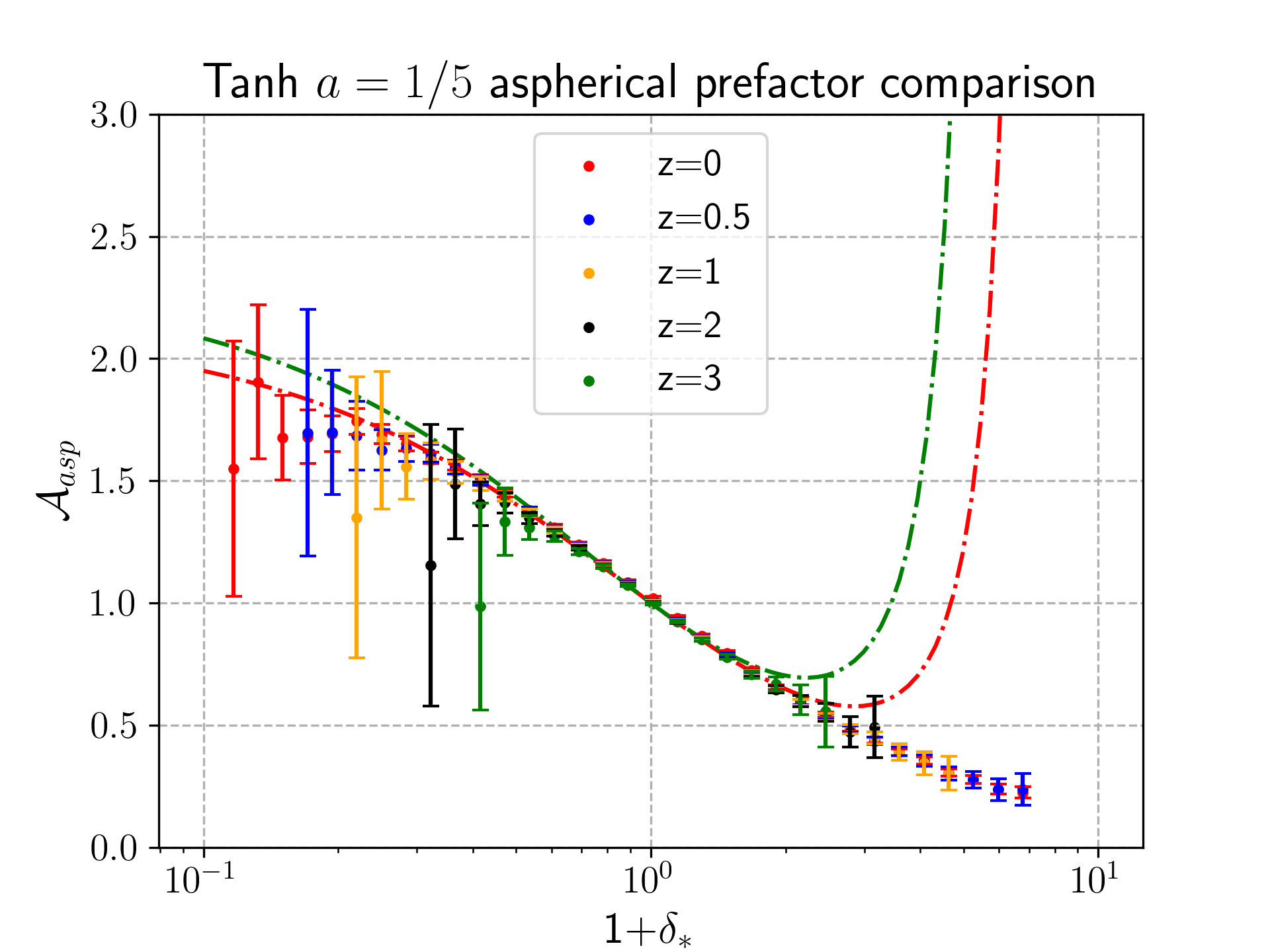}
\hspace*{-1cm}                                            
\includegraphics[width=0.4\paperwidth]{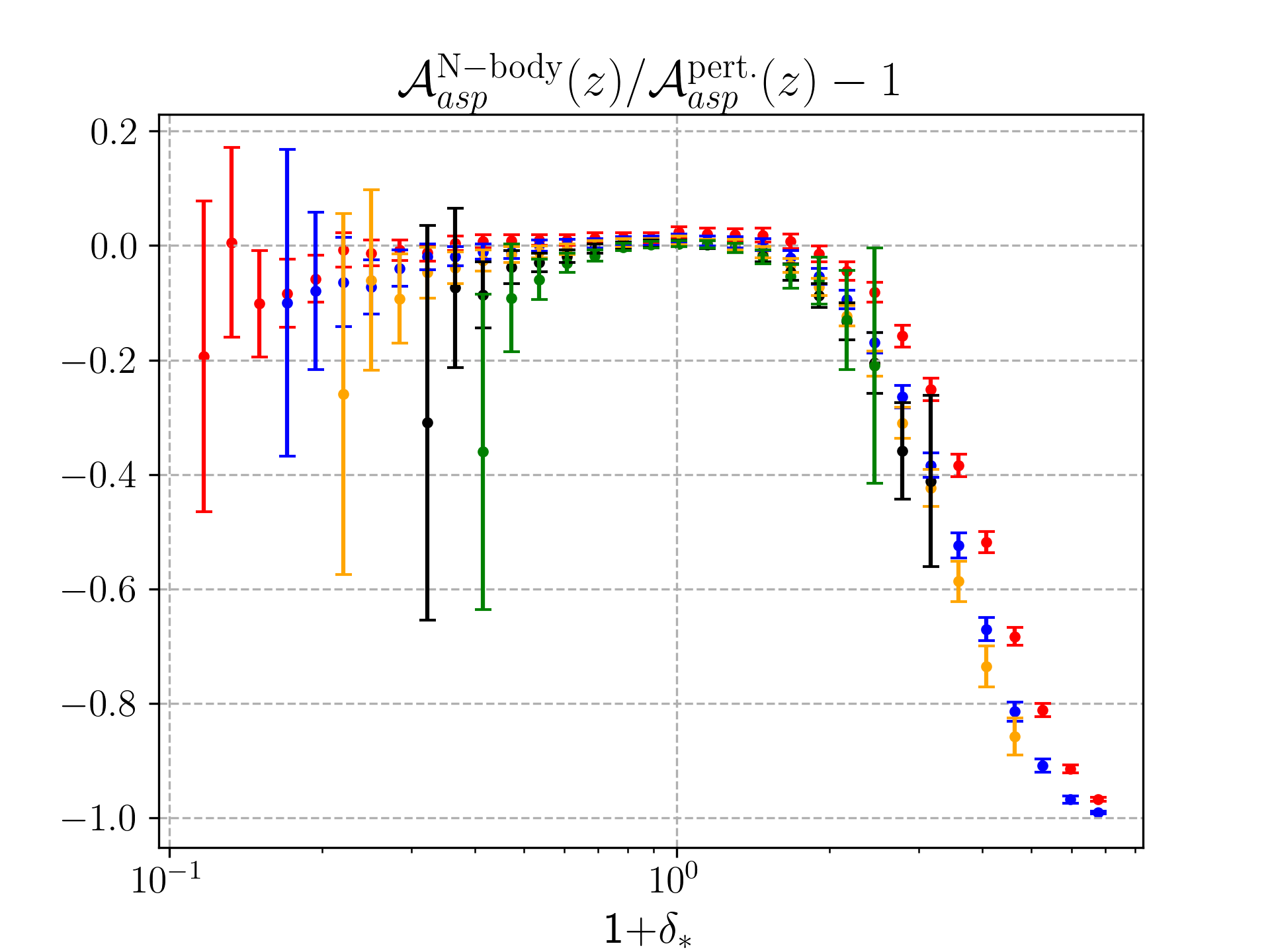}
\hspace*{-1cm}                                            
\includegraphics[width=0.4\paperwidth]{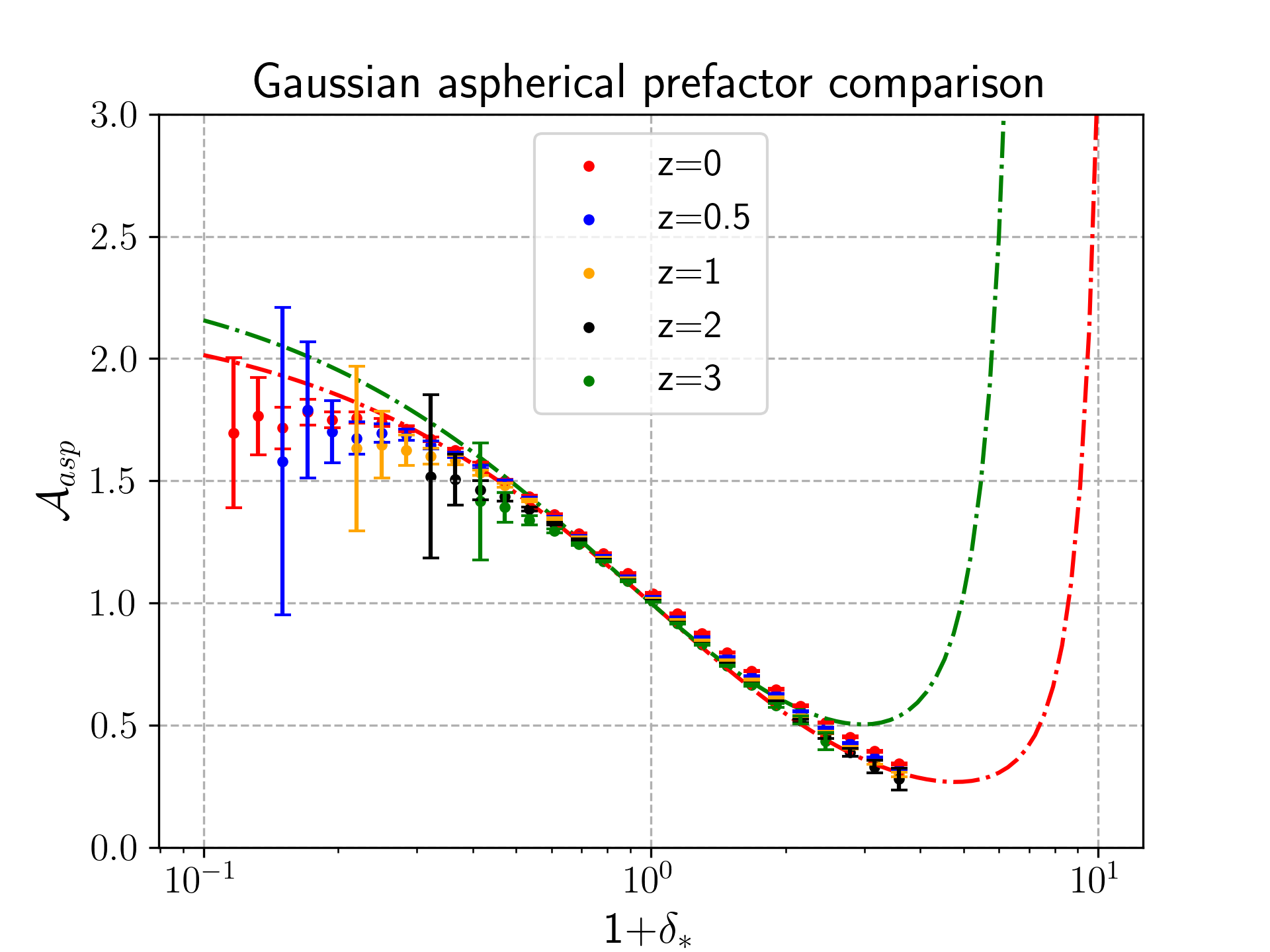}
\hspace*{-1cm}                                            
\includegraphics[width=0.4\paperwidth]{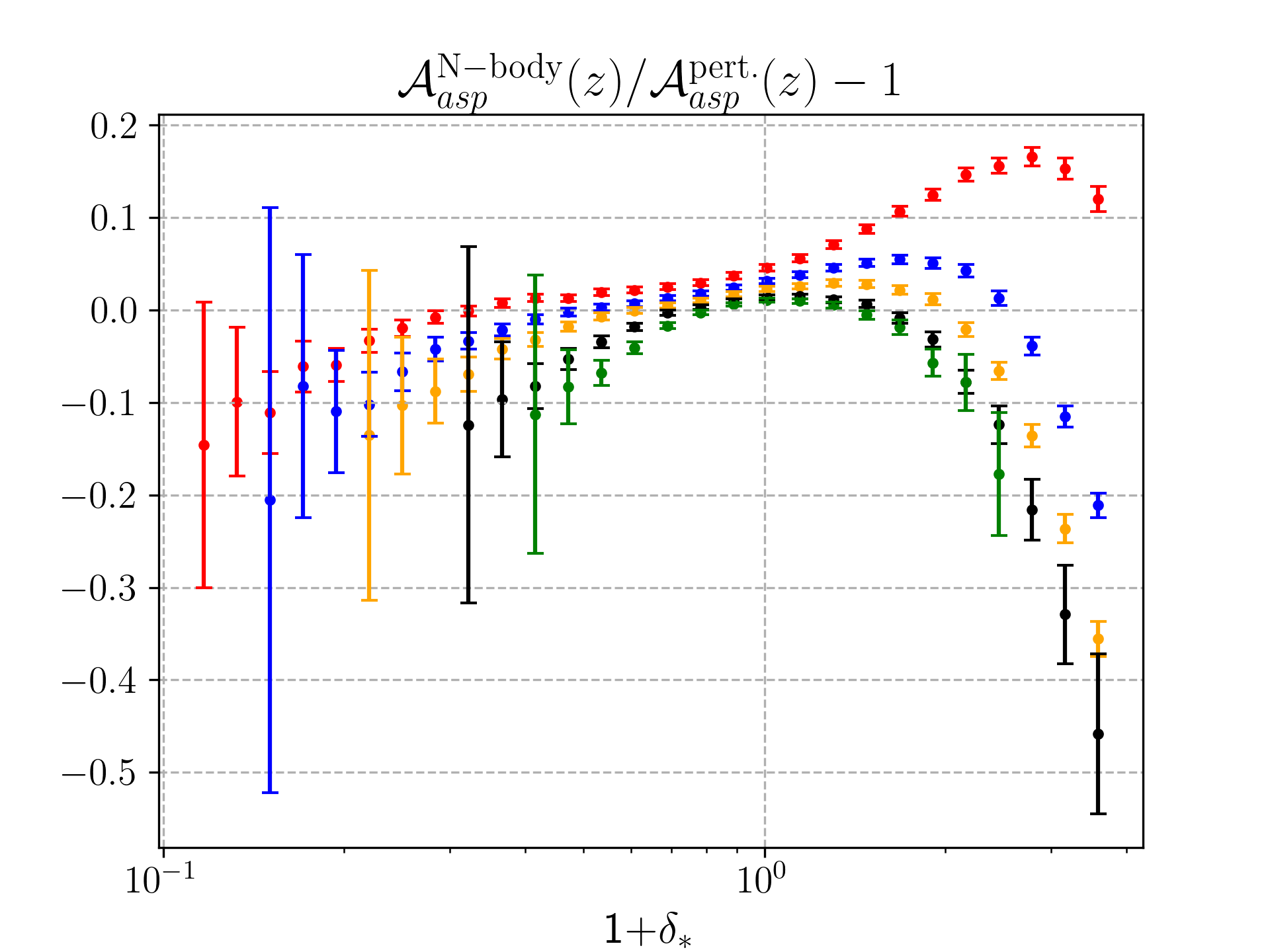}
\caption{ \textit{Left panels:} Aspherical prefactor measured from the \texttt{Quijote} N-body simulations for the Tanh ($a=1/5$) and Gaussian filters at various redshifts (dots) along with the perturbative expression (dot-dash). The perturbative expressions are shown for 
$z=0$ (red) and $z=3$ (green). The parameters of the filters are listed in Table~\ref{table:ar}.
\textit{Right panel:} Residuals between the N-body and perturbative prefactors at different redshifts. }
\label{fig:asp_comps}
\end{figure}

It is instructive to compare the aspherical prefactor (\ref{Asp}) extracted from the N-body simulations with the perturbative expression (\ref{asp_ren1}). 
Note that the latter depends on the redshift through the counterterm $\gamma(z)$.
We present such comparison in Fig.~\ref{fig:asp_comps} for the Tanh ($a=1/5$) and Gaussian filters. We observe that the perturbative expression works remarkably well at $\delta_*\lesssim 1$. In particular, it describes the aspherical prefactor with the precision $\lesssim 15\%$ across the whole range of underdensities, $\delta_*<0$. We have observed the same accuracy for other filters considered in this work. 
We do not know if this property will hold for other sizes and filter shapes. It would be interesting to investigate this question and understand if this is a general property of the cosmological perturbations theory, or just a numerical coincidence arising from the specific $\Lambda$CDM parameters considered in this work.

On a more refined level, for the Gaussian filter we see significant (of order a few per cent) residuals between the N-body aspherical prefactor and the perturbative expression at $\delta_*=0$. The magnitude of these residuals monotonically increases towards lower redshifts, supporting their interpretation as 2-loop corrections (not included in Eq.~(\ref{asp_ren1})). The fact that these residuals are stronger for the Gaussian ($r_*=4.7\,{\rm Mpc}/h$) than for the Tanh ($r_*=9.1\,{\rm Mpc}/h$) filter, is consistent with our earlier observation that 2-loop corrections are more relevant for smaller cells.  

For high redshifts $z=2,3$ we also observe a quadratic $\delta_*^2$ term in the residuals. This cannot be attributed to the EFT counterterm, which has been included in the perturbative expression, or the 2-loop corrections, which decrease with the redshift. A possible origin of this behavior is the numerical error in the value of the action, which propagates into ${\cal A}_{asp}^\text{N-body}$ through an error in the spherical PDF. Since the action enters into ${\cal P}_{sp}(\delta_*)$ exponentially and is divided by the square of the growth factor $g^2(z)$, see Eq.~(\ref{SPDF}), even a small error in it can lead to sizable errors in the PDF at high redshifts. As mentioned in Sec.~\ref{sec:results}, the accuracy of the action can in principle be improved with more computer power. Before undertaking such an effort, however, we believe it is important to get control of other uncertainties arising in the connection of the PDF model to the actual observational data.
We leave this task for future research.  

\section{Discussion}\label{sec:discuss}

In this work, we developed a model for the one-point PDF of matter densities averaged with an arbitrary window function. Using the analytic path integral approach to the PDF and considering the linear density variance as an expansion parameter, we factorized the PDF into the exponential part and a prefactor. The latter was further factorized into the {\it monopole prefactor} due to spherically symmetric density perturbations and the remaining {\it aspherical prefactor}. Such split has several advantages. First, different building blocks encapsulate different physical processes. The exponent and monopole prefactor are determined by the dynamics of spherical collapse and, for broad enough filters, can be found exclusively within SPT. On the other hand, the aspherical prefactor is sensitive to the non-perturbative physics of short-scale clustering which is accounted for by the inclusion of the EFTofLSS counterterms. 
Second, each building block has a fixed redshift dependence. Thus, computing them once one can reconstruct the PDF at an arbitrary redshift. Finally, different blocks contribute with different strength into the dependence of the PDF on the density contrast, with the dominant contribution coming from the exponential part, followed by the monopole prefactor, and the aspherical prefactor concluding the list. This allows one to compute the building blocks separately using different degrees of approximation. 

We devised and implemented a numerical algorithm to compute the exponential part of the PDF and the monopole prefactor. The key step of the algorithm is solution of a multidimensional constrained minimization problem which yields a discrete-lattice version of the most probable density profile corresponding to a given averaged density contrast. We validated the pipeline using known analytic results for the TopHat filter. For the product of the exponent and the monopole prefactor, which we called {\it spherical PDF}, we found a sub-per cent agreement for redshift $z=0$, cell radius ${r_*=10\,{\rm Mpc}/h}$, and density contrast $\delta_*\gtrsim -0.6$. The error somewhat deteriorates at the extreme underdense tail reaching $\sim 6\%$ at $\delta_*=-0.9$. It appears possible to further improve the accuracy of the code by increasing the number of lattice points. This, however, would raise the dimensionality of the constrained minimization problem, and solving it with our present \texttt{Python} code would be too time-consuming. It will be interesting to explore the ways in which the code can be accelerated.  
One could rewrite the minimization scheme in C/C\texttt{++} and use Nvidia’s CUDA platform to GPU accelerate the process. Similarly, \texttt{JAX} \cite{jax} can be used too as it provides GPU accelerated options while using a \texttt{Python}-like syntax. 

We applied our pipeline to construct spherical PDFs for several sample window functions, including the Gaussian filter, a family of smooth monotonic filters interpolating between Gaussian and TopHat, and a non-monotonic filter peaking in the outer part of the cell. We found significant differences between the respective most probable linear density profiles. At the non-linear level, the behavior depends on whether one considers over- or underdensities. For overdensities, $\delta_*>0$, the non-linear most probable profiles strongly depend on the choice of the filter. On the other hand, for underdensities, $\delta_*<0$, the non-linear profiles collapse, up to minor variations, to a single universal curve, common to all filters sharing the same linear density variance. It will be interesting to explore the implications of this universality for predicting the distribution of dark matter within voids \cite{voids,Contarini:2026yfv}. An accurate modeling of dark matter void profiles is important for increasing the sensitivity of indirect decaying dark matter searches \cite{DM_void}.
 
Despite the differences in the linear density profiles, we have found that spherical PDFs for all filters closely resemble each other, once the widths of the filters are adjusted to yield the same linear density variance. Sizable variations (up to a factor of a few) occur only at the tails of the PDF, at extreme over- and underdensities. It is unclear, however, if they can be detected in a realistic survey: given the rarity of extreme density configuration, the differences in their probabilities are likely to get buried under the statistical and systematic uncertainties. This question deserves further study.

From an alternative viewpoint, the weak dependence of the PDF on the filter presents a prediction of the $\Lambda$CDM cosmology and can be used as a null test of the consistency between the observational data and $\Lambda$CDM. It would be interesting to understand if this PDF universality is peculiar to $\Lambda$CDM or persists also in extended cosmological models, such as models of modified gravity and models with non-gravitational dark matter interactions. 

We did not attempt to construct a full semi-analytic model for the aspherical prefactor, which requires intensive computations even in the TopHat case \cite{ivanov,anton}. Instead, we derived an analytic perturbative 1-loop estimate for it valid at small density contrasts. This allowed us to probe the sensitivity of the PDF to the EFTofLSS counterterms. Contrary to a naive expectation, PDFs for smooth filters feature essentially the same sensitivity, as the PDF for the sharp TopHat filter with the same linear density variance. The only way to reduce the sensitivity of the PDF to the uncontrollable short-distance physics is by considering bigger cells with smaller linear variance.

We compared our PDF model with the results of pure dark matter N-body simulations. To this end, we constructed PDFs for different filters using the snapshots from the high-resolution 
\texttt{Quijote} suite \cite{quijote} at several redshifts. We observed that the full N-body PDFs exhibit similar weak dependence on the filter as the spherical PDFs found with our pipeline. For each filter, we further obtained an estimate of the aspherical prefactor by taking the ratio of the full N-body PDF over the spherical PDF. We verified that the aspherical prefactor has weak redshift dependence, as predicted by our model, and is well approximated at $\delta_*\approx 0$ by the perturbative expression. In fact, we found  that the perturbative expression provides a surprisingly good approximation to the aspherical prefactor (with $\lesssim 15\%$ error) up to moderate overdensities $\delta_*\leq 1$ and in the whole range of considered underdensities $-0.9\leq \delta_*\leq 0$.  

At a finer level, we found small --- up to a few per cent --- but statistically significant discrepancies between the aspherical prefactor extracted from N-body and its 1-loop approximation even at $\delta_*=0$. The fact that these discrepancies grow with decreasing redshift suggests to associate them with 2-loop corrections. A precision modeling of the PDF for general filters will require  either explicitly including these corrections into the expression for the aspherical prefactor, or obtaining a reliable upper estimate on their magnitude, to be treated as a theoretical error along the lines of \cite{Baldauf:2016sjb}. We leave this study for future.    

There are more directions in which our work can be extended. Other window functions, such as momentum-space TopHat (Sharp-k) filter used in the 
excursion set theory \cite{bond}, 
can be implemented in our code. Computing PDF coming from the Sharp-k filter, however, is expected to be computationally expensive. Due to the slow falloff and oscillatory behavior of the filter in position space, it is expected that a larger lattice size and number of lattice points would be necessary to ensure the convergence of the density profiles to the most probable configuration. A potential gain that can warrant such effort is establishing connection between the one-point PDF and excursion set approach to the halo mass function. Since halos correspond to extreme virialized overdensities, achieving this goal would also require extending the PDF model beyond its current domain of applicability which is limited to moderate overdensities that do not experience collapse.

It will be important to relate the matter PDF, which is not directly observable, to the statistics of galaxies. This will require modeling the galaxy bias \cite{desjacques}. The approach of \cite{kokron} appears promising in this regard. Constructing an observable galaxy PDF would also involve taking into account the effects of redshift-space distortions \cite{EFT_ivanov}. We plan to address these challenges in our future work.

Other potential generalizatons of our method include  
modeling one-point PDF for the weak lensing convergence field using spherical and cylindrical collapse functions \cite{cylinders,Barthelemy:2023mer}, as well as connecting PDF to other statistics, such as $k-$Nearest Neighbor ($k$NN) \cite{knn}. 
 With semi-analytical method capable of describing the density field in the non-perturbative regime, one can go beyond the $\Lambda$CDM model to search for signatures of interacting dark matter, primordial non-Gaussanity, and other cosmological phenomena that could provide evidence for new fundamental physics.

\section*{Acknowledgments}
We thank Caio Bastos de Senna Nascimento, Cliff Burgess, Neal Dalal, Elena Pinetti, Beatriz Tucci, Cora Uhlemann and James Wadsley 
for insightful discussions. AC acknowledges funding from the Swiss National Science
Foundation. The work of AK and SS is supported by the
Natural Sciences and Engineering Research Council (NSERC) of Canada. Research at Perimeter
Institute is supported in part by the Government of Canada through the Department of Innovation,
Science and Economic Development Canada and by the Province of Ontario through the Ministry
of Colleges and Universities. This research
was enabled in part by support provided by Compute Ontario (www.computeontario.ca)
and Digital Research Alliance of Canada (alliancecan.ca).

\appendix

\section{Conventions} \label{app:convention}
Following the same conventions as in \cite{ivanov}, we define the Fourier transform of a field $\delta(\textbf{x})$ as 
\begin{equation}
    \delta(\textbf{x})=\int_\textbf{k} \delta(\textbf{k}) e^{i \textbf{k}\cdot \textbf{x}}
\end{equation}
where we define the integral measure over $k-$space as
\begin{equation}
    \int_\textbf{k} = \int \frac{d^3k}{(2\pi)^3}. 
\end{equation}
Similarly, we define the radial integral measure in $k-$space to be 
\begin{equation}
\label{radial_measure}
    \int [dk] = \int_0^\infty \frac{dk k^2}{(2\pi)^3}, 
\end{equation}
and its generalization to multiple wavenumbers as
\begin{equation}
    \int [dk]^n = \int_0^\infty \prod_{i=1}^n \frac{dk_i k_i^2}{(2\pi)^3}. 
\end{equation}
In the scenario where we work with spherically symmetric fields, the Fourier transform turns into the following sine transform, 
\begin{equation}\label{FT_sin}
    k\delta(k)=4\pi \int dx\,x\delta(x) \sin (kx), 
\end{equation}
where $x,k$ are the magnitudes of $\textbf{x}, \textbf{k}$ respectively. 
The power spectrum is defined as, 
\begin{equation}
    \left<  \delta(\textbf{k})\delta(\textbf{k}')\right> = (2\pi)^3 P(k) \delta_D(\textbf{k}+\textbf{k}') 
\end{equation}
where $\delta_D$ is the Dirac delta function. \\
\indent We also apply the following definition for the spherical harmonics: 
\begin{subequations}
    \begin{equation}
        Y_0(\theta, \phi)=1,
    \end{equation} 
    \begin{equation}
        Y_{\ell m}(\theta, \phi)= \frac{(-1)^{\ell+m}}{2^\ell \ell! } \left[ \frac{2\ell+1}{4\pi} \frac{(\ell-|m|)!}{(\ell+|m|)!}\right]^{1/2} e^{i m \phi } (\sin \theta)^{|m|} \left( \frac{d}{ d \cos \theta} \right)^{\ell+|m|}(\sin \theta)^{2\ell}
    \end{equation} 
\end{subequations}
for $\ell>0$ and $-\ell<m<\ell$. They obey the following relations, 
\begin{equation}
    \Delta_{\Omega}Y_{\ell m}=-\ell(\ell+1)Y_{\ell m}, \quad Y_{\ell m}(-\textbf{n})=(-1)^\ell Y_{\ell m}, \quad Y^*_{\ell m}(\textbf{n})=Y_{\ell,-m}(\textbf{n}),
\end{equation}
where $\Delta_\Omega$ is the Laplacian on the 2D sphere. The spherical harmonics also obey the following orthogonal and normalization conditions when integrated over the 2D sphere
\begin{equation}\label{Yid}
    \int d\Omega Y_{\ell m} Y^*_{\ell' m'} = (4\pi)^{\delta_{0,\ell}}\delta_{\ell \ell'}\delta_{m m'}
\end{equation}
where $\delta_{ij}$ is the Kronecker delta.\\
\indent Fields in position and momentum space can be decomposed over spherical harmonics as, 
\begin{subequations}\label{Yexpand}
    \begin{equation}
        \delta(\textbf{x})=\delta_0(r)+\sum_{\ell>0} \sum_{m=-\ell}^\ell \delta_{\ell m}(r)Y_{\ell m}(\textbf{x}/r), 
    \end{equation}

    \begin{equation}
        \delta(\textbf{k})=\delta_0(k)+\sum_{\ell>0} \sum_{m=-\ell}^\ell (-i)^\ell \delta_{\ell m}(k)Y_{\ell m}(\textbf{k}/k). 
    \end{equation}
\end{subequations}
Using (\ref{Yid}), the fields have the properties
\begin{equation}
    (\delta_{\ell m}(r))^*=\delta_{\ell, -m}(r), \quad (\delta_{\ell m}(k))^* = \delta_{\ell, -m}(k). 
\end{equation}
The position and momentum coefficients in \ref{Yexpand} are related by, 
\begin{equation}
    \delta_{\ell m}(r)=4\pi \int [dk] j_\ell (kr) \delta_{\ell m}(k)
\end{equation}
where $j_\ell(x)$ is the spherical Bessel function of order $\ell$. It is related to the Bessel function of the first kind as 
\begin{equation}
    j_\ell (x)=\sqrt{\frac{\pi}{2x}} J_{\ell+1/2}(x). 
\end{equation}
Spherical Bessel functions with product arguments of the form $j_\ell(kr)$ obey an orthogonal relation on the half-line given by 
\begin{equation}
    \int_0^\infty dr r^2 j_\ell(k'r) j_\ell (kr) = \frac{\pi}{2k^2}\delta_D(k-k'). 
\end{equation}

\section{Gaussian fluctuations around the saddle-point}\label{app:kernal}

\subsection{Kernels}
As discussed in \cite{ivanov}, contributions to the prefactor are found by examining small fluctuations around the saddle-point configuration. Writing the general fluctuation as
$\delta_L(\textbf{k})=\hat{\delta}_L(k)+\delta_L^{(1)}(\textbf{k})$, the perturbation to the filtered density contrast, up to second order, will take the form 
\begin{equation}\label{ddw}
    \bar{\delta}_W = \delta_* + \int_{\textbf{k}} S(\textbf{k}) \delta_L^{(1)}(\textbf{k}) + \int_{\textbf{k}_1}\int_{\textbf{k}_2} Q_{tot} (\textbf{k}_1, \textbf{k}_2)\delta_L^{(1)}(\textbf{k}_1)\delta_L^{(1)}(-\textbf{k}_2), 
\end{equation}
where we define 
\begin{subequations}\label{SQ_vec}
\begin{equation}\label{S_def}
    S(\textbf{k})= \left.\frac{\partial \bar{\delta}_W }{\partial \delta_L(\textbf{k})}\right|_{\hat{\delta}_L}, 
\end{equation} 
\begin{equation}\label{Qtotapp}
     Q_{tot}(\textbf{k}_1, \textbf{k}_2)=\frac{1}{2} \left.\frac{\partial ^2\bar{\delta}_W }{\partial \delta_L(\textbf{k}_1)\partial \delta_L(-\textbf{k}_2)}\right|_{\hat{\delta}_L}. 
\end{equation}
\end{subequations}
Expanding the fluctuations into spherical harmonics
\begin{equation}\label{dlY}
    \delta_L^{(1)}(\textbf{k})=\delta_{L,0}^{(1)}(k)+\sum_{\ell>0} \sum_{m=-\ell}^{\ell} (-i)^\ell \delta_{L,\ell m}^{(1)}(k)Y_{\ell m}(\textbf{k}/k),   
\end{equation}
and using the orthogonality condition of $Y_{\ell m}$, we find that the monopole contributions coming from (\ref{dlY}) can be rewritten as
\begin{equation}\label{dw0}
    \bar{\delta}_{W} = \delta_* + \int [dk] 4\pi S(k) \delta_{L,0}^{(1)}(k) + \int [dk]^2 16 \pi^2 \left< Q_{tot}(\textbf{k}_1,\textbf{k}_2) \right>_{\Omega_1, \Omega_2}\delta_{L,0}^{(1)}(k_1)\delta_{L,0}^{(1)}(k_2)+\ldots,
\end{equation}
where $\left<. \right>_{\Omega_1, \Omega_2}$ denotes the spherical average over $\Omega_1$ and $\Omega_2$ and dots stand for the contributions of higher multipoles. Here we have
used that the spherical symmetry of $\hat\delta_L(k)$ implies spherical symmetry of $S(k)$  \cite{ivanov}.
Also the kernel 
$\left< Q_{tot}(\textbf{k}_1,\textbf{k}_2) \right>_{\Omega_1, \Omega_2}$ now depends only on the magnitudes $k_1$ and $k_2$, but not on their directions.

Alternatively, we can expand 
$\bar{\delta}_W$ over spherically symmetric configurations from the start, 
\begin{equation}\label{dw_sp}
    \bar{\delta}_{W,0}=\delta_*+\int[dk]4\pi S(k)\delta^{(1)}_{L,0}(k)+ \int [dk]^2 4\pi  Q_0(k_1,k_2) \delta_{L,0}^{(1)}(k_1)\delta_{L,0}^{(1)}(k_2), 
\end{equation}
for some spherically symmetric kernel $Q_0(k_1,k_2)$. 
Comparing Eqs.~(\ref{dw0}) and (\ref{dw_sp}) we conclude that 
\begin{equation}\label{Q0}
    Q_0(k_1,k_2)=4\pi \left< Q_{tot}(\textbf{k}_1,\textbf{k}_2) \right>_{\Omega_1, \Omega_2}
\end{equation}
In order to find the expressions for $S(k)$ and $Q_0(k_1,k_2)$, we use Eq.~(\ref{dw2}). Expanding its right-hand side to the second order in the perturbation of the averaged linear density $\bar\delta_L^{(1)}(R)$, we get:
\begin{equation}\label{dw_var}
    \begin{split}            
    \bar{\delta}_W=&\delta_*-\frac{4\pi}{3r_*^4}\int dR\, R^3 \,\Tilde{W}' \frac{f'}{(1+f)^{4/3}}\,\bar\delta_L^{(1)}(R) \\
     &+ \frac{2\pi}{3r_*^4}\int dR\, R^3 \bigg[ \frac{R\Tilde{W}^{''}}{3r_*} \frac{(f')^2}{(1+f)^{8/3}} + \Tilde{W}'\frac{4(f')^2}{3(1+f)^{7/3}}
     -\Tilde{W}'\frac{f''}{(1+f)^{4/3}} \bigg]\,\big(\bar\delta_L^{(1)}(R) \big)^2
    \end{split}
\end{equation}
where primes denote differentiation with respect to the argument of the function,
$f$ and its derivatives are evaluated at $\bar{\hat\delta}_L(R)$, while
$\Tilde W$ and its derivatives are evaluated at $R(1+f)^{-1/3}/r_*$.
Next, we use the relation,
\begin{equation}
\label{deltaLav}
\bar\delta_L(R)=\int_\k W_{\rm th}(k R) \,\delta_L(\k)~~\Longrightarrow~~
    \bar\delta^{(1)}_L(R)=4\pi\int [dk]\, W_{\rm th}(kR) \,\delta_{L,0}^{(1)}(k)\;.
\end{equation}
Substituting into (\ref{dw_var}) and comparing with the form (\ref{dw_sp}), we read off the kernels:
\begin{subequations}\label{SQint}
\begin{align}
\label{Skeq}
&S(k)=-\frac{4\pi}{3r_*^4}\int dR\, R^3\, \Tilde{W}' \frac{f'}{(1+f)^{4/3}} W_{\rm th}(kR)\,,\\
\label{Q0_full}
&Q_0(k_1,k_2)=\frac{8\pi^2}{3r_*^4}\int dR\, R^3  \bigg[ \frac{R\Tilde{W}^{''}(f')^2}{3r_*(1\!+\!f)^{8/3}}  + \frac{4\Tilde{W}'(f')^2}{3(1\!+\!f)^{7/3}}
     -\frac{\Tilde{W}'f''}{(1\!+\!f)^{4/3}} \bigg]   W_{\text{th}}(k_1R)W_{\text{th}}(k_2R)\,.
    \end{align}
\end{subequations}
An alternative formula for $S(k)$ is obtained by substituting Eq.~(\ref{dw_sp}) into Eq.~(\ref{EL1}), giving us 
\begin{equation}\label{S_fin}
    S(k)=-\frac{\hat{\delta}_L(k)}{\hat{\lambda}P(k)}\;. 
\end{equation}
By combining this with Eq.~(\ref{Skeq}) we get an integral equation for the saddle-point configuration, 
\begin{equation}
\label{dl_int}
    \hat{\delta}_L(k)=\hat{\lambda}P(k) \frac{4\pi}{3r_*^4} \int dR\, R^3\, \Tilde{W}' \frac{f'}{(1+f)^{4/3}} W_{\rm th}(kR)\;.
\end{equation}

\subsection{Monopole prefactor}

To compute the prefactor of the PDF, corresponding to the term $\alpha_1$ in the expansion (\ref{PDFexpansion}), we substitute the quadratic expression (\ref{ddw}) into Eq.~(\ref{PDFfull}) and perform Gaussian integral over $\delta_L^{(1)}(\k)$. Due to the spherical symmetry of the saddle-point solution, the integrals over perturbations with different multipole numbers factorize \cite{ivanov} leading to the expression (\ref{pertb_pdf}). Here we focus on the monopole contribution which has the form,
\begin{equation}
\begin{split}
    \mathcal{A}_0&=\mathcal{N}_0^{-1}\int_{-i\infty}^{i\infty} \frac{d\lambda^{(1)}}{2\pi i g^2}\int \mathcal{D}\delta^{(1)}_{L,0} \exp \biggl\{ -\frac{4\pi}{g^2}\left[ \int \frac{[dk]}{2P(k)} \big(\delta^{(1)}_{L,0}(k)\big)^2 \right. \\
    &\qquad \left. +\lambda^{(1)}\int[dk]S(k)\delta^{(1)}_{L,0}(k)+\hat{\lambda}\int [dk]^2 Q_0(k_1,k_2) \delta^{(1)}_{L,0}(k_1)\delta^{(1)}_{L,0}(k_2 ) \right] \biggr\}, 
\end{split}
\end{equation}
where $\mathcal{N}_0$ is the normalization constant,
\begin{equation}
    \mathcal{N}_0=\int \mathcal{D}\delta^{(1)}_{L,0} \exp \biggl\{ -\frac{4\pi}{g^2} \int \frac{[dk]}{2P(k)} \big(\delta^{(1)}_{L,0}(k)\big)^2\biggr\} ,
\end{equation}
and we have introduced the perturbation of the Lagrange multiplier, $\lambda=\hat\lambda+\lambda^{(1)}$. 
Note that the above expressions do not depend on the choice of filter, except in determining the explicit forms of the kernels $S(k)$, $Q_0(k_1,k_2)$. 
Rescaling of the integration variables,
\begin{equation}
    \tilde \delta_{L,0}^{(1)}(k)=\frac{4\pi}{g}\delta_{L,0}^{(1)}(k)~,~~~~~~~
    \tilde \lambda^{(1)}=\frac{\lambda^{(1)}}{g}\;,
\end{equation}
leads to
\begin{align}
    &\mathcal{A}_0\!=\tilde{\mathcal{N}}_0^{-1}\!\!\!\int\limits_{-i\infty}^{i\infty} \!\!\frac{d\tilde\lambda^{(1)}}{2\pi i g}\int \!\mathcal{D}\tilde\delta^{(1)}_{L,0} \exp \biggl\{ -\frac{1}{2} 
    \int [dk_1][dk_2] {\cal O}(k_1,k_2) \tilde\delta^{(1)}_{L,0}(k_1)\tilde\delta^{(1)}_{L,0}(k_2) 
    -\!\int [dk] S(k) \tilde\delta^{(1)}_{L,0}(k)\tilde\lambda^{(1)}\biggr\}, \\
    &\tilde{\mathcal{N}}_0=\int \mathcal{D}\tilde\delta^{(1)}_{L,0} \exp \biggl\{ -\frac{1}{2} \int \frac{[dk]}{P(k)} \big(\tilde\delta^{(1)}_{L,0}(k)\big)^2\biggr\} ,
\end{align}
where ${\cal O}(k_1,k_2)$ is the kernel introduced in (\ref{Okernel}). Finally, using the standard expression for the Gaussian integrals and taking into account that the integral in ${\cal A}_0$ has one more variable than the integral in $\tilde{\cal N}_0$, bringing an extra factor $\sqrt{2\pi}$, we arrive at Eq.~(\ref{monopre}) from the main text.

Let us discuss the value of ${\cal A}_0$ at $\delta_*=0$. In this case we have $\hat\delta_L(R)=\hat\lambda=0$, so that ${\cal O}(k_1,k_2)=\frac{\mathbbm{1}(k_1,k_2)}{4\pi P(k_1)}$. Then, using the formula for the determinant of a block matrix,
\begin{equation}
    \det
    \begin{pmatrix} 
    A&B\\
    C&D
    \end{pmatrix}=\det D\,\det(A-BD^{-1}C)\;,
\end{equation}
we obtain,
\begin{equation}
\label{detHS}
    \det \mathcal{H} = \det \bigg(\frac{\mathbbm{1}(k_1,k_2)}{4\pi P(k_1)} \bigg)\bigg(-\int [dk] 4\pi P(k) \big(S(k)\big)^2\bigg)\;.
\end{equation}
Further, Eqs.~(\ref{SCs}) imply $f(0)=0$, $f'(0)=1$. Integrating by parts in Eq.~(\ref{Skeq}) and using the identity 
\begin{equation}
    \frac{d}{dz}\bigg(\frac{z^3}{3}W_{\rm th}(z)\bigg)=z^2j_0(z)
\end{equation}
we find 
\begin{equation}
    S(k)=\frac{4\pi}{r_*^3} \int dR\,R^2\tilde W(R/r_*)j_0(kR)=W(kr_*)\;.
\end{equation}
Substituting into Eq.~(\ref{detHS}) and then into Eq.~(\ref{monopre}), we obtain 
\begin{equation}
\label{A00}
{\cal A}_0(0)=(2\pi g^2\sigma_{L,r_*}^2)^{-1/2}\;,    
\end{equation}
 with $\sigma_{L,r_*}^2$ being the linear variance defined in (\ref{var_L}). This leads to Eq.~(\ref{PDF0}) from the main text.

\section{Taylor expansion of aspherical prefactor at small $\delta_*$}
\label{app:C}

\subsection{Consistency relations}
\label{app:C1}
The first two terms in the Taylor expansion of the aspherical prefactor at small density contrasts can be found directly from the identities (\ref{norms}) satisfied by the PDF. 
The idea is to Taylor expand all factors in the expression $\mathcal{P}(\delta_*)=\mathcal{A}_{asp}(\delta_*)\mathcal{A}_0(\delta_*)e^{-\alpha_0(\delta_*)/g^2}$ around $\delta_*=0$ and evaluate the integrals at the leading order in $g^2$. 

For the first condition in (\ref{norms}) we have at the zeroth order in $g^2$:
\begin{align}
    1=\mathcal{A}_{asp}(0)\mathcal{A}_{0}(0) \int e^{- \frac{\delta_*^2}{2g^2 \sigma_{L,r_*}^2}}\, d\delta_*=\mathcal{A}_{asp}(0) \mathcal{A}_{0}(0) \sqrt{2\pi g^2\sigma_{L,r_*}^2}\;, 
\end{align}
where we have used
\begin{equation}
\label{alphaTexp}
\alpha_0(\delta_*)=\frac{\delta_*^2}{2\sigma_{L,r_*}^2}+{\cal O}(\delta_*^3)\;.
\end{equation}
Using further Eq.~(\ref{A00}), we obtain ${\cal A}_{asp}(0)=1$, which is the first of Eqs.~(\ref{Aaspprops}). 
In the second condition (\ref{norms}) the leading non-trivial terms are of order $g^2$ which come from expanding the prefactors up to linear terms in $\delta_*$ and the exponent --- to the cubic term. We have 
\begin{align*}
    0&=\int\delta_* \left(1+\mathcal{A}'_{asp}(0)\,\delta_*\right) {\cal A}_0(0)
    \left( 1 +(\log {\cal A}_0)'(0)\,\delta_*\right) \exp \left\{ -\frac{\delta_*^2}{2\sigma_{L,r_*}^2 g^2} - \frac{\alpha_0'''(0)}{6g^2}\delta_*^3\right\} \, d\delta_*\\
    &=g^2\sigma_{L,r_*}^2\left[\mathcal{A}'_{asp}+(\log\mathcal{A}_0)'(0) -\frac{\alpha_0'''(0)}{2}\sigma_{L,r_*}^2\right], 
\end{align*}
which gives the second equation in (\ref{Aaspprops}).

We cannot extend this tactics to get the ${\cal O}(\delta_*^2)$ term in ${\cal A}_{asp}$ since that would require considering subleading terms in $g^2$. These, in turn, acquire contributions from the higher-order term in the saddle-point expansion of the PDF (the term $\alpha_2g^4$ in Eq.~(\ref{PDFexpansion})), which we have neglected in this paper. We thus employ a different approach based on the perturbative expansion the PDF in the density contrast.

\subsection{Perturbative calculation}
\label{asp_pert}
 
In this Appendix we fill the details of the perturbative calculation of ${\cal A}_{asp}$ up to quadratic order in $\delta_*$ described in Sec.~\ref{sec:pert}. First we write down the expansion of the function $f(x)$ from the spherical collapse map (\ref{SCmap}), which we need to the cubic order. In the EdS approximation, we obtain from~(\ref{SCs}):
\begin{equation}
\label{fcub}
    f(x)=x+\frac{17}{21}x^2+\frac{341}{567}x^3+\ldots\;.
\end{equation}
Next, we determine the saddle-point solution $\hat\delta_{L}(k)$ to the second order in $\delta_*$. To this aim, we use the integral equation (\ref{dl_int}). Expanding its r.h.s. to the second order and using the form (\ref{lam_exp}) of the Lagrange multiplier, we have 
\begin{equation}
\label{deltakeq2}
\begin{split}
 \hat\delta_L(k)=&\frac{\delta_*}{\sigma^2_{L,r_*}}P(k)W(kr_*)-\delta_*^2 a_2 P(k)W(kr_*)\\
 -&\frac{\delta_*}{\sigma^2_{L,r_*}}\frac{4\pi}{3r_*^4}P(k)
 \int dR\,R^3\,\bar\delta_L(R)\bigg(-\frac{R}{3r_*}\tilde W''(R/r_*)+\frac{2}{7} \tilde W'(R/r_*)\bigg) W_{\rm th}(kR)\;,
 \end{split}
\end{equation}
where the coefficient $a_2$ is yet to be determined. In deriving this equation, we have integrated by parts and simplified the result using the identities for the spherical Bessel functions. The linear in $\delta_*$ term on the r.h.s. already has an explicit form. It gives an expression for $\hat\delta_L(R)$ and its TopHat-averaged version at linear order,
\begin{equation}
\label{deltak1}
    \hat\delta_L(R)=\frac{\delta_*}{\sigma_{L,r_*}^2}\xi(R)+{\cal O}(\delta_*^2)\;,\qquad
    \bar{\hat\delta}_L(R)=\frac{\delta_*}{\sigma_{L,r_*}^2}\bar\xi(R)+{\cal O}(\delta_*^2)\;,
\end{equation}
where 
\begin{equation}
\label{deltastand}
 {\xi}(R)=4\pi\int [dk]P(k)W(kr_*)j_0(kR)\;,\qquad 
 \bar{\xi}(R)=4\pi\int [dk]P(k)W(kr_*)W_{\rm th}(kR)\;.
\end{equation}
Substituting these results into the last term of Eq.~(\ref{deltakeq2}) and integrating by parts to eliminate the second derivatives of the window function we obtain,
\begin{align}
\label{deltak2}
\begin{split}
    \hat{\delta}_L(k)= & \frac{\delta_*}{\sigma_{L,r_*}^2} P(k)W(kr_*) -\delta_*^2 a_2 P(k)W(kr_*) \\ 
    -&\frac{\delta_*^2}{\sigma_{L,r_*}^4}
    \frac{4\pi}{3r_*^4} P(k)\!\! \int \!\! dR \, R^3 \tilde{W}'(R/r_*) \left[ W_{\rm th}(kR){\xi}(R) + j_0(kR)\bar\xi(R) -\frac{8}{21}W_{\rm th}(kR)\bar\xi(R) \right].
    \end{split}
\end{align}
We still need to find $a_2$. The shortest route is to insert Eq.~(\ref{deltak2}) into the action (\ref{expalpha}), which we get to the cubic order,
\begin{align}
\label{alphacub}
    \alpha_0(\delta_*)= & \frac{\delta_*^2}{2\sigma^2_{L,r_*}} -\delta_*^3 a_2 
    -\frac{\delta_*^3}{ \sigma^6_{L,r_*} }\frac{8\pi}{3r_*^4} \int dR \, R^3 \tilde{W}'(R/r_*) \bar\xi(R) \left( {\xi}(R) - \frac{4}{21}\bar\xi(R) \right)\;. 
\end{align}
Finally, Eq.~(\ref{lam}) relates it to the Lagrange multiplier, whence we extract:
\begin{align}
    a_2 =   -\frac{4\pi}{\sigma^6_{L,r_*}r_*^4} \int dR \, R^3 \tilde{W}'(R/r_*) \bar\xi(R) \left(\xi(R) - \frac{4}{21}\bar\xi(R) \right)\;. 
    \label{a2app}
\end{align}

We proceed to the expression for the kernel $Q_0(k_1,k_2)$. The general formula for it is given in Eq.~(\ref{Q0_full}). We expand it to linear order and substitute the expressions (\ref{deltak1}). This yields, 
\begin{equation}
\begin{split}
    &Q_0(k_1,k_2)=\frac{8\pi^2}{3r_*^4}\int dR\, R^3\left[ \frac{R}{3r_*}\Tilde{W}''(R/r_*) -\frac27 \Tilde{W}'(R/r_*)\right]W_{\text{th}}(k_1R)W_{\text{th}}(k_2R)\\
     &+ \frac{\delta_*}{\sigma^2_{L,r_*}}\frac{8\pi^2}{3r_*^4}\!\int \!\! dR \, R^3\bar\xi(R) \left[ -\frac{R^2}{9r_*^2} \tilde{W}'''(R/r_*) + \frac{2R}{7r_*}\tilde{W}''(R/r_*)
   -\frac{46}{189}\tilde{W}'(R/r_*) \right] W_{\rm th}(k_1R)W_{\rm th}(k_2R)\;.
\end{split}
\end{equation}
Now it is straightforward to evaluate the traces (\ref{Taus}), with the result,
\begin{subequations}
\label{t_eqs}
\begin{align}
    \mathcal{T}_{10}=&\frac{2\pi}{3r_*^4}\int dR\, R^3\left[ \frac{R}{3r_*}\Tilde{W}''(R/r_*) -\frac27 \Tilde{W}'(R/r_*) \right]\xi_{\rm th}(R,R)\;, \\
    \mathcal{T}_{11}=&\frac{1}{\sigma^2_{L,r_*}}\frac{2\pi}{3r_*^4}\int dR \, R^3\left[ -\frac{R^2}{9r_*^2}\tilde{W}'''(R/r_*) + \frac{2R}{7r_*}\tilde{W}''(R/r_*)
  -\frac{46}{189}\Tilde{W}'(R/r_*) \right] \bar\xi(R) \xi_{\text{th}}(R,R), \\
    \mathcal{T}_{20}=&\frac{4\pi^2}{9r_*^8}\int dR_1 \, dR_2 \, R_1^3 R_2^3\left[ \frac{R_1}{3r_*}\Tilde{W}''(R_1/r_*) -\frac27\Tilde{W}'(R_1/r_*) \right] \notag\\
&\qquad \qquad \qquad \qquad~\times \left[ \frac{R_2}{3r_*}\Tilde{W}''(R_2/r_*) -\frac27\Tilde{W}'(R_2/r_*) \right]\big(\xi_{\text{th}}(R_1,R_2)\big)^2\;, 
\end{align}
\end{subequations}
with 
\begin{equation}
    \xi_{\text{th}}(R_1,R_2)=4\pi \int \left[ dk \right] P(k)W_{\text{th}}(kR_1)W_{\text{th}}(kR_2)\;.
\end{equation}

To find $Q_{tot}({\bf k}_1,{\bf k}_2)$, we adopt a different strategy. We use the definition (\ref{Qtotapp}) and evaluate the non-linear density contrast entering in $\bar\delta_W$ using the cosmological perturbation theory. We start with SPT and perform the EFT renormalization in the end. The SPT expression reads,
\begin{equation}
\label{deltaSPT}
    \delta(\textbf{k}) = \delta_L(\textbf{k}) + \sum_{n=2}^\infty \int_{\textbf{k}_1}...\int_{\textbf{k}_n} (2\pi)^3 \delta_D\Big( \textbf{k}-\sum_{i=1}^n \textbf{k}_{i} \Big) F_n(\textbf{k}_1, ..., \textbf{k}_n) \prod_{i=1}^n \delta_L(\textbf{k}_i)\;, 
\end{equation}
where $F_n$ are the SPT kernels whose explicit form can be found e.g. in  \cite{bernardeauSPT}. Averaging this expression with the filter and computing the variational derivatives, we obtain the first two terms in the expansion of $Q_{tot}$ in the saddle-point density,
\begin{equation}
\label{Qtot_2nd}
    Q_{tot}(\textbf{k}_1, \textbf{k}_2)= F_2(\textbf{k}_1, -\textbf{k}_2)W(|\textbf{k}_{1}-\textbf{k}_2|r_*)+3\int_\textbf{q}F_3(\textbf{k}_1, -\textbf{k}_2, \textbf{q})W(|\textbf{k}_1-\textbf{k}_2+\textbf{q}|r_*)\hat{\delta}_L(\textbf{q})\;.
\end{equation}
We now substitute here the linear term from Eq.~(\ref{deltak2}) and compute the traces
\begin{subequations}
\label{Tracestot}
\begin{align}
        &\Tr(Q_{tot}P)=\frac{3\delta_*}{\sigma^2_{L,r_*}}\int_{\textbf{k}} \int_{\textbf{q}}F_3(\textbf{q},-\textbf{q}, \textbf{k}) |W(kr_*)|^2 P(q)P(k)+{\cal O}(\delta_*^2)\;, \\
      &\Tr(Q_{tot}PQ_{tot}P) = \int_{\textbf{k}_1}\int_{\textbf{k}_2} F_2^2(\textbf{k}_1, \textbf{k}_2) P(k_1) P(k_2) |W(|\textbf{k}_{12}|r_*)|^2+{\cal O}(\delta_*)\;, 
\end{align}
\end{subequations}
where we have used the identities 
\[
F_2({\bf k},-{\bf k})=0\;,\qquad 
F_{2}({\bf k}_1,{\bf k}_2)=F_{2}({\bf k}_2,{\bf k}_1)\;,\qquad
F_{2}({\bf k}_1,{\bf k}_2)=F_{2}(-{\bf k}_1,-{\bf k}_2)\;.
\]
Substituting this into the expression
\begin{align}
\label{Dtotmass}
    \sqrt{D_{tot}} \approx \exp \left[\hat{\lambda}\Tr(Q_{tot}P) -  \hat{\lambda}^2\Tr(Q_{tot}PQ_{tot}P)\right]\;,
\end{align}
and combining with Eq.~(\ref{lam_exp}) we find the expressions (\ref{Dtot_pert}), (\ref{sig-1loop_main}) from the main text, with the 1-loop correction to the power spectrum given by the SPT expression,
    \begin{equation}
    \label{sig_1loop}
         P_{\text{1-loop}}^{\rm SPT}(k)=\int_{\textbf{q}}\left( 6F_3(\textbf{k}, -\textbf{q}, \textbf{q}) P(q)P(k) + 2F_2^2(\textbf{k}-\textbf{q}, \textbf{q})P(|\textbf{k}-\textbf{q}|)P(q) \right)\; . 
    \end{equation}
Equation (\ref{Dtot_pert}) shows that, similarly to the case of the TopHat filter \cite{ivanov,anton}, the determinant $D_{tot}$ depends on 1-loop variance of the filtered density. 

We now need to take into account that the SPT expression for the non-linear density (\ref{deltaSPT}) is incomplete: it must be corrected by the EFT counterterm responsible for non-perturbative short-distance physics. This leads to a counterterm in $D_{tot}$. Repeating the general arguments of \cite{ivanov,anton} relating the counterterms in the PDF and in the correlation functions, one concludes that at quadratic order in $\delta_*$ we just need to replace $P_{\text{1-loop}}^{\rm SPT}$ in the 1-loop variance by its renormalized expression (\ref{Pcntr}). This brings us to the final expression for the aspherical prefactor (\ref{asp_ren1}).

\section{Code validation}\label{app:figs}

We have tested the convergence of the numerical algorithm described in Sec.~\ref{sec:numeric} for various lattice parameters $(N,D)$ by comparing its output with the exact analytic expressions for the spherical PDF for the TopHat filter. We have observed that the same choice of $(N,D)$ leads to a slower convergence of 
the PDF for strongly underdense regions, $\delta_*\lesssim -0.4$, than for overdense and moderately underdense cells. This is likely due to the shape of the saddle-point density profile: as seen from Fig.~\ref{fig:prof_comp}, the amplitude and the width of the saddle-point configuration grow when $\delta_*$ gets large and negative. One thus expects that getting accurate results at the underdense tail of the PDF will require a larger lattice size. We find that a good accuracy is achieved for the lattice size parameter taking the values $D=20$ for $\delta_*>-0.4$ and $D=30$ for $\delta_*\leq -0.4$. To preserve the same density of lattice points across all values of $\delta_*$, we rescale $N$, taking $N=300$ for $\delta_*>-0.4$ and $N=450$ for $\delta_*\leq -0.4$. These fiducial values are listed in the second column of Table~\ref{table:conv_test}. 
Note that, according to Eq.~(\ref{lattice}), the fiducial values of $D$ imply that the radius of the lattice is $20$ to $30$ times bigger than the Lagrangian radius of the cell.

\begin{figure}[t]
\centering
\hspace*{-1cm}
\includegraphics[width=0.38\paperwidth]{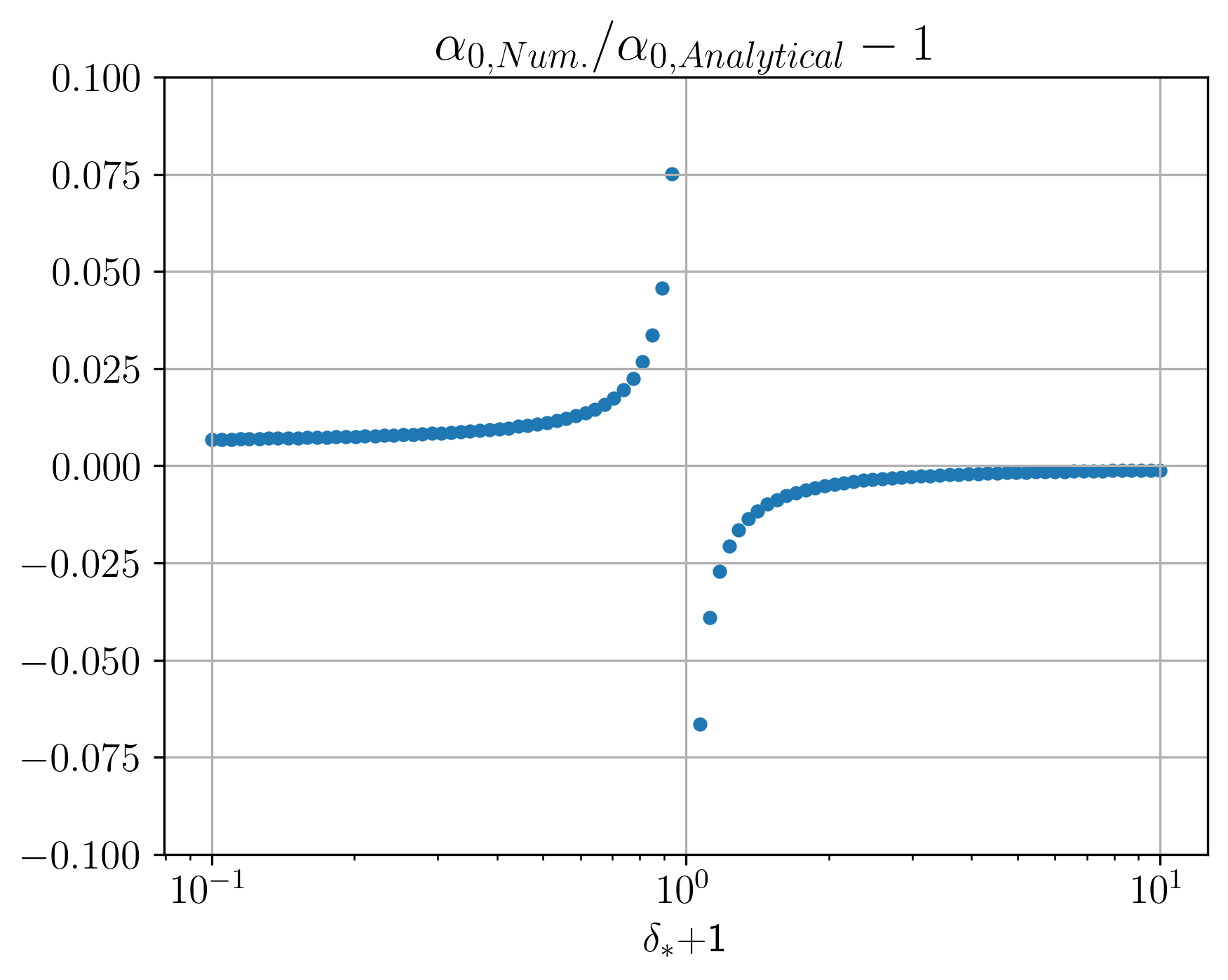}
\includegraphics[width=0.387\paperwidth]{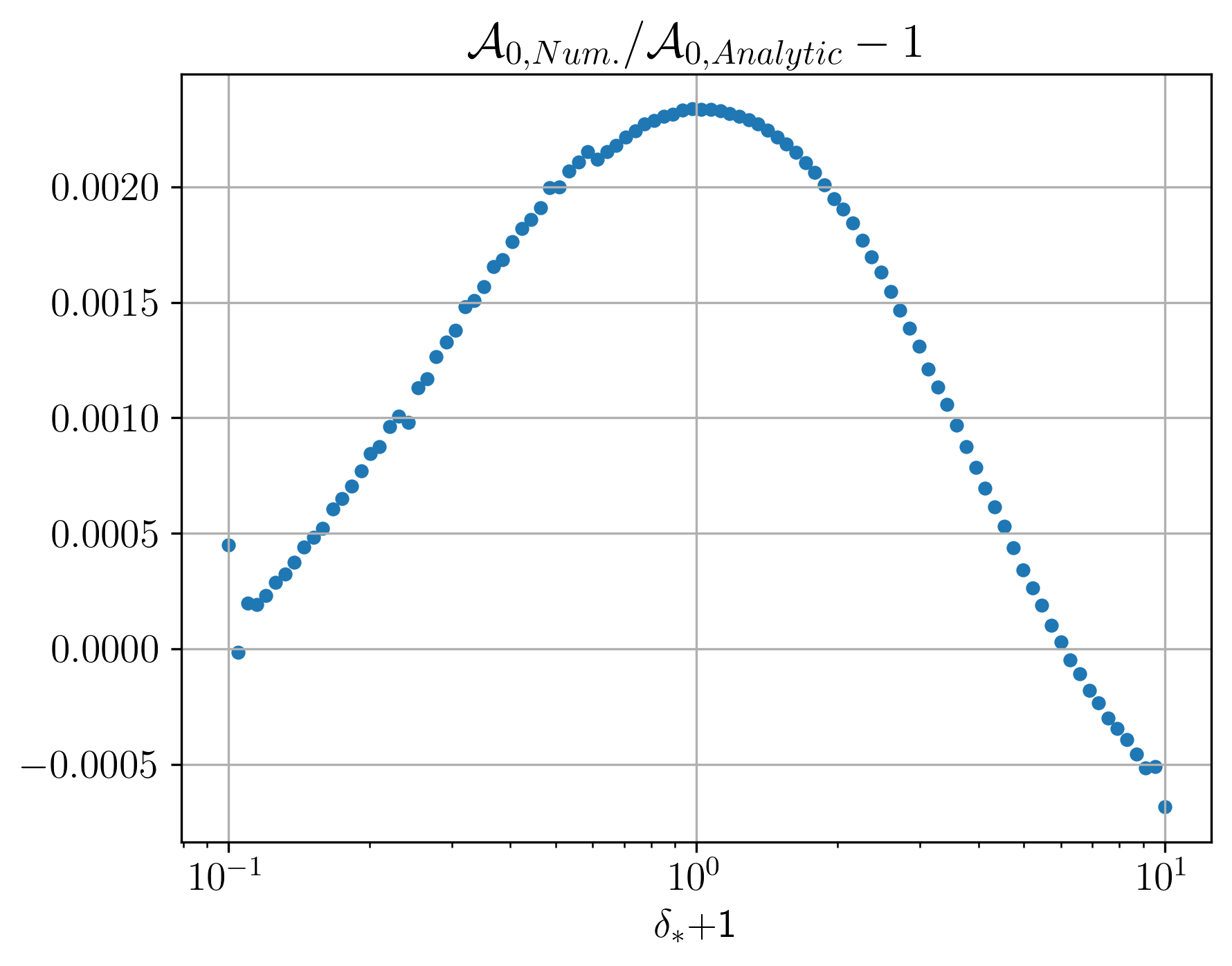}
\caption{Residuals between the numerical and analytical values of the action ({\it left}) and monopole prefactor ({\it right}) for the TopHat filter for the fiducial lattice parameters. }
\label{fig:THvalid}
\end{figure}

In Fig.~\ref{fig:THvalid} we show the residuals of the saddle-point action and monopole prefactor obtained with the fiducial choice of $(N,D)$. We observe relatively large residuals in the action near $\delta_*=0$. This is, however, not an issue, since the action vanishes at the origin, so large residuals come merely from the division by a small number. More important is the $\sim 0.5\%$ residual of the action persisting throughout the underdense region. Since the action enters into the exponent, this error gets enhanced leading to a strong increase of the systematic PDF error at extreme underdensities, see Fig.~\ref{fig:TH_Psp} from the main text. 
At the same time, the error in the prefactor plotted on the right panel of Fig.~\ref{fig:THvalid} is always below $0.3\%$. It gives a subdominant contribution into the PDF error budget.

\begin{table}[ht]
\centering
\begin{tabular}{|c|c|c|c|}
\hline
      & Fiducial & Test 1   & Test 2   \\ [1ex]  \hline
$(N,D)_<$ & (450,30) &  (360,30) & (360,24) \\ [1ex] \hline
$(N,D)_>$ & (300,20) &  (240,20) & (225,15) \\ [1ex] \hline 
\end{tabular} 
\caption{Lattice parameters used in convergence tests for non-TopHat filters. Here $(N,D)_<$ and $(N,D)_>$ denote the lattice parameters when $\delta_*\leq-0.4$ and $\delta_*>-0.4$ respectively.}
\label{table:conv_test}
\end{table}

To assess the precision of the code for non-TopHat filters, where analytic results are unavailable, we vary the lattice parameters away from their fiducial values and compare the results for the action and monopole prefactor. If the difference stays below 
$\sim 2\%$ across the range of considered $\delta_*$, we accept the results as converged. We consider two tests. In Test~1 we decrease the number of the lattice points keeping lattice size fixed. This corresponds to larger lattice spacing and thus a poorer spatial resolution. In Test~2 we decrease the lattice size, while keeping the lattice spacing fixed. 
Table \ref{table:conv_test} summarizes the parameter choices made for each case. Note that in each case the density of points $N/D$ is kept the same across all $\delta_*$. 

\begin{figure}[t]
\centering
\hspace*{-1cm}                                                           
\includegraphics[width=0.38\paperwidth]{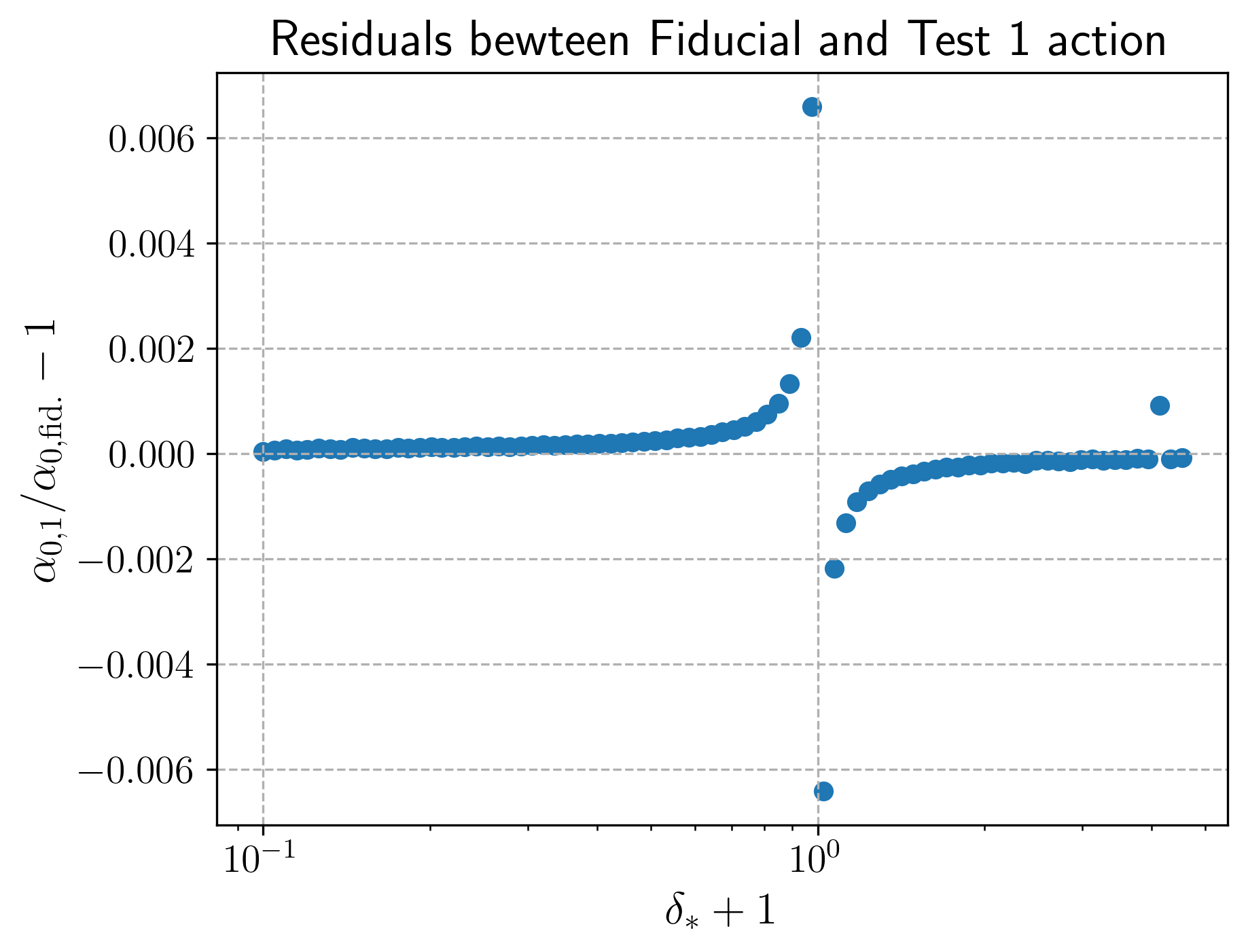}
\hspace*{0cm}                                                           
\includegraphics[width=0.4\paperwidth]{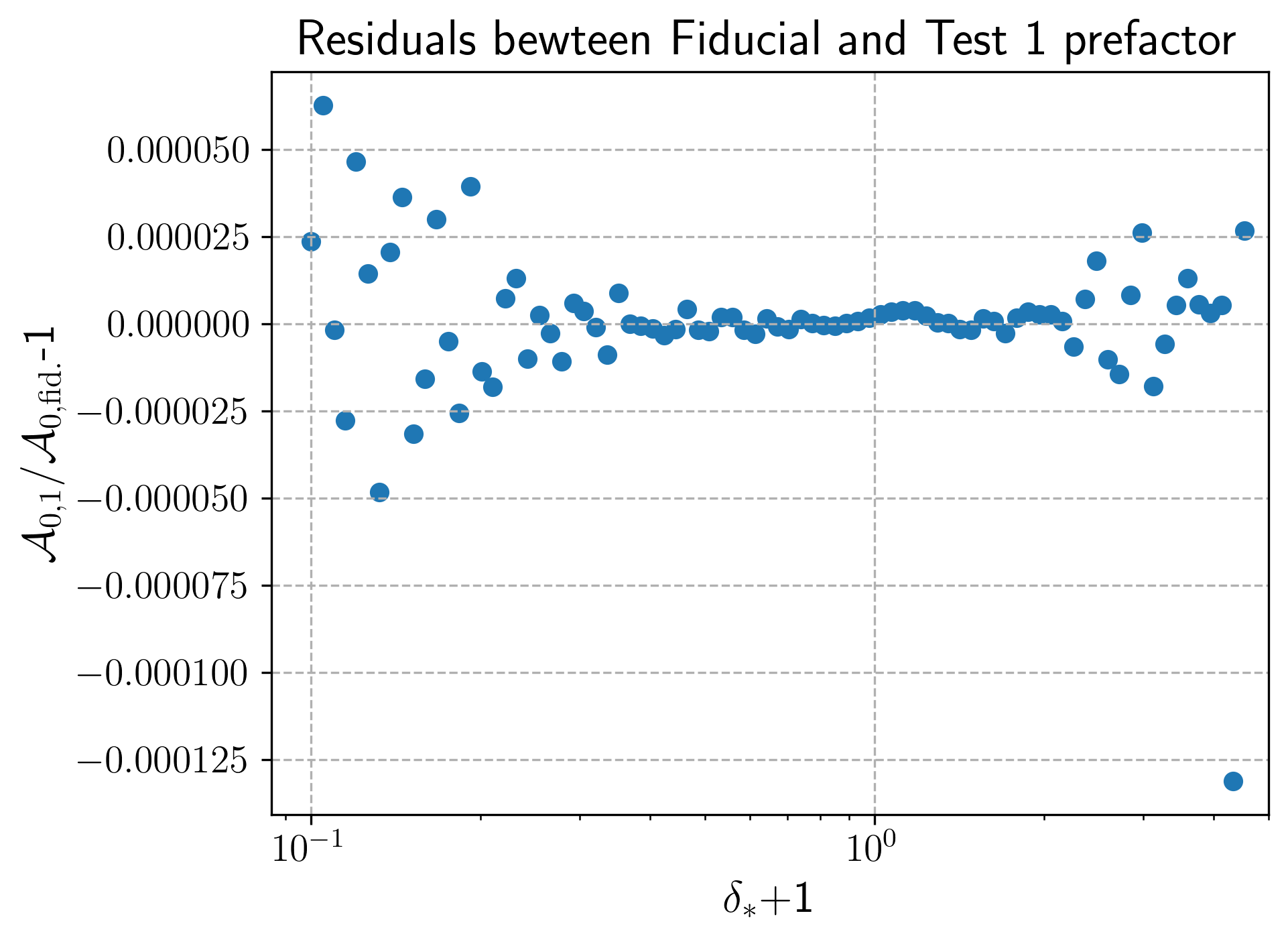}
\hspace*{-1cm}                                                           
\includegraphics[width=0.38\paperwidth]{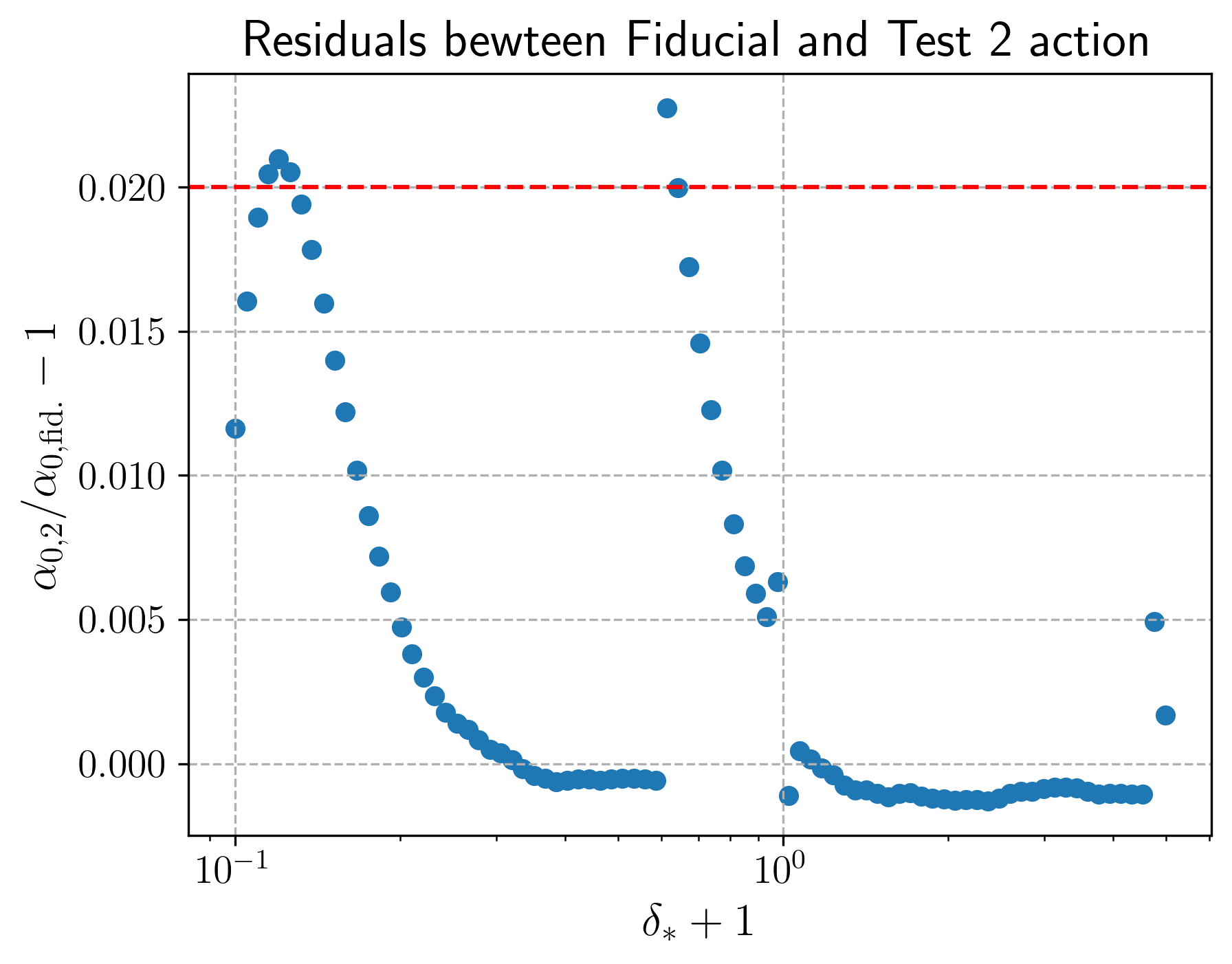}
\hspace*{0cm}                                                           
\includegraphics[width=0.4\paperwidth]{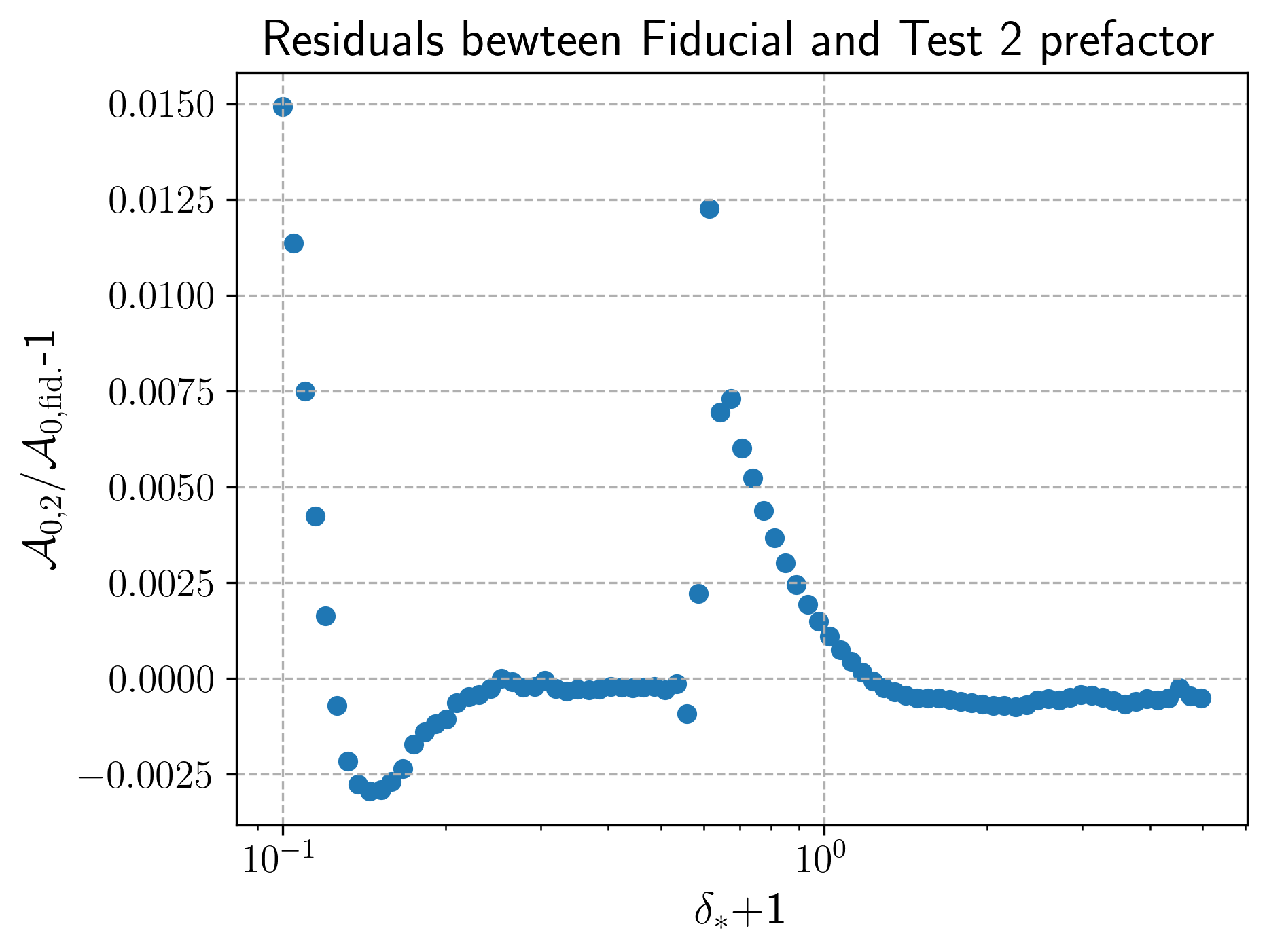}
\caption{Residuals of the action and monopole prefactor between several test cases listed in Table~\ref{table:conv_test}. Red dashed line displays the $2\%$ error threshold. }
\label{fig:convergence}
\end{figure}

Figure~\ref{fig:convergence} shows an example of the convergence tests for the Gaussian filter. Looking at the comparison with Test~1, we see that the difference between cases is well below $1\%$ across all $\delta_*$ for both the action and prefactor. This implies that we use sufficiently high number of lattice points and increasing it further will not improve the precision. 
Comparison with Test~2, however, shows larger differences in the action term and prefactor. 
Still, these remain below the $\sim 2\%$ threshold and thus, according to our criterion, the pipeline has converged. Larger discrepancies than in Test~1 imply that 
the lattice size has stronger effect on the accuracy of the code than the lattice spacing. 
We also see that the error significantly grows when the density contrast approaches the value $\delta_* = -0.4$ from above. This motivates the increase of $D$ below $\delta_* = -0.4$.

\section{Hockey-Stick filter: Analytic results}
\label{app:DWell}

In this Appendix we present analytical derivation of the spherical PDF for the Hockey-Stick filter (\ref{DWmain}). 
Computing $\bar{\delta}_W$ for this filter, we find
\begin{align}
    \bar{\delta}_W&= \frac{1}{(\alpha -1)\beta^3 +1} \left[ (\alpha -1)\beta^3 \bar{\delta}(\beta r_*) + \bar{\delta}(r_*)\right] \equiv n_1 \bar\delta_1 +n_2 \bar\delta_2\;, \label{DW_dw}
\end{align}
where we have introduced the notations
\begin{equation}
\label{HS_notations}
\bar\delta_{1,2}=\bar \delta(r_{1,2})\;,\qquad r_1=\beta r_*\;,~~~r_2=r_*\;,
\end{equation}
and $\bar\delta(r)$ stands for the TopHat-averaged density contrast (\ref{d_avg}).
Thus $\bar\delta_W$ is a linear combination of density contrasts inside two concentric cells. Note that for $0<\alpha<1$, the coefficient $n_1$ is negative, whereas $n_2$ is always positive. 

The probability that the two density contrasts equal $\delta_{*1}$ and $\delta_{*2}$ is given by the path integral over linear density fluctuations of the type (\ref{PDFfull}),
\begin{equation}
\label{DW_PDF}
    \begin{split}
    \mathcal{P}(\delta_{*1},\delta_{*2})=\mathcal{N}^{-1}\!\!\!\int\limits_{-i\infty}^{+i\infty} 
    \!\frac{d\lambda_1d\lambda_2}{(2\pi i g^2)^2}
    \int \!\mathcal{D}\delta_{L} \exp{ -\frac{1}{g^2}\left[\int_{\textbf{k}}\frac{|\delta_{L}({\bf k})|^2}{2P(k)} 
    -\lambda_1 (\delta_{*1}\!-\!\bar\delta_1[\delta_L])-\lambda_2 (\delta_{*2}\!-\!\bar\delta_2[\delta_L])\right]},
    \end{split}
\end{equation}
where $\lambda_{1,2}$ are the respective Lagrange multipliers. If we restrict the path integral to spherically symmetric configurations (i.e. we focus on the {\it spherical PDF}), we can profit from the one-to-one map (\ref{SCmap}) and rewrite the constraints on the non-linear density as the constraints on the linear density,
\begin{equation}
    \delta_D\big(\delta_{*i}-\bar\delta_i[\delta_L]\big)=C_i[\delta_L]\cdot\delta_D\big(F(\delta_{*i})-\bar\delta_L(R_i)\big)\;,
\end{equation}
where $R_i=r_i(1+\delta_{*i})^{1/3}$, $i=1,2$, and the factors
\begin{equation}\label{DW_C}
    C_i[\delta_{L}]=F'(\delta_{*i}) + \frac{1}{1+\delta_{*i}}\big(\bar{\delta}_{L}(R_i) - \delta_{L}(R_i)\big)
\end{equation}
come from the requirement of the preservation of measure. Introducing new Lagrange multipliers we write,
\begin{equation}
\label{DW_PDF2}
\begin{split}
    \mathcal{P}_{sp}(\delta_{*1},\delta_{*2})=\mathcal{N}^{-1} \!\!\!
    \int\limits_{-i\infty}^{i\infty} \!&\frac{d\mu_1d\mu_2}{(2\pi ig^2)^2}\;
    e^{\frac{1}{g^2}\boldsymbol{\mu}^{\rm T}{\bf F}} \!\!\!\!\!\int\limits_{\text{spherical}} \!\!\!\!\!
    \mathcal{D} \delta_{L} \,C_1[\delta_{L}]C_2[\delta_{L}] \\
    &\times
    \exp{-\frac{1}{g^2} \left[\int_{\textbf{k}} \frac{|\delta_{L}(k)|^2}{2P(k)} + \mu_1 \bar\delta_L(R_1) + \mu_2 \bar\delta_L(R_2)  \right]},
\end{split}
\end{equation}
where we have denoted
\begin{equation}
    \boldsymbol{\mu}=\begin{pmatrix}
    \mu_1\\
    \mu_2
    \end{pmatrix}\;,\qquad
    {\bf F}=\begin{pmatrix}
    F(\delta_{*1})\\
    F(\delta_{*2})
    \end{pmatrix}\;.
\end{equation}
Recalling that
\begin{equation}
    \bar\delta_L(R_i)=\int_{\bf k} \delta_L(k) W_{\rm th}(kR_i)\;,
\end{equation}
it is straightforward to evaluate all integrals in the saddle-point approximation. This yields
\begin{equation}
   \mathcal{P}_{sp}(\delta_{*1},\delta_{*2})=\frac{C_{*1}C_{*2}}{2\pi g^2\sqrt{\det\mathbf{\Sigma}}}
   \exp{-\frac{1}{2g^2}{\bf F}^{\rm T}\mathbf{\Sigma}^{-1} {\bf F}}\;,
\end{equation}
where 
\begin{equation}
    \mathbf{\Sigma}= \begin{pmatrix}
\sigma^2_{11} & \sigma^2_{12} \\[6pt]
\sigma^2_{21} & \sigma^2_{22} 
\end{pmatrix}  \;,\qquad 
\sigma^2_{ij}=\int_{\textbf{k}} P(k) W_{\text{th}}(kR_i)
     W_{\text{th}}(kR_j)\;,
\end{equation}
is the covariance matrix and 
\begin{equation}
    C_{*i}=F'(\delta_{*i})+\frac{1}{1+\delta_{*i}}\big(F(\delta_{*i})-\hat\delta_L(R_i)\big)\;. 
\end{equation}
The saddle-point configuration entering in the last expression reads,
\begin{equation}
\label{DW_Rprof}
        \hat{\delta}_L(R)={\bf F}^{\rm T}\mathbf{\Sigma}^{-1}\boldsymbol{\xi}(R)\;,
\end{equation}
with 
\begin{equation}
\boldsymbol{\xi}(R)=\begin{pmatrix}
\xi_1(R)\\
\xi_2(R)
\end{pmatrix}\;,\qquad
    \xi_i(R)=\int_\textbf{k} P(k)W_{\text{th}}(kR_i)j_0(kR)\;. 
\end{equation}

Finally, to obtain the Hockey-Stick PDF, we perform integration over $\delta_{*1}$ and $\delta_{*2}$ subject to the constraint (\ref{DW_dw}),
\begin{equation}
    {\cal P}_{sp}(\delta_*)=\int_{-1}^\infty d\delta_{*1}d\delta_{*2}\;\delta_D(\delta_*-n_1\delta_{*1}-n_2\delta_{*2})\;{\cal P}_{sp}(\delta_{*1},\delta_{*2})\;.
\end{equation}
Here we first take the integral over $\delta_{*2}$ exactly and then integrate over $\delta_{*1}$ in the saddle-point approximation. We arrive at
\begin{equation}
\label{DW_PDF3}
    \mathcal{P}_{sp}(\delta_*)=\frac{C_{*1}C_{*2}}{n_2\sqrt{\pi g^2\det \mathbf{\Sigma}}}\left( \frac{\partial^2 G}{\partial \delta_{*1}^2}\right)^{-1/2}e^{-\frac{1}{2g^2}G(\delta_*,\delta_{*1})}\bigg|_{\delta_{*1}=\hat\delta_{*1}}\;, 
\end{equation}
where 
\begin{equation}
   G(\delta_*,\delta_{*1})\equiv\mathbf{F}^\mathrm{T} \mathbf{\Sigma}^{-1}\mathbf{F}\Big|_{\delta_{*2}=\frac{\delta_*-n_1\delta_{*1}}{n_2}}\;, 
\end{equation}
and $\hat\delta_{*1}$ is determined from the equation
\begin{equation}
\label{DW_sp}
    \frac{\partial G}{\partial\delta_{*1}}\bigg|_{\delta_{*1}=\hat\delta_{*1}}=0\;.
\end{equation}
For given values of $r_*$, $\alpha$, $\beta$ the function $G(\delta_*,\delta_{*1})$ can be easily constructed 
and the solution to Eq.~(\ref{DW_sp}) found 
numerically. The exponent and prefactor in Eq.~(\ref{DW_PDF3}) are then straightforwardly evaluated. Note that while it is possible to obtain an analytic expression for the second derivative $\partial^2G/\partial\delta_{*1}^2$, the resulting expressions are rather long and unwieldy. We find it more efficient to take the second derivative numerically. The PDF we obtain in this way is shown as the curve labeled ``Analytic'' in the left panel of Fig.~\ref{fig:dw_pdfs} from Sec.~\ref{subsec:res_DWell}. In addition, we evaluate the saddle-point density profiles (\ref{DW_Rprof}) for several values of $\delta_*$. These are shown by solid lines in the left panels of Fig.~\ref{fig:dw_prof}.

\bibliographystyle{JHEP}
\bibliography{ref}

\providecommand{\href}[2]{#2}\begingroup\raggedright\begin{thebibliography}{10}

\bibitem{DES:2021wwk}
{\scshape DES} collaboration, \emph{{Dark Energy Survey Year 3 results:
  Cosmological constraints from galaxy clustering and weak lensing}},
  \href{https://doi.org/10.1103/PhysRevD.105.023520}{\emph{Phys. Rev. D}
  {\bfseries 105} (2022) 023520}
  [\href{https://arxiv.org/abs/2105.13549}{{\ttfamily 2105.13549}}].

\bibitem{Dalal:2023olq}
R.~Dalal et~al., \emph{{Hyper Suprime-Cam Year 3 results: Cosmology from cosmic
  shear power spectra}},
  \href{https://doi.org/10.1103/PhysRevD.108.123519}{\emph{Phys. Rev. D}
  {\bfseries 108} (2023) 123519}
  [\href{https://arxiv.org/abs/2304.00701}{{\ttfamily 2304.00701}}].

\bibitem{Wright:2025xka}
A.H.~Wright et~al., \emph{{KiDS-Legacy: Cosmological constraints from cosmic
  shear with the complete Kilo-Degree Survey}},
  \href{https://doi.org/10.1051/0004-6361/202554908}{\emph{Astron. Astrophys.}
  {\bfseries 703} (2025) A158}
  [\href{https://arxiv.org/abs/2503.19441}{{\ttfamily 2503.19441}}].

\bibitem{DESI:2024hhd}
{\scshape DESI} collaboration, \emph{{DESI 2024 VII: cosmological constraints
  from the full-shape modeling of clustering measurements}},
  \href{https://doi.org/10.1088/1475-7516/2025/07/028}{\emph{JCAP} {\bfseries
  07} (2025) 028} [\href{https://arxiv.org/abs/2411.12022}{{\ttfamily
  2411.12022}}].

\bibitem{DESI:2025zgx}
{\scshape DESI} collaboration, \emph{{DESI DR2 results. II. Measurements of
  baryon acoustic oscillations and cosmological constraints}},
  \href{https://doi.org/10.1103/tr6y-kpc6}{\emph{Phys. Rev. D} {\bfseries 112}
  (2025) 083515} [\href{https://arxiv.org/abs/2503.14738}{{\ttfamily
  2503.14738}}].

\bibitem{turner}
M.S.~Turner, \emph{The road to precision cosmology},
  \href{https://doi.org/https://doi.org/10.1146/annurev-nucl-111119-041046}{\emph{Annual
  Review of Nuclear and Particle Science} {\bfseries 72} (2022) 1}
  [\href{https://arxiv.org/abs/2201.04741}{{\ttfamily 2201.04741}}].

\bibitem{baumann}
D.~Baumann, A.~Nicolis, L.~Senatore and M.~Zaldarriaga, \emph{{Cosmological
  Non-Linearities as an Effective Fluid}},
  \href{https://doi.org/10.1088/1475-7516/2012/07/051}{\emph{JCAP} {\bfseries
  07} (2012) 051} [\href{https://arxiv.org/abs/1004.2488}{{\ttfamily
  1004.2488}}].

\bibitem{Carrasco:2012cv}
J.J.M.~Carrasco, M.P.~Hertzberg and L.~Senatore, \emph{{The Effective Field
  Theory of Cosmological Large Scale Structures}},
  \href{https://doi.org/10.1007/JHEP09(2012)082}{\emph{JHEP} {\bfseries 09}
  (2012) 082} [\href{https://arxiv.org/abs/1206.2926}{{\ttfamily 1206.2926}}].

\bibitem{EFT_ivanov}
M.M.~Ivanov, \emph{{Effective Field Theory for Large-Scale Structure}},
  \href{https://arxiv.org/abs/2212.08488}{{\ttfamily 2212.08488}}.

\bibitem{bernardeauSPT}
F.~Bernardeau, S.~Colombi, E.~Gaztanaga and R.~Scoccimarro, \emph{{Large scale
  structure of the universe and cosmological perturbation theory}},
  \href{https://doi.org/10.1016/S0370-1573(02)00135-7}{\emph{Phys. Rept.}
  {\bfseries 367} (2002) 1}
  [\href{https://arxiv.org/abs/astro-ph/0112551}{{\ttfamily
  astro-ph/0112551}}].

\bibitem{snowmass}
G.~Cabass, M.M.~Ivanov, M.~Lewandowski, M.~Mirbabayi and M.~Simonovi{\'c},
  \emph{{Snowmass white paper: Effective field theories in cosmology}},
  \href{https://doi.org/10.1016/j.dark.2023.101193}{\emph{Phys. Dark Univ.}
  {\bfseries 40} (2023) 101193}
  [\href{https://arxiv.org/abs/2203.08232}{{\ttfamily 2203.08232}}].

\bibitem{renorm1}
E.~Pajer and M.~Zaldarriaga, \emph{{On the Renormalization of the Effective
  Field Theory of Large Scale Structures}},
  \href{https://doi.org/10.1088/1475-7516/2013/08/037}{\emph{JCAP} {\bfseries
  08} (2013) 037} [\href{https://arxiv.org/abs/1301.7182}{{\ttfamily
  1301.7182}}].

\bibitem{renorm2}
A.A.~Abolhasani, M.~Mirbabayi and E.~Pajer, \emph{{Systematic Renormalization
  of the Effective Theory of Large Scale Structure}},
  \href{https://doi.org/10.1088/1475-7516/2016/05/063}{\emph{JCAP} {\bfseries
  05} (2016) 063} [\href{https://arxiv.org/abs/1509.07886}{{\ttfamily
  1509.07886}}].

\bibitem{twoloop}
T.~Baldauf, L.~Mercolli and M.~Zaldarriaga, \emph{{Effective field theory of
  large scale structure at two loops: The apparent scale dependence of the
  speed of sound}},
  \href{https://doi.org/10.1103/PhysRevD.92.123007}{\emph{Phys. Rev. D}
  {\bfseries 92} (2015) 123007}
  [\href{https://arxiv.org/abs/1507.02256}{{\ttfamily 1507.02256}}].

\bibitem{Carrasco:2013mua}
J.J.M.~Carrasco, S.~Foreman, D.~Green and L.~Senatore, \emph{{The Effective
  Field Theory of Large Scale Structures at Two Loops}},
  \href{https://doi.org/10.1088/1475-7516/2014/07/057}{\emph{JCAP} {\bfseries
  07} (2014) 057} [\href{https://arxiv.org/abs/1310.0464}{{\ttfamily
  1310.0464}}].

\bibitem{Baldauf:2015aha}
T.~Baldauf, L.~Mercolli and M.~Zaldarriaga, \emph{{Effective field theory of
  large scale structure at two loops: The apparent scale dependence of the
  speed of sound}},
  \href{https://doi.org/10.1103/PhysRevD.92.123007}{\emph{Phys. Rev. D}
  {\bfseries 92} (2015) 123007}
  [\href{https://arxiv.org/abs/1507.02256}{{\ttfamily 1507.02256}}].

\bibitem{Angulo:2014tfa}
R.E.~Angulo, S.~Foreman, M.~Schmittfull and L.~Senatore, \emph{{The One-Loop
  Matter Bispectrum in the Effective Field Theory of Large Scale Structures}},
  \href{https://doi.org/10.1088/1475-7516/2015/10/039}{\emph{JCAP} {\bfseries
  10} (2015) 039} [\href{https://arxiv.org/abs/1406.4143}{{\ttfamily
  1406.4143}}].

\bibitem{Baldauf:2014qfa}
T.~Baldauf, L.~Mercolli, M.~Mirbabayi and E.~Pajer, \emph{{The Bispectrum in
  the Effective Field Theory of Large Scale Structure}},
  \href{https://doi.org/10.1088/1475-7516/2015/05/007}{\emph{JCAP} {\bfseries
  05} (2015) 007} [\href{https://arxiv.org/abs/1406.4135}{{\ttfamily
  1406.4135}}].

\bibitem{Steele:2020tak}
T.~Steele and T.~Baldauf, \emph{{Precise Calibration of the One-Loop Bispectrum
  in the Effective Field Theory of Large Scale Structure}},
  \href{https://doi.org/10.1103/PhysRevD.103.023520}{\emph{Phys. Rev. D}
  {\bfseries 103} (2021) 023520}
  [\href{https://arxiv.org/abs/2009.01200}{{\ttfamily 2009.01200}}].

\bibitem{Bertolini:2016bmt}
D.~Bertolini, K.~Schutz, M.P.~Solon and K.M.~Zurek, \emph{{The Trispectrum in
  the Effective Field Theory of Large Scale Structure}},
  \href{https://doi.org/10.1088/1475-7516/2016/06/052}{\emph{JCAP} {\bfseries
  06} (2016) 052} [\href{https://arxiv.org/abs/1604.01770}{{\ttfamily
  1604.01770}}].

\bibitem{Steele:2021lnz}
T.~Steele and T.~Baldauf, \emph{{Precise Calibration of the One-Loop
  Trispectrum in the Effective Field Theory of Large Scale Structure}},
  \href{https://doi.org/10.1103/PhysRevD.103.103518}{\emph{Phys. Rev. D}
  {\bfseries 103} (2021) 103518}
  [\href{https://arxiv.org/abs/2101.10289}{{\ttfamily 2101.10289}}].

\bibitem{Chudaykin:2025aux}
A.~Chudaykin, M.M.~Ivanov and O.H.E.~Philcox, \emph{{Reanalyzing DESI DR1. I.
  {\ensuremath{\Lambda}}CDM constraints from the power spectrum and
  bispectrum}}, \href{https://doi.org/10.1103/qsnt-dppc}{\emph{Phys. Rev. D}
  {\bfseries 113} (2026) 063502}
  [\href{https://arxiv.org/abs/2507.13433}{{\ttfamily 2507.13433}}].

\bibitem{Chudaykin:2025lww}
A.~Chudaykin, M.M.~Ivanov and O.H.E.~Philcox, \emph{{Reanalyzing DESI DR1. II.
  Constraints on dark energy, spatial curvature, and neutrino masses}},
  \href{https://doi.org/10.1103/42f4-grkj}{\emph{Phys. Rev. D} {\bfseries 113}
  (2026) 123506} [\href{https://arxiv.org/abs/2511.20757}{{\ttfamily
  2511.20757}}].

\bibitem{Chudaykin:2025vdh}
A.~Chudaykin, M.M.~Ivanov and O.H.E.~Philcox, \emph{{Reanalyzing DESI DR1. III.
  Constraints on inflation from galaxy power spectra and bispectra}},
  \href{https://doi.org/10.1103/fhj3-6q4x}{\emph{Phys. Rev. D} {\bfseries 113}
  (2026) 063552} [\href{https://arxiv.org/abs/2512.04266}{{\ttfamily
  2512.04266}}].

\bibitem{Ivanov:2026dvl}
M.M.~Ivanov, J.M.~Sullivan, S.-F.~Chen, A.~Chudaykin, M.~Maus and
  O.H.E.~Philcox, \emph{{Reanalyzing DESI DR1: 4. Percent-Level Cosmological
  Constraints from Combined Probes and Robust Evidence for the Normal Neutrino
  Mass Hierarchy}},  \href{https://arxiv.org/abs/2601.16165}{{\ttfamily
  2601.16165}}.

\bibitem{Chudaykin:2026nls}
A.~Chudaykin, M.M.~Ivanov and O.H.E.~Philcox, \emph{{Reanalyzing DESI DR1: 5.
  Cosmological Constraints with Simulation-Based Priors}},
  \href{https://arxiv.org/abs/2602.18554}{{\ttfamily 2602.18554}}.

\bibitem{NovellMasot:2025fju}
S.~Novell~Masot et~al., \emph{{Full-Shape analysis of the power spectrum and
  bispectrum of DESI DR1 LRG and QSO samples}},
  \href{https://doi.org/10.1088/1475-7516/2025/06/005}{\emph{JCAP} {\bfseries
  06} (2025) 005} [\href{https://arxiv.org/abs/2503.09714}{{\ttfamily
  2503.09714}}].

\bibitem{peebles}
P.J.E.~{Peebles}, \emph{{The Large-Scale Structure of the Universe}} (1981).

\bibitem{ivanov}
M.M.~Ivanov, A.A.~Kaurov and S.~Sibiryakov, \emph{{Non-perturbative probability
  distribution function for cosmological counts in cells}},
  \href{https://doi.org/10.1088/1475-7516/2019/03/009}{\emph{JCAP} {\bfseries
  03} (2019) 009} [\href{https://arxiv.org/abs/1811.07913}{{\ttfamily
  1811.07913}}].

\bibitem{anton}
A.~Chudaykin, M.M.~Ivanov and S.~Sibiryakov, \emph{{Renormalizing one-point
  probability distribution function for cosmological counts in cells}},
  \href{https://doi.org/10.1088/1475-7516/2023/08/079}{\emph{JCAP} {\bfseries
  08} (2023) 079} [\href{https://arxiv.org/abs/2212.09799}{{\ttfamily
  2212.09799}}].

\bibitem{Valageas:2001zr}
P.~Valageas, \emph{{Dynamics of gravitational clustering. 2. Steepest-descent
  method for the quasi-linear regime}},
  \href{https://doi.org/10.1051/0004-6361:20011663}{\emph{Astron. Astrophys.}
  {\bfseries 382} (2002) 412}
  [\href{https://arxiv.org/abs/astro-ph/0107126}{{\ttfamily
  astro-ph/0107126}}].

\bibitem{Valageas:2001td}
P.~Valageas, \emph{{Dynamics of gravitational clustering v. subleading
  corrections in the quasi-linear regime}},
  \href{https://doi.org/10.1051/0004-6361:20011584}{\emph{Astron. Astrophys.}
  {\bfseries 382} (2002) 477}
  [\href{https://arxiv.org/abs/astro-ph/0109408}{{\ttfamily
  astro-ph/0109408}}].

\bibitem{Uhlemann:2015npz}
C.~Uhlemann, S.~Codis, C.~Pichon, F.~Bernardeau and P.~Reimberg, \emph{{Back in
  the saddle: Large-deviation statistics of the cosmic log-density field}},
  \href{https://doi.org/10.1093/mnras/stw1074}{\emph{Mon. Not. Roy. Astron.
  Soc.} {\bfseries 460} (2016) 1529}
  [\href{https://arxiv.org/abs/1512.05793}{{\ttfamily 1512.05793}}].

\bibitem{Uhlemann:2019gni}
C.~Uhlemann, O.~Friedrich, F.~Villaescusa-Navarro, A.~Banerjee and S.~Codis,
  \emph{{Fisher for complements: Extracting cosmology and neutrino mass from
  the counts-in-cells PDF}},
  \href{https://doi.org/10.1093/mnras/staa1155}{\emph{Mon. Not. Roy. Astron.
  Soc.} {\bfseries 495} (2020) 4006}
  [\href{https://arxiv.org/abs/1911.11158}{{\ttfamily 1911.11158}}].

\bibitem{Boyle:2020bqn}
A.~Boyle, C.~Uhlemann, O.~Friedrich, A.~Barthelemy, S.~Codis, F.~Bernardeau
  et~al., \emph{{Nuw CDM cosmology from the weak-lensing convergence PDF}},
  \href{https://doi.org/10.1093/mnras/stab1381}{\emph{Mon. Not. Roy. Astron.
  Soc.} {\bfseries 505} (2021) 2886}
  [\href{https://arxiv.org/abs/2012.07771}{{\ttfamily 2012.07771}}].

\bibitem{Friedrich:2021xff}
O.~Friedrich, A.~Halder, A.~Boyle, C.~Uhlemann, D.~Britt, S.~Codis et~al.,
  \emph{{The PDF perspective on the tracer-matter connection: Lagrangian bias
  and non-Poissonian shot noise}},
  \href{https://doi.org/10.1093/mnras/stab3703}{\emph{Mon. Not. Roy. Astron.
  Soc.} {\bfseries 510} (2022) 5069}
  [\href{https://arxiv.org/abs/2107.02300}{{\ttfamily 2107.02300}}].

\bibitem{generalW}
F.~Bernardeau and P.~Reimberg, \emph{{Large deviation principle at play in
  large scale structure cosmology}},
  \href{https://doi.org/10.1103/PhysRevD.94.063520}{\emph{Phys. Rev. D}
  {\bfseries 94} (2016) 063520}
  [\href{https://arxiv.org/abs/1511.08641}{{\ttfamily 1511.08641}}].

\bibitem{bond}
J.R.~Bond, S.~Cole, G.~Efstathiou and N.~Kaiser, \emph{{Excursion set mass
  functions for hierarchical Gaussian fluctuations}},
  \href{https://doi.org/10.1086/170520}{\emph{Astrophys. J.} {\bfseries 379}
  (1991) 440}.

\bibitem{zentner}
A.R.~Zentner, \emph{{The Excursion Set Theory of Halo Mass Functions, Halo
  Clustering, and Halo Growth}},
  \href{https://doi.org/10.1142/S0218271807010511}{\emph{Int. J. Mod. Phys. D}
  {\bfseries 16} (2007) 763}
  [\href{https://arxiv.org/abs/astro-ph/0611454}{{\ttfamily
  astro-ph/0611454}}].

\bibitem{halo_gauss}
T.~Baldauf, S.~Codis, V.~Desjacques and C.~Pichon, \emph{{Nonperturbative halo
  clustering from cosmological density peaks}},
  \href{https://doi.org/10.1103/PhysRevD.103.083530}{\emph{Phys. Rev. D}
  {\bfseries 103} (2021) 083530}
  [\href{https://arxiv.org/abs/2012.14404}{{\ttfamily 2012.14404}}].

\bibitem{BH_mass}
K.~Tokeshi, K.~Inomata and J.~Yokoyama, \emph{{Window function dependence of
  the novel mass function of primordial black holes}},
  \href{https://doi.org/10.1088/1475-7516/2020/12/038}{\emph{JCAP} {\bfseries
  12} (2020) 038} [\href{https://arxiv.org/abs/2005.07153}{{\ttfamily
  2005.07153}}].

\bibitem{galaxy_window}
T.~Karim, M.~Rezaie, S.~Singh and D.~Eisenstein, \emph{{On the impact of the
  galaxy window function on cosmological parameter estimation}},
  \href{https://doi.org/10.1093/mnras/stad2210}{\emph{Mon. Not. Roy. Astron.
  Soc.} {\bfseries 525} (2023) 311}
  [\href{https://arxiv.org/abs/2305.11956}{{\ttfamily 2305.11956}}].

\bibitem{Bernardeau:1992zw}
F.~Bernardeau, \emph{{The Gravity induced quasi-Gaussian correlation
  hierarchy}}, \href{https://doi.org/10.1086/171398}{\emph{Astrophys. J.}
  {\bfseries 392} (1992) 1}.

\bibitem{Valageas:1998xr}
P.~Valageas, \emph{{Structure formation: A Spherical model for the evolution of
  the density distribution}}, {\emph{Astron. Astrophys.} {\bfseries 337} (1998)
  655} [\href{https://arxiv.org/abs/astro-ph/9807033}{{\ttfamily
  astro-ph/9807033}}].

\bibitem{Matarrese:2000iz}
S.~Matarrese, L.~Verde and R.~Jimenez, \emph{{The Abundance of high-redshift
  objects as a probe of non-Gaussian initial conditions}},
  \href{https://doi.org/10.1086/309412}{\emph{Astrophys. J.} {\bfseries 541}
  (2000) 10} [\href{https://arxiv.org/abs/astro-ph/0001366}{{\ttfamily
  astro-ph/0001366}}].

\bibitem{CLASS}
D.~Blas, J.~Lesgourgues and T.~Tram, \emph{{The Cosmic Linear Anisotropy
  Solving System (CLASS) II: Approximation schemes}},
  \href{https://doi.org/10.1088/1475-7516/2011/07/034}{\emph{JCAP} {\bfseries
  07} (2011) 034} [\href{https://arxiv.org/abs/1104.2933}{{\ttfamily
  1104.2933}}].

\bibitem{quijote}
F.~Villaescusa-Navarro et~al., \emph{{The Quijote simulations}},
  \href{https://doi.org/10.3847/1538-4365/ab9d82}{\emph{Astrophys. J. Suppl.}
  {\bfseries 250} (2020) 2} [\href{https://arxiv.org/abs/1909.05273}{{\ttfamily
  1909.05273}}].

\bibitem{Contarini:2026yfv}
S.~Contarini, G.~Verza and A.~Pisani, \emph{{The era of precision cosmology
  with voids}}, \href{https://doi.org/10.1007/s00159-026-00166-x}{\emph{Astron.
  Astrophys. Rev.} {\bfseries 34} (2026) 1}
  [\href{https://arxiv.org/abs/2601.14362}{{\ttfamily 2601.14362}}].

\bibitem{Chudaykin:2020aoj}
A.~Chudaykin, M.M.~Ivanov, O.H.E.~Philcox and M.~Simonovi{\'c},
  \emph{{Nonlinear perturbation theory extension of the Boltzmann code CLASS}},
  \href{https://doi.org/10.1103/PhysRevD.102.063533}{\emph{Phys. Rev. D}
  {\bfseries 102} (2020) 063533}
  [\href{https://arxiv.org/abs/2004.10607}{{\ttfamily 2004.10607}}].

\bibitem{Pylians}
F.~{Villaescusa-Navarro}, ``{Pylians: Python libraries for the analysis of
  numerical simulations}.''
  \href{https://github.com/franciscovillaescusa/Pylians}{https://github.com/franciscovillaescusa/Pylians},
  2018.

\bibitem{jax}
J.~Bradbury, R.~Frostig, P.~Hawkins, M.J.~Johnson, C.~Leary, D.~Maclaurin
  et~al., ``{JAX}: composable transformations of {P}ython+{N}um{P}y programs.''
  \href{http://github.com/jax-ml/jax}{http://github.com/jax-ml/jax}, 2018.

\bibitem{voids}
R.~Voivodic, H.~Rubira and M.~Lima, \emph{{The Halo Void (Dust) Model of Large
  Scale Structure}},
  \href{https://doi.org/10.1088/1475-7516/2020/10/033}{\emph{JCAP} {\bfseries
  10} (2020) 033} [\href{https://arxiv.org/abs/2003.06411}{{\ttfamily
  2003.06411}}].

\bibitem{DM_void}
S.~Arcari, E.~Pinetti and N.~Fornengo, \emph{{Got plenty of nothing: cosmic
  voids as a probe of particle dark matter}},
  \href{https://doi.org/10.1088/1475-7516/2022/11/011}{\emph{JCAP} {\bfseries
  11} (2022) 011} [\href{https://arxiv.org/abs/2205.03360}{{\ttfamily
  2205.03360}}].

\bibitem{Baldauf:2016sjb}
T.~Baldauf, M.~Mirbabayi, M.~Simonovi{\'c} and M.~Zaldarriaga, \emph{{LSS
  constraints with controlled theoretical uncertainties}},
  \href{https://arxiv.org/abs/1602.00674}{{\ttfamily 1602.00674}}.

\bibitem{desjacques}
V.~Desjacques, D.~Jeong and F.~Schmidt, \emph{{Large-Scale Galaxy Bias}},
  \href{https://doi.org/10.1016/j.physrep.2017.12.002}{\emph{Phys. Rept.}
  {\bfseries 733} (2018) 1} [\href{https://arxiv.org/abs/1611.09787}{{\ttfamily
  1611.09787}}].

\bibitem{kokron}
N.~Kokron, J.~DeRose, S.-F.~Chen, M.~White and R.H.~Wechsler, \emph{{Priors on
  red galaxy stochasticity from hybrid effective field theory}},
  \href{https://doi.org/10.1093/mnras/stac1420}{\emph{Mon. Not. Roy. Astron.
  Soc.} {\bfseries 514} (2022) 2198}
  [\href{https://arxiv.org/abs/2112.00012}{{\ttfamily 2112.00012}}].

\bibitem{cylinders}
C.~Uhlemann, C.~Pichon, S.~Codis, B.~L'Huillier, J.~Kim, F.~Bernardeau et~al.,
  \emph{{Cylinders out of a top hat: counts-in-cells for projected densities}},
  \href{https://doi.org/10.1093/mnras/sty664}{\emph{Mon. Not. Roy. Astron.
  Soc.} {\bfseries 477} (2018) 2772}
  [\href{https://arxiv.org/abs/1711.04767}{{\ttfamily 1711.04767}}].

\bibitem{Barthelemy:2023mer}
A.~Barthelemy, A.~Halder, Z.~Gong and C.~Uhlemann, \emph{{Making the leap.
  Part~I. Modelling the reconstructed lensing convergence PDF from cosmic shear
  with survey masks and systematics}},
  \href{https://doi.org/10.1088/1475-7516/2024/03/060}{\emph{JCAP} {\bfseries
  03} (2024) 060} [\href{https://arxiv.org/abs/2307.09468}{{\ttfamily
  2307.09468}}].

\bibitem{knn}
A.~Banerjee and T.~Abel, \emph{{Nearest neighbour distributions: New
  statistical measures for cosmological clustering}},
  \href{https://doi.org/10.1093/mnras/staa3604}{\emph{Mon. Not. Roy. Astron.
  Soc.} {\bfseries 500} (2020) 5479}
  [\href{https://arxiv.org/abs/2007.13342}{{\ttfamily 2007.13342}}].

\end{thebibliography}\endgroup

\end{document}